\documentclass[aps,prd,twocolumn,amsmath,groupedaddress,nofootinbib]{revtex4-1}   

\pdfoutput=1                                   
\RequirePackage[colorlinks=true            
,urlcolor=blue
,anchorcolor=blue
,citecolor=blue
,filecolor=blue
,linkcolor=blue
,menucolor=blue
,pagecolor=blue
,linktocpage=true
,pdfproducer=medialab
,pdfa=true
]{hyperref}	

\usepackage{graphicx}	%
\usepackage{amssymb}	
\usepackage{bm}			
\usepackage{dcolumn}	
\usepackage{alltt}	    

\newcommand{\sct}[1]{Sec.~\ref{#1}}          
\newcommand{\Sct}[1]{Sec.~\ref{#1}}          
\newcommand{\scts}[1]{Secs.~\ref{#1}}       
\newcommand{\Scts}[1]{Secs.~\ref{#1}}       
\newcommand{\fig}[1]{Fig.~\ref{#1}}           
\newcommand{\Fig}[1]{Fig.~\ref{#1}}           
\newcommand{\figs}[1]{Figs.~\ref{#1}}        
\newcommand{\apx}{Appendix}                  
\newcommand{\apxs}{Appendices}             

\newcommand{\eqva}{\!&\equiv&\!}      
\newcommand{\eqa}{\!\!&=&\!\!}           

\newcommand{\be}{\begin{eqnarray}}           \newcommand{\ee}{\end{eqnarray}}   
\newcommand{\ba}[1]{\begin{array}{#1} \vph}  \newcommand{\ea}{\vph \end{array}} 
\newcommand{\bs}{\begin{subequations}}  \newcommand{\es}{\end{subequations}} 
\newcommand{\ben}{\begin{enumerate}}	\newcommand{\een}{\end{enumerate}} 
\newcommand{\itm}{\item}
\newcommand{\rf}[1]{~(\ref{#1})}  	
\newcommand{\rfd}[2]{~(\ref{#1},\,\ref{#2})}
\newcommand{\rfs}[2]{~(\ref{#1}--\ref{#2})}
\newcommand{\rft}[3]{~(\ref{#1},\,\ref{#2},\,\ref{#3})}

\newcommand{\ct}[1]{~\cite{#1}} 
\newcommand{\lb}[1]{\label{#1}}  
\newcommand{\nn}{\nonumber}
\newcommand{\lf}{\left}     \newcommand{\rt}{\right}

\newcommand{\fr}{\frac}     

\newcommand{\nbrk}[1]{\mbox{$#1$}}

\newcommand{\pd}{\partial}   \newcommand{\D}{{\mathcal{D}}}  
\newcommand{\R}{R}    
         \newcommand{\g}{\gamma}

\newcommand{\Rset}{\mathbb{R}}	\newcommand{\Cset}{\mathbb{C}}
\newcommand{\dist}{\mathcal{V}}
\newcommand{\Nobj}{{\cal N}}
\newcommand\cL{{\cal L}}    \newcommand\cH{{\cal H}}
\newcommand\Op{{\mathcal O}}
\newcommand\W{{\mathcal W}}
 	\newcommand\veps{\varepsilon}
\newcommand\xv{{\bm x}}    	\newcommand\yv{{\bm y}}      
\newcommand\kv{{\bm k}}		   
\newcommand\nv{{\bm n}} 	\newcommand\mv{{\bm m}}   
\newcommand\Nv{{\bm N}}
\newcommand\Av{{\bm A}}
\newcommand\lapse{N} 	 	\newcommand\shift{N}    
\newcommand\var{\xi} 	 
\newcommand{\Hub}{H_{\rm Hub}} 
\newcommand{\cl}{_{\rm c}}
\newcommand{\prob}{P}
\newcommand{\gaus}{_{\rm gaus}}
\newcommand{\ext}{{\rm ext}}
\newcommand\nf{{\tt n}}
\newcommand\acom[1]{\{#1\}}

\newcommand\pert[1]{\bm\tilde{\unboldmath #1}}

\newcommand{\dbar}{\,\mathchar'26\mkern-13mu d}
\newcommand{\deltabar}{\mathchar'26\mkern-10mu \delta}

\newcommand\pdot{{\cdot}}

\newcommand\vph{\vphantom{\fr{\hat I}{\hat I}}}

\newcommand\diff{\stackrel{\rm diff}{\to}}

\newcommand{\const}{{\rm const}}

\newcommand{\trace}{\mathop{\rm tr}}
\newcommand{\re}{\mathop{\rm Re}}

\begin{document}   
\title{\Large             
            Realistic quantum fields with gauge and gravitational  \\   
            interaction emerge in the generic static structure}

\author{Sergei Bashinsky}     
\affiliation{Moscow, Russia}
\homepage{http://bashinsky.edicypages.com}

\date{released April 7, 2015;~ last revised August 16, 2018}   

\begin{abstract}    
We describe how physical universes that are composed of
gauge and gravitationally interacting bosonic and fermionic quantum fields arise from the generic discrete distribution of many quantifiable properties of arbitrary static entities.  Alternate presentations of the smooth coarse-graining~(fit) for this discrete distribution compose probability-related evolving wave function of the fields' dynamical modes. Their gauge modes, being symmetry transformations, and constrained modes require no additional material structure. We prove that evolution of any origin for which the quantum superposition principle is absolute cannot be governed by specific laws. In contrast, locally supersymmetric quantum fields that emerge as described from the basic \emph{discrete} distribution evolve by unchanging and closed physical laws. The emergent quantum evolution is many-world; yet its Everett's branches whose norm diminishes below a positive threshold cease to exist. Then some experiments that for the standard Everett view would seem safe are instead fatal for the participants. The Born rule arises dynamically in emergent systems with extended regular past. It and, consequently, quasi-deterministic macroscopic evolution emerge in systems that allow cosmological inflation but not in typical random ones. This resolves the Boltzmann brain problem. We explain how inflation creates new physical degrees of freedom around the Planck scale. Quantum entanglement for the emergent fields is trivial because their wave function, up to its representation, is material.
\end{abstract}   

\maketitle
\tableofcontents   

\section{Introduction and sketch of the results}
\label{sec_intro}

\subsection{Fundamental questions resolved}
\label{subsec_questions}

Today quantum mechanics---including quantum field theory as its relativistic generalization---enjoys the status of one of the most impeccable and yet the most counterintuitive theory of the natural world. It has flawlessly passed a century of accurate experimental tests. So deeply has it permeated the contemporary physical picture that in the search for the ultimate ``theory of everything'' some physicists regard its principles as fundamental postulates. Popular approaches to unifying the Standard Model of particle physics and general relativity (e.g., by string theory, supergravity, or loop quantum gravity) usually impose them axiomatically.

Yet ever since its inception, quantum mechanics has seemed to defy logic. To this day,  broad disagreement persists even on its consistent formulation. Many of its various ``interpretations'' are not alternative formulations of the same theory but are nonequivalent physical theories. They imply different fundamental organization of nature and different results for certain, in principle, plausible experiments.  The mechanism for wave function collapse, origin of the probability for predicted alternate outcomes and their physical status, or quantum-mechanical description of gravity remain debated.

Everett's Ph.D.~thesis\ct{Everett} partly demystified quantum mechanics. It demonstrated that the standard Copenhagen interpretation can abandon the awkward postulate of probabilistic (and, by Bell's theorem\ct{Bell_theorem}, nonlocal and superluminal) collapse of a wave function during measurement. In the Everett's picture, the perceived collapse of the wave function is the natural decoherence of its separate terms (\emph{branches})\ct{Zeh_70,Zurek_81,Joos_Zeh_84,Zurek_91,Zurek_03}.
Nonetheless, the assignment of probabilities to measurement outcomes in~it remains a postulate. The Born rule cannot be derived only from the other standard postulates of quantum mechanics.\footnote{
   \label{ftn_Gleason}
    Indeed, this paper will describe scenarios where the Born rule fails but the other postulates of quantum mechanics hold. (E.g., see the end of \sct{subsec_Born_rule}.) This does not contradict Gleason's theorem\ct{Gleason_theorem}, proving that the Born rule is the only option for a probability measure on a Hilbert space, because then probability is not a measure on the Hilbert space in the sense required by\ct{Gleason_theorem}. Namely, then the probability is defined for the alternate outcomes of a quantum process but not for \emph{any} set of orthogonal subspaces of the Hilbert space.
} 
The world's consequent splitting into numerous co-existing branches of ``alternate realities" is also worrisome or objectionable to many people. Different approaches continue to be explored\ct{Bohm_51a,Bohm_51b,consistent_histories_84,Ghirardi_Rimini_Weber_86,TIQM_86,tHooft_90,Penrose_98,Adler_02,QBism_10,tHooft_cell_autom_interp_14}.

The origin of many other fundamental physical principles has likewise remained unclear. 
The fundamental questions addressed by the presented work include:
\ben
\itm 
Why does nature obey physical laws? Why does it obey the laws that we observe\ct{Tegmark_mathematical_universe,Linde_foundational_conference,tHooft_superdet_chall_17}?
Why do the core numerical parameters of the Standard Model of particle physics remain constant\ct{Uzan_couplings_variation_02_review} in time and space?
\itm 
What is the dynamics on the short scales (e.g., the Planck scale) where the Standard Model and canonically quantized general relativity\ct{DeWitt} break down?
\itm 
Why do time and space unify in the classical relativistic description of nature but they conceptually differ in quantum description even of relativistic systems? 
\itm
Why is the physical dynamics observed to be local in spacetime?
\itm 
Do the alternate Everett worlds exist?
If yes, why do we find ourselves in a given branch with probability proportional to the squared norm of the branch's wave function\ct{Deutsch_probability_decisions99,Barnum_etal_99,Saunders_02,Wallace_02, Weinberg_11,Carroll_Sebens_14}?
\itm 
Do the quantum-mechanical postulates and the physical Lagrangian follow from deeper, more elementary and natural principles?
\itm 
Why is our observable universe consistent with inflationary past, rather than is one of the more generic anthropically suitable configurations of matter and metric fields\ct{Penrose_Difficulties_88,Hollands_Wald_02,Carroll_Tam_10,Steinhardt_infl_Sc_Am,Carroll_Fine-Tuning_14}, including ``Boltzmann brain'' worlds\ct{Dyson_Kleban_Sussk_02,BB_DeSimone_etal_08,BB_Carroll_17}?
\itm 
Does and, if yes, how does eternal cosmological inflation\ct{Guth_inflation,Linde_eternal} create new physical degrees of freedom? Can inflation ultimately ``run out'' of the new small-scale degrees of freedom? 
\itm 
What is the precise initial state of the short-scale modes during inflation\ct{vac_choices_infl_03}?
\itm 
How to reconcile unitarity of quantum evolution with apparent loss of information during the Hawking evaporation of a black hole\ct{Hawking_Breakdown_of_Predictability_76,Preskill_92}? What would an observer who falls into a black hole see at its horizon\ct{firewall,Mathur_09,Braunstein_09}? What happens at its center and at the end of the evaporation?
\een

Below and in a companion paper\ct{my_bh}, studying black holes, we suggest a surprisingly simple and natural resolution of all these questions. We show that elementary and generic static structures give rise to materially existing evolving wave functions of phenomenologically viable quantum-field worlds. The Born rule indeed relates these emergent wave functions to the probability for their internal inhabitants to observe various branches of their quantum evolution. These worlds are bound to evolve by dynamical laws and from initial conditions that resemble and for some of the worlds apparently coincide with those that govern our universe. The generality and omnipresence of the basic constituents for the emergent quantum worlds suggest that our physical environment has similar origin. For the emergent physical worlds, possibly including our observed universe, we answer the listed above and several other important fundamental questions. 

\subsection{Sketch of the results}

This subsection is an overview of the entire natural picture developed in the paper. The reader should remember that this part is only a qualitative sketch of the results obtained systematically and described rigorously in the main sections.

\subsubsection{Emergent wave function}
\label{subsubsec_basic}

Consider a light field during cosmological inflation. Its modes of frequency much higher than the Hubble rate of the inflationary expansion are almost in the ground state. (Otherwise, pressure of the modes' excitations would prevent inflation: eqs.\rfs{inflation_rad_pressure}{inflation_necessary_condition} and the text between them.) The wave function of the ground state for a free field is almost Gaussian: for its $N$~modes
\be
\psi(q)\approx \prod_{m=1}^N \lf(\pi^{-\fr14}\,e^{-\fr12\,q^2_m}\rt)
= \pi^{-\fr{N}4}\,e^{-\fr12\sum_m q^2_m}
\lb{psi_0_intro}
\ee
where~$q_m$ are the appropriately normalized amplitudes of the modes. The physically viable emergent fields studied in the main sections are interacting. Their ground state is hence not expected to resemble the Gaussian form\rf{psi_0_intro}. Yet let us continue the illustration by assuming negligible interaction and the modes' initial wave function\rf{psi_0_intro}.

\begin{figure*}[t]
\centering    
\includegraphics[width=1.0\textwidth]{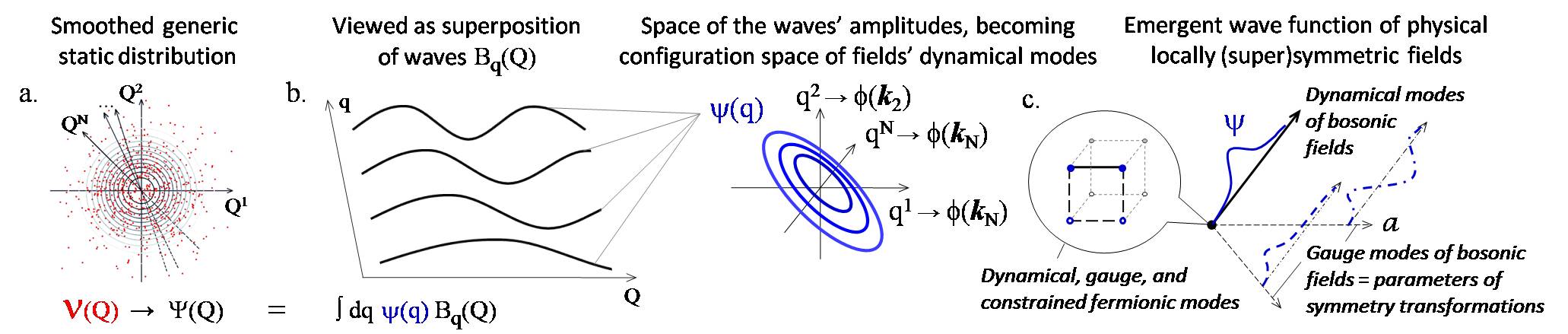}    
\caption{(a).~The discrete distribution~$\dist(Q)$ of the generic quantifiable properties $\nbrk{Q \equiv (Q^1,\dots, Q^N)}$ at a finite resolution can be approximated by a smooth function~$\Psi(Q)$. (b).~View this~$\Psi(Q)$ as a superposition of smooth basis functions~$B_q(Q)$ multiplied by coefficients~$\psi(q)$. (c).~The wave function of the emergent physical fields is a continuous set of the alternative, equivalent presentations~$\psi(q)$ of the smoothed basic distribution~$\Psi(Q)$. I.e., the fields' evolving wave function is composed of $\{\psi(q)\}$~for a continuous set of choices of the basis~$\{B_q(Q)\}$. The appropriately rescaled independent arguments $\nbrk{(q^1,\dots,q^N)\equiv q}$ of~$\psi$ are identified with the amplitudes $(\phi(\kv_1),\dots,\phi(\kv_N))$ of the fields' \emph{dynamical} sub-Planckian modes. Then the local field operators~$\hat\phi^\iota(x)$ for the elementary particles of various types~$\iota$ are linear combinations of the operators~$\hat q^{\,m}$, where $\nbrk{\hat q^m \psi(q)\equiv q^m \psi(q)}$ [e.g.\rf{phi_standing_harmonics}]. (Continued in \fig{fig_summary}.)
}
\label{fig_outline}
\end{figure*}

Are there elementary structures that are described by eq.\rf{psi_0_intro} and that can \emph{materially represent}\ct{PBR_theorem} this wave function, unchanged in the Heisenberg picture of quantum mechanics? The Gaussian dependence\rf{psi_0_intro} arises in almost any collection of many objects, regardless of what the word ``object'' stands for. It is sufficient that the objects or their groups possess numerous independent quantifiable properties.
Of course, properties~$Q^p$, $\nbrk{p=1,2\dots}$, of many familiar physical or mathematical objects have non-Gaussian distribution. Nevertheless, general linear combinations of many independent properties, $\nbrk{Q^n = \sum_p c^n_p\, Q^p}$, are distributed by the universal Gaussian law, at least, under the conditions of the central limit theorem of probability theory. Thus $N$~typical uncorrelated and appropriately normalized ``generic properties''~$Q^n$ are indeed distributed by\rf{psi_0_intro}.

Consider a large but finite collection of objects~$\{a\}$. Select randomly $N$~their generic properties $\nbrk{(Q^1,\dots, Q^N)\equiv Q}$. The distribution density~$\dist(Q)$ of the selected $N$~properties is a sum of Dirac's delta functions:
\be
\dist(Q)= \sum_a\delta^{(N)}(Q-Q_a)\,.
\lb{distrib_discrete_intro}
\ee  
Here $\nbrk{Q_a\equiv(Q^1_a,\dots, Q^N_a)}$ are the values that the $N$~selected properties take for the $a$'th object. At a sufficiently coarse resolution~$\Delta Q$ the discrete distribution~$\dist(Q)$ appears as a smooth distribution~$\Psi(Q)$ that is compatible with the Gaussian form. Expand the smooth fit~$\Psi(Q)$ into linear combinations of linearly independent, smooth basis functions~$B_q(Q)$:
\be
\Psi(Q)= \int\! dq~\psi(q)\,B_q(Q)\,.
\lb{psi_general_intro}
\ee
\Fig{fig_outline} illustrates this, with more details in \sct{sec_structure} and its \fig{fig_1d}.
The linear space of linear combinations of~$B_q(Q)$ will become the Hilbert space of quantum states (\sct{sec_evolution}). The complex structure of the Hilbert space and its unique Hermitian product that is related to \emph{objective}\footnote{
\lb{footnote_BR_is_objective}
	  Refs.\ct{Deutsch_probability_decisions99,Barnum_etal_99,Saunders_02,Wallace_02} suggest that probability may be \emph{subjective}. 
	  They propose that the branches where the Born rule for outcomes of repeated experiments is strongly violated develop on a par with the branches abiding by it. They then propose to account for only the branches where repeated experiments are consistent with the Born rule because only these branches contain ``rational'' internal observers. However, for example, our lives would proceed normally even when all the Stern-Gerlach experiments returned substantial disagreement with the Born rule as long as the rest of the world, save the Stern-Gerlach experiments in every laboratory, runs as usual. This and similar examples show that the proposal of  Refs.\ct{Deutsch_probability_decisions99,Barnum_etal_99,Saunders_02,Wallace_02} contradicts to our observations.
	  We find below that in the emergent systems the probability of following various branches of quantum evolution is objective and does not depend on the presence of an observer.
}
probability in the emergent physical world will appear as explained in \sct{sec_evolution}. 

The possibilities for continuous linear transformation  $\nbrk{\Psi(Q)\to\psi(q)}$ for various sets of smooth basis functions~$B_q(Q)$ in\rf{psi_general_intro} seemingly vastly outnumber the transformations of quantum evolution with a specific Hamiltonian or specific Lagrangian density. Yet the identical concern applies to the standard axiomatic quantum mechanics. \Sct{sec_first_observation} will prove that if the quantum superposition principle is an absolute law then a state~$\psi$ can evolve into a state $\nbrk{\psi'=\exp(-i\int \hat H'dt)\,\psi}$ with any Hamiltonian~$\hat H'(t)$ that continuously transforms the system's pointer states\ct{Zurek_81,Zurek_03}, stable to decoherence, into other pointer states. Then a typical internal observer of a quantum field system would see that the particle masses and couplings change arbitrarily and inherently unpredictably (\sct{sec_first_observation}). 

The observer would then perceive the dynamical laws changing unpredictably over the shortest conceivable timescale. Subsystems of such a world cannot develop into internal intelligent observers by natural biological evolution. To prove it, let us define intelligence broadly as the ability to predict future consequences of given conditions. Only by successfully anticipating future outcomes, an organism can choose an action that will benefit it, whether in distant or immediate future. If the laws of dynamics changed unpredictably on all scales, predicting future outcomes would be impossible. In particular, adaption to the current environment would not help a prospective biological subsystem and its replicas to survive the new, unpredictably changed fitness criteria. The random change of dynamical laws thus disallows biological evolution and evolutionary development of intelligent observers, even those that might differ drastically from the familiar chemistry-based life.

Why nonetheless, as far as experimental data\ct{Uzan_couplings_variation_02_review} and our daily experience indicate, do the fundamental physical laws remain constant in time and space? A proposition:
``Because only this dynamics has produced intelligent life," is \emph{unsatisfactory}. Even if the anthropic principle uniquely restricted physical evolution since the beginning of the Big Bang up to, e.g., development of the human civilization, that alone would fail to explain why the experiments \emph{continue} to find the core physical laws inflexible. Why do ``fundamental" dimensionless physical constants not vary even within limits safe for our continuing existence? 

\subsubsection{Origin of physical laws}
\lb{subsubsec_laws_emergence}

A key to understanding why nature evolves by unchanged and, incidentally, highly symmetric dynamical laws may be the following. Many of the known \emph{local} symmetries apply not only to the action, specifying the dynamics, but also to the wave function of the system's instantaneous state. This has been long recognized in canonical quantum gravity\ct{DeWitt} but rarely emphasized in elementary particle physics, where calculations are typically performed in a fixed gauge and for fields' correlation functions rather than the wave function. We will however find that the symmetries of the wave function, not only those of the action, are crucial for fundamental understanding of the physical world. 

The number of independent degrees of freedom for locally symmetric fields is enormously smaller than for similar fields without the symmetry. Neither gauge nor constrained modes of locally symmetric fields are their physical degrees of freedom. The amplitudes of the gauge or constrained modes are not independent dimensions of the system's wave function. The wave function of the system with local symmetry can hence arise from a basic structure of considerably smaller dimensionality than that necessary to produce a wave function of superficially similar non-symmetric fields. 

The second key ingredient for evolution of the emergent fields proceeding by unchanged physical laws is the \emph{discreteness} of the distribution~$\dist(Q)$ that underlies their continuous wave function. Transformation of the wave function that breaks its inherent symmetries increases suddenly and substantially the number of the fields' dynamical modes and, hence, the dimensionality of the configuration space. The resulting continuous wave function of substantially larger dimensionality cannot be represented by the same fundamental discrete distribution~$\dist(Q)$ (\sct{subsec_locality}). Thus the discreteness of the underlying basic structure~$\dist(Q)$ preserves the inherent symmetry of the emergent wave function during evolution.

The third ingredient is the requirement that the inherent local symmetry of the wave function of the fields uniquely relates their current state to  their dynamics. This applies, e.g., to local supersymmetry.   Details of how a locally supersymmetric wave function at a fixed time encodes the dynamical laws are described in \sct{subsec_fixing}. Other symmetries that fix quantum dynamics might as well be contemplated. 
Akin to local supersymmetry, they require physical fields beyond the Standard Model or its typical extensions for Grand Unification. Their analysis is postponed to future work.  \Scts{sec_first_observation},~\ref{sec_gauge_fields}, and~\ref{sec_gravity} show that pure gauge- and diffeomorphism-symmetric quantum field theories, whether emergent or fundamental, cannot be logically consistent complete description of nature.

A discrete structure behind the continuous wave function of physical fields
limits the number of their independent dynamical variables. It thus protects the local symmetry that fixes the dynamical laws. There is another compelling motivation to expect the physical wave function to originate from a discrete structure: this naturally explains the existence of objective probabilities of quantum outcomes that are governed by the Born rule. Introductory part~\ref{subsubsec_branches} will outline how the probabilities and the Born rule arise. This will be described rigorously in \scts{sec_structure} and~\ref{sec_probabilities}.

\subsubsection{Mapping the basic structure to quantum fields, spacetime, and physical objects}

Let us review how the generic static distribution~$\dist(Q)$ of\rf{distrib_discrete_intro} represents the standard elements of quantum field theory and the physical objects of an emergent world. Their rigorous identification will follow in the main sections.

Decompose the studied matter and metric fields into nonlocal modes, e.g., the harmonic modes. Assign every mode to one of the three classes:\\ 
(a).~\emph{Dynamical modes}. They permit independent initial conditions and evolve by specific equations of motion. Their amplitudes are in one-to-one correspondence with physically distinct field configurations.\\  
(b).~\emph{Gauge modes}. Their amplitudes can be set by symmetry transformation to any value without affecting the physical observables.\\  
(c).~\emph{Constrained modes}. Their amplitudes are determined by dynamical and gauge modes via constraint equations.

For example, the electromagnetic abelian gauge field $A_\mu(x)$ has for every wavevector~$\kv$: two dynamical modes (transverse photons with $\nbrk{\kv\pdot{\bm A}=0}$), a gauge mode (with $\nbrk{A_\mu \propto k_\mu}$), and a constrained mode (whose amplitude is determined by the Gauss law).

In compliance with PBR~theorem\ct{PBR_theorem}, the wave function of the \emph{dynamical modes} of emergent fields~$\phi^\iota(\xv)$ exists materially. Their wave function is the smoothed generic basic distribution~$\Psi(Q)$ in some of its various presentations~$\psi(q)$, eq.\rf{psi_general_intro}. When $\psi(q)$ is identified with the wave function of a system then $\psi$'s arguments~$q$ should be identified with the dynamical variables of the system. The dynamical variables for the emergent fields may be chosen as the amplitudes of their dynamical modes with various spatial wavevectors~$\kv_m$. These amplitudes are then identified with appropriately rescaled independent arguments~$q^m$ of the presentation~$\psi(q)$, $\nbrk{q\equiv (q^1,\dots, q^N)}$.  In any specific gauge, the local field operators $\hat\phi^\iota(\xv)$ are then linear combinations [e.g.\rf{phi_standing_harmonics}] of the operators~$\hat q^m$, defined by $\nbrk{\hat q^m \psi(q)\equiv q^m \psi(q)}$.

\begin{figure*}[t]
\centering
\includegraphics[width=0.8\textwidth]{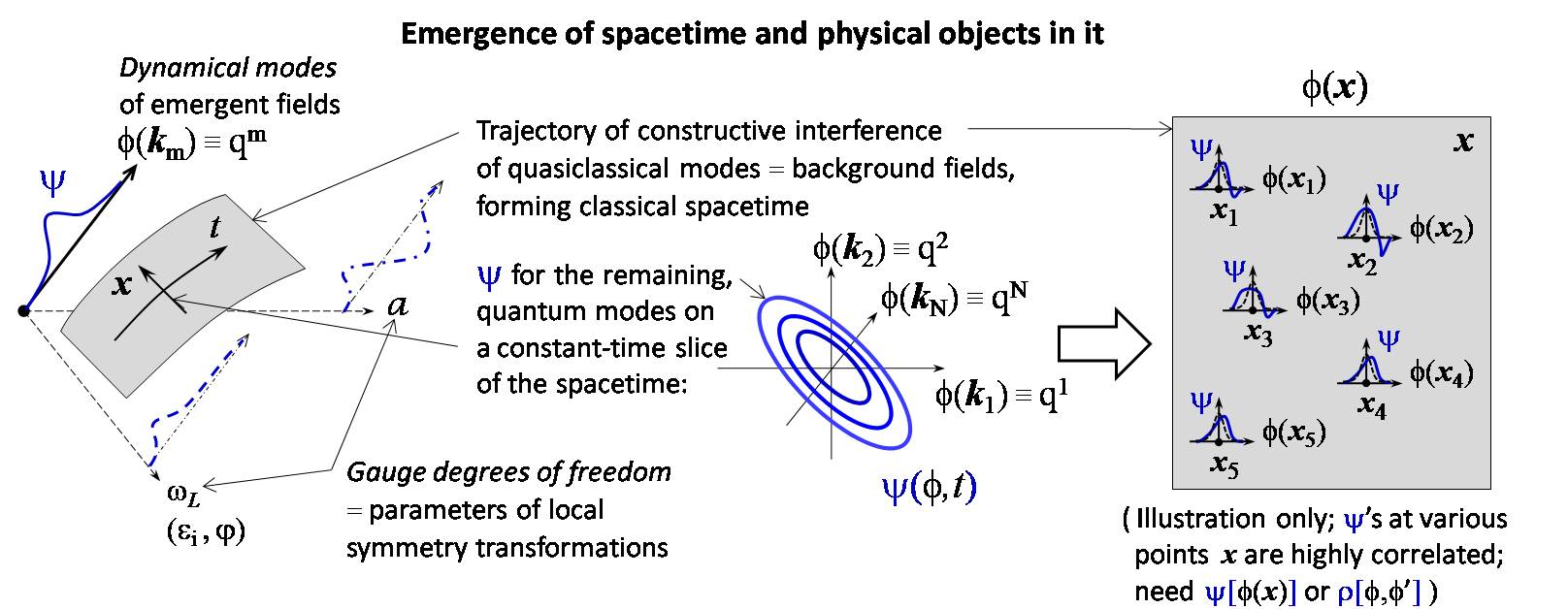}
\caption{
Left: Physical spacetime for the emergent quantum fields arises by constructive interference of their quasiclassical large-scale modes\ct{DeWitt,Gerlach} as discussed in \sct{sec_physical_world}. The remaining, shorter-scale modes of the fields are described by a reduced wave function, \sct{subsec_tilde_Hamiltonian} and\ct{LapchRubakov}.  \newline
Right: Integrate out all the dynamical variables but the fields~$\phi^\iota(\xv)$ at a single spatial point~$\xv$.
Familiar macroscopic objects correspond to the deviations of the resulting space-dependent density matrix from its vacuum value. The fields~$\nbrk{\phi(\xv)\equiv\{\phi^\iota(\xv)\}}$ at various spatial points are highly correlated. Their pure state is completely described by a wave function~$\psi[\phi(\kv_m)]$ for the fields' dynamical sub-Planckian modes. Our observed world is represented by mixed states, with a density matrix~$\rho[\phi(\kv_m),\phi'(\kv'_{m'})]$, \sct{sec_probabilities}.
}
\label{fig_summary}
\end{figure*}

The amplitudes of the \emph{gauge modes} of the emergent fields are \emph{not coordinates}~$q^m$ of a presentation~$\psi(q)$ of the basic material structure. 
The gauge degrees of freedom are ``created'' by transforming the materially represented wave function of the dynamical modes along the symmetry orbits of the eventual, covering every gauge, wave function of the fields. The amplitudes of the gauge modes are the respective transformation parameters~$\var$. In other words, the fields' gauge degrees of freedom~$\var$ are coordinates for a continuous set of the equivalent presentations~$\{\psi_\var(q)\}$ of the smoothed generic distribution~$\Psi(Q)$. The rightmost panel~c.\ in \fig{fig_outline} illustrates this. 

Finally, the amplitudes of the \emph{constrained modes} are determined by the constraint equations for the local symmetries of the fields.  

The identified quantum field operators~$\hat\phi(\xv)$
and the Hamiltonian for their evolution involve no fundamental material structure other than the basic distribution~$\dist(Q)$. In particular, there is no external spacetime for the emergent physical fields. Likewise, the underlying basic objects~$\{a\}$ of part~\ref{subsubsec_basic} are not assumed to be embedded in any space or time. The generic basic distribution~$\dist(Q)$ is everything that is sufficient to exist fundamentally in order to produce the evolving quantum fields, physical objects made of them, their possible internal intelligent observers, and the emergent physical spacetime where those are located and evolve. The field operators and composed of them operators for observables fundamentally are merely artificial \emph{rules} that help to describe the physically relevant alternate presentations~$\psi(q)$ of the smoothed generic discrete distribution. 

We can similarly identify \emph{fermionic} field operators, which obey the canonical anticommutation relations. Wave functions of emergent systems that contain bosonic as well as fermionic quantum fields can be identified with multi-component presentations~$\psi(q,\nf)$ of the same smoothed generic static distribution~$\dist(Q)$ (\sct{sec_fermions}). These presentations carry bit indices $\nbrk{\nf\equiv (\nf^1,\nf^2,\dots)}$, $\nbrk{\nf^m\in\{0,1\}}$, to be associated with the occupation numbers of the dynamical modes of fermionic fields. The only basic material structure involved in this representation of fermions is the same generic distribution~$\dist(Q)$.

\Sct{subsec_supersym} identifies emergent wave functions of bosonic and fermionic fields with \emph{local supersymmetry}. Unlike the wave functions of fields with gauge and diffeomorphism but no other local symmetry, a wave function of locally supersymmetric fields permits no freedom for its dynamics (aside from the freedom of setting spacetime coordinates). Different locally supersymmetric field systems with physically different Lagrangian densities generically emerge. In each of them, the Lagrangian density does not change during evolution of the system. \Sct{sec_why_the} proves that an emergent world with local supersymmetry cannot continuously evolve into another world that has a different Lagrangian. We thus expect emergence of different physical worlds that evolve by their own unchanged dynamical laws.

The described emergent quantum fields with gauge and gravitational interaction are not merely a mathematical construction. Their wave function objectively exists as a tangible distinctive entity. For example, it emerges from the surrounding us material macroscopic objects, in the role of the basic objects~$\{a\}$ of part~\ref{subsubsec_basic}. Yet, as discussed in \sct{sec_conclusion}, our physical world is not expected to arise from its own internal objects. A more natural option is that the wave function of our world is produced by a static material structure that exists independently of any spacetime. 

\emph{Physical spacetime} emerges\ct{DeWitt,Gerlach,LapchRubakov,Kiefer_88,Kim_95} for any system whose large-scale degrees of freedom evolve quasiclassically and whose wave function obeys the Wheeler-DeWitt equation. The physical spacetime for the emergent fields arises from constructive interference\ct{DeWitt,Gerlach} of their quasiclassical large-scale modes. \Sct{sec_physical_world} will describe comprehensively the emergence of spacetime and of the Schrodinger equation for quantum evolution in it.  \Fig{fig_summary}, continuing \fig{fig_outline}, sketches the key steps of emergence of the familiar structure of the physical world: the general relativistic background spacetime and physical objects that evolve quantum-mechanically in it. 

The rightmost part of \fig{fig_summary} illustrates the fundamental nature of physical objects in the emergent world.
A spatial point with coordinates~$\xv$ is associated with the operators of particle fields $\hat\phi^\iota(\xv)$ at the respective value of their continuous parameter~$\xv$, cf.~eq.\rf{phi_standing_harmonics}. The same physical configuration has, of course, many gauge-equivalent presentations.
Spatial arrangement of familiar physical objects corresponds to the deviation of the expectation values of space-dependent physical operators---constructed of~$\hat\phi^\iota(\xv)$ and canonically conjugate momenta operators~$\hat\pi_\iota(\xv)$, cf.\rf{pi_standing_harmonics}---from their vacuum expectation values. 
A pure state of the system would be completely described by the wave function~$\psi[\phi(\kv_m)]$, where $\nbrk{\phi\equiv\{\phi^\iota\}}$. Its arguments~$\phi(\kv_m)$ are the amplitudes of the fields' \emph{sub-Planckian} modes, as outlined in the next part~\ref{subsubsec_Planck}. \Sct{sec_probabilities} argues that the state of a physical universe such as ours is fundamentally mixed. It is fully described by a density matrix~$\rho[\phi(\kv_m),\phi'(\kv'_n)]$.

\subsubsection{Physics near the Planck scale}
\lb{subsubsec_Planck}

The physical degrees of freedom of the studied generically emergent fields are limited to their sub-Planckian modes.  With comprehensive discussion to follow in \sct{subsec_Planck_scale}, let us outline the main points. 

Phenomenological evidence for unification of the gauge couplings suggests that at least up to the energy scale of grand unification our world is described by $\nbrk{3+1}$ dimensional quantum field theory. The grand unification scale is only two orders of magnitude below the Planck energy. \Sct{subsec_Planck_scale} or part~\ref{subsubsec_renorm} next argue that, while not inevitable, it is natural to expect that below the Planck energy the spacetime metric evolves by the standard Hilbert action of general relativity. Then any \emph{excited} state of the field modes of wavelength comparable to the Planck length gravitationally collapses over about the Planck time to trans-Planckian energy density. 

We consider emergent fields whose evolution (fixed by local supersymmetry or another mechanism) has no regular continuation whenever Planckian energy density is reached. This characterizes at least the typical general relativistic systems with the Hilbert gravitational action. Then, of various alternate presentations of~$\Psi(Q)$ that could compose emergent wave functions~$\psi[\phi(\kv_m)]$, only those presentations that place the fields' modes of wavelength comparable to the Planck length to the ground state can evolve further over a physically meaningful time. The other presentations do not correspond to evolving physical systems. This confines the degrees of freedom near the Planck scale to the ground state. Evolution of sub-Planckian modes of the emergent fields can then be described by an effective quantum field theory.

Gravitational collapse on spatial scales much larger than the Planck length is analyzed in detail in a companion paper\ct{my_bh}. In short, if gravitational self-attraction of matter overcomes its pressure and the collapse becomes irreversible then a black hole forms. It slowly converts the energy of the collapsed matter into an equal amount of energy in the Hawking radiation\ct{Hawking_radiation}. The black hole information paradox\ct{Hawking_Breakdown_of_Predictability_76} does not arise because quantum information in the collapsed matter \emph{abandons} the emergent physical system. Ref.\ct{my_bh} proves that under local generally covariant dynamics of any fundamental origin the Hawking radiation cannot return the full information about the collapsed matter. In the emergent physical world information vanishes inside the black hole as the matter under a formed event horizon evolves to the physical singularity. (Of course, the information in the underlying basic distribution~$\dist(Q)$, which is static, remains unchanged.)  

The density matrix of the fields outside the event horizon evolves regularly through the complete evaporation of the black hole\ct{my_bh}. The same applies to the fields under the horizon at a sufficient distance from the central singularity. The degrees of freedom for the outgoing modes of the Hawking radiation emerge from the Planck scale in an initially trivial state of the vacuum of short-scale modes in a locally Minkowski frame around the horizon\ct{my_bh}. Throughout the evaporation, including its final stage, the distance from the event horizon to the physical singularity substantially exceeds the Planck length\ct{my_bh}.

The loss of physical information and consequent breakdown of unitarity for a black hole do not violate the energy-momentum conservation\ct{my_bh}. Full (non-perturbative) quantum evolution of the emergent fields is unitary in regular regions, free of black holes much heavier than the Planck mass.

\subsubsection{Regularization of ultraviolet loops}
\lb{subsubsec_renorm}

As outlined above, the modes of the studied emergent fields of wavelength comparable to the Planck length in regular spatial regions occupy the ground state. The fields can be described by an effective field theory where the influence of the near-Planck-scale degrees of freedom is quantified by counterterms that maintain the inherent local symmetries of the emergent fields. However, the local symmetry transformations are now restricted to those that affect only sub-Planckian modes. 

In particular, transformations of physical evolution, belonging to the diffeomorphism symmetry group, become restricted to temporal intervals much greater than the Planck time. Evolution of the emergent fields for spatial and temporal  scales \emph{smaller than the Planck length is thus physically meaningless and undefined} (\sct{subsec_Planck_scale}).

Although the considered generically emergent quantum fields are non-renormalizable, evolution of all their physical degrees of freedom is well defined. It is thus possible and may be expected that \emph{gravity remains non-renormalizable} up to the Planck scale, and that no dynamical theory objectively describes our world on the shorter scales. The physical fields of spin less than two could evolve by an action that contains renormalizable and non-renormalizable terms.  Wilson's treatment of renormalization\ct{Wilson_renorm_71a,Wilson_renorm_71b,Wilson_renorm_75}, with a formal proof in\ct{Polchinski_renorm_84}, shows that only renormalizable terms are generic for the non-gravitational part of the effective action below the Planck energy. 

In the discussed scenario it is physically meaningless to specify the Hamiltonian for Planck-scale and trans-Planckian modes. Below the Planck energy, actions that contain only renormalizable or, for gravity, the closest to renormalizable terms are concentration points in the set of all the actions of emergent systems. We can expect that one of such generic actions describes our observed world. The available experimental data conforms to these expectations.

\subsubsection{Probability, entanglement, locality}
\lb{subsubsec_branches}

\Sct{sec_Planck_scale}  will show that the smoothed generic Gaussian distribution~$\Psi(Q)$ of\rf{psi_general_intro} can be transformed into a multi-branch Schrodinger wave function of an inflating universe. The wave function complicates further when in some pockets of space inflation ends and nonlinear cosmic structure develops. This corresponds to the well understood transformation of the simple ground-state wave function of the inflaton and coupled particle fields into their today's global wave function. Our perceived world is one of its numerous decoherent branches\ct{Starob_branches_86}. 

A wave function that fundamentally is a presentation of a \emph{smoothed discrete} basic distribution~$\dist(Q)$ has an intrinsic uncertainty. Indeed, the discrete distribution~$\dist(Q)$ can be approximated by slightly different but equally statistically significant smooth fits~$\Psi(Q)$. This inherent uncertainty in~$\Psi(Q)$ translates into an uncertainty in its equivalent presentations~$\psi(q)$.  An analogy is the uncertainty in the smooth boundary of any familiar macroscopic physical object that below certain resolution is composed of almost point-like elementary particles. 

\begin{table*}[t]
\begin{center}
{\renewcommand{\arraystretch}{0}
\begin{tabular}{|c|c|c|}
\hline
\strut                                                                            & \textbf{Features that}   & \textbf{Discussed}  \\
\strut \raisebox{1.4ex}[0pt]{\textbf{Requirement}} & \textbf{enable it} & \textbf{in}  \\
\hline
\rule{0pt}{2pt} & & \\
\hline
\strut   Specific dynamics of  & \,Discreteness of underlying structure\, &  \\
\strut   wave function~$\psi$  &  $\&$  Local supersymmetry  &   
                                \raisebox{1.2ex}[0pt]{\Sct{sec_why_the}} \\
\hline
\strut   Relation of~$\psi$ to & Discreteness of underlying structure  &  \\
\strut   objective probability &  $\&$ Extended regular past  &  
                \raisebox{1.2ex}[0pt]{~\Scts{sec_structure},~\ref{sec_probabilities}~}\\
\hline
\strut   ~Concrete, low-entropy &  Inflation &  \\
\strut    initial conditions   &  $\&$ Singular gravitational collapse   & \raisebox{1.2ex}[0pt]{~\Scts{sec_init_conds},~\ref{sec_Planck_scale}~}\\
\hline
\end{tabular}
}
\caption{\lb{tab_phys_syst}
    Column~1: Requirements to a quantum system that is a physical world with internal life.  \newline
    Column~2: Features of the studied emergent field systems that naturally produce the required properties. }
\end{center}
\end{table*}

\Sct{sec_structure} constructs a bilinear product~$\nbrk{\langle\cdot|\cdot\rangle}$ with the following property.
Let $\{\psi_i(q)\}$~be the decoherent (Everett's) branches of an emergent wave function $\nbrk{\psi=\sum_i\psi_i}$. Then both $\psi$~and $\nbrk{\psi-\psi_i}$, for some picked~$i$, are acceptable smooth presentations of the  basic discrete distribution if and only~if
\be
\langle\psi_i|\psi_i\rangle\lesssim \delta\chi^2_{\rm min}\,.
\lb{norm_intro}
\ee
Here $\delta\chi^2_{\rm min}$ is a fixed positive number that depends on the basic distribution~$\dist(Q)$ and dimensionality of the configuration space of the emergent system but not on the presentation of its wave function. 

Thus when a branch~$\psi_i$ that satisfies\rf{norm_intro} is removed from the overall~$\psi$, the remainder $\nbrk{\psi-\psi_i=\sum_{j\neq i}\psi_j}$ continues to be a smooth presentation of~$\dist(Q)$. Hence the branch~$\psi_i(q)$, 
with $\nbrk{\langle\psi_i|\psi_i\rangle\lesssim \delta\chi^2_{\rm min}}$, is not an objectively existing component of the considered smooth presentation of the discrete $\dist(Q)$.
Since the squared norm of every Everett's branch that has objective material presentation, i.e.,  that physically exists, should exceed the positive value~$\delta\chi^2_{\rm min}$ of\rf{norm_intro}, the number of such Everett's branches~$\{i\}$ is finite. This leads to objective and unambiguous \emph{probability} of various macroscopic outcomes of quantum processes in the emergent system (\scts{sec_structure} and~\ref{sec_probabilities}). 

Unlike the earlier formulations of quantum mechanics, including Everett's, we do not impose the \emph{Born rule} as a postulate. Instead, we establish that it arises as a result of dynamical equilibrium, \sct{sec_probabilities}.

For the emergent quantum worlds, the phenomenon of \emph{quantum entanglement} across any distance is trivial. Their dynamical variables fundamentally are the variables $\nbrk{q\equiv(q^1,\dots, q^N)}$ of a presentation~$\psi(q)$ of the smoothed basic distribution. An entangled quantum state of two dynamical variables, e.g., $q^1$~and~$q^2$, is a dynamically isolated (decoherent) term~$\psi_i(q^1,q^2,\dots)$ of the emergent wave function. The variables~$q^1$ and~$q^2$ are entangled when the domain of non-neligible probability $\nbrk{|\psi_i(q^1,q^2,\dots)|^2}$ is localized in both~$q^1$ and~$q^2$ so that specifying one of the variables restricts the other.

\emph{Locality of the dynamics} of the field operators in the emergent spacetime likewise follows from the first principles, \sct{sec_conclusion}. 

\subsubsection{Resolution of the Boltzmann brain problem}
\lb{subsubsec_ic}

Consider prospective wave functions for the dynamical modes of quantum-field worlds with internal intelligent observers. Apparently, the most typical of such randomly constructed wave functions would be Boltzmann-brain worlds\ct{Dyson_Kleban_Sussk_02,BB_DeSimone_etal_08,BB_Carroll_17}. Outside a region just large enough to currently harbor an internal observer, they would differ drastically from the low-energy, low-entropy environment suitable for life. Cosmological observations starkly contradict this picture.

\Sct{sec_init_conds} shows that the described emergent systems avoid the Boltzmann brain problem. In short, for them, the Born rule develops upon establishing dynamical equilibrium in the evolving ensemble of emergent states (\sct{sec_probabilities}). Anthropically suitable states of emergent generally covariant worlds that did not inflate lack extended regular past. Therefore, their ensemble may not have sufficient time to achieve the equilibrium. Without it, all the alternate quantum outcomes, including ``weird'' ones, become practically possible. Macroscopic evolution in such universes, despite their wave functions  evolving by fixed laws, is unpredictable. Then biological evolution is impossible in principle (\sct{subsubsec_basic}).

On the other hand, emergent systems that undergo inflation can evolve regularly for arbitrarily long duration. Then the detailed equilibrium\rf{Fw_equilibrium} inevitably arises and so does the Born rule (\sct{sec_probabilities}). It results in a physical world with reasonably predictable quasiclassical macroscopic evolution by unchanged laws.
 
Table~\ref{tab_phys_syst} summarizes requirements to a system that can contain a viable physical world with internal life. The generically existing structures identified in the paper fulfill these requirements owing to their features listed in the second column of Table~\ref{tab_phys_syst}. The third column refers to the paper sections that describe these features and their role.

\subsection{Paper outline}

The rest of the paper is organized as follows. The last part~\ref{subsec_notations} of this Introduction specifies the notations. \Sct{sec_first_observation} considers an arbitrary quantum theory where the superposition principle applies absolutely to the evolution, at least, on the scales presently accessible to experiments. 
The section proves that its any quantum state develops physically real paths of the subsequent evolution by any conceivable, generally time- and space-dependent dynamical laws.  This raises the question of why the dynamics of our world does not vary randomly in time and space. 

\Sct{sec_structure} presents the basics for showing how an evolving, probability-related wave function of particle fields emerges from the generic discrete static distribution of arbitrary quantities.
\Sct{sec_evolution} describes the emergence of a \emph{complex} wave function, Hamiltonian for its evolution, and probability-related Hermitian product for its branches. 

\Sct{sec_gauge_fields} proves that if the dynamics (action) of quantum fields is symmetric under a group of local transformations then the fields' wave function~$\psi[\phi(\xv),t]$ is also invariant under changing its field arguments by transformations from this group. The section then constructs wave functions of gauge-symmetric fields. \Sct{sec_gravity} studies emergent quantum fields with diffeomorphism symmetry and their gravitational degrees of freedom.
\Sct{sec_physical_world} describes how the static wave function of diffeomorphism-symmetric fields represents a universe with quasiclassical spacetime. 

\Sct{sec_fermions} introduces generically emergent fermionic quantum fields. \Sct{sec_emergent_SUSY} identifies emergent wave functions of field systems with local supersymmetry. They, unlike the prospective wave functions of the preceding sections, evolve by concrete dynamical laws. 
\Sct{sec_why_the} analyzes the mechanisms that fix the dynamical laws for the locally supersymmetric emergent quantum field systems.

\Sct{sec_Planck_scale} describes evolution of the studied emergent fields close to the Planck scale.  
\Sct{sec_probabilities} demonstrates how different macroscopic outcomes of quantum processes in the generically emergent worlds acquire objective frequentist probabilities that are governed by the Born rule. 
\Sct{sec_init_conds} shows that the typical physical universes that emerge as described avoid the Boltzmann brain problem and naturally experience cosmological inflation. 

\Sct{sec_conclusion} summarizes the results. The paper has three technical \apxs, which derive and summarize some formulas for the main sections.  Notably, \apx~\ref{apx_fermions} summarizes the structure of the general local (super)\,symmetry and construction of the symmetric Hamiltonians.

The paper describes the emergent physical worlds up to the most basic level. It thus lets us tackle challenges that require the knowledge of physics at the fundamental Planck scale.  \Sct{sec_Planck_scale} discusses how new short-distance physical degrees of freedom appear during inflationary or other cosmological expansion.
Opposite to inflationary expansion, gravitational collapse and the entire Hawking evaporation of the formed black hole likewise become fully tractable, and its ``information paradox''\ct{Hawking_Breakdown_of_Predictability_76} understood\ct{my_bh}. The companion paper\ct{my_bh} describes the complete evaporation of gravitational black holes, their central singularity, and the final moments of their evaporation. 

\subsection{Notations}
\lb{subsec_notations}

We employ the \emph{units} with  $\nbrk{\hbar=c=m_P=\ell_P=1}$, where the Planck mass~$m_P$ and Planck length~$\ell_P$ are related to the Newton gravitational constant~$G$ as
\be
\ell_P\equiv m_P^{-1}= (8\pi G)^{1/2}\, .
\lb{m_P_def}
\ee

We use Greek indices  $\mu, \nu, \dots$ for the components of spacetime tensors and Latin indices  $i, j, \dots$ for the components of spatial tensors. 
In a locally inertial frame, or in tangent Lorentz spacetime, we label the components of spacetime tensors with Latin indices $a, b, \dots$ and the components of spinors with Greek indices $\alpha, \beta, \dots$\,.

We apply the spacetime metric signature $(-,+,+,+)$ and parameterize the metric by the Arnowitt-Deser-Misner (ADM) decomposition\ct{ADM_59,ADM_62_republished}
\be
ds^2 = - (N dt)^2 + \g_{ij} (dx^i+\shift^i dt)(dx^j+\shift^j dt).
\lb{ADM}
\ee
The respective metric tensor and its inverse are
\be
g_{\mu\nu} \eqa  \lf(\ba{cc} - \shift^2 + \lapse_i\lapse^i &  \lapse_i  \\
                                                      ~\, \lapse_i &  \g_{ij}
                \ea\rt)\!,
\lb{gADM}  \\
g^{\mu\nu} \eqa  \fr1{\lapse^2} \lf(\ba{cc} -1  &  \shift^i  \\
                                                                ~   \shift^i  &  \lapse^2\g^{ij} - \shift^i\shift^j
                \ea\rt)\!.
\lb{ginvADM}
\ee
We lower and raise indices of spatial tensors by the spatial metric~$\g_{ij}$ and its inverse~$\g^{ij}$ respectively.
We denote differentiation by a comma,  covariant differentiation based on the spacetime metric~$g_{\mu\nu}$ by a semicolon, and covariant differentiation based on the spatial metric~$\g_{ij}$ by a vertical bar~``$|$".

We label the components of general vectors or coordinates in \emph{configuration space} with indices $n$ and~$m$. We denote the components of general field multiplets with indices  $\iota, \kappa, \dots$. However, we distinguish the multiplets---including symmetry generators, transformation parameters, and gauge fields---that transform by the adjoint representation of the local symmetry group as follows: they carry indices $r, s, \dots$ for Yang-Mills gauge symmetries; indices $a, b, \dots$ for the local Poincare symmetry; spinor indices $\alpha, \beta, \dots$ for (local)~supersymmetry transformations, and indices $A, B, \dots$ for the general local symmetry transformations. 

Bosonic and fermionic field operators $\nbrk{\hat F(x)\equiv\{\hat F^\iota(x)\}}$ at a spacetime point~$x$ change under symmetry transformation as
\be
\hat F(x) \to \hat {\tilde F}(x)= \hat U^{-1} \hat F(x) \hat U\,. 
\lb{F_transformation_into}
\ee 
For an infinitesimal vicinity of~$x$ in a locally inertial frame
\be
\hat U = \exp(-\var^A \hat T_A)\,
\ee
where $\var^A$~are transformation parameters and $\hat T_A$~are symmetry generators. We will consider
\be
\var^A\eqa (\veps^a,\veps^{ab},\xi^\alpha,\varphi^r)
\lb{e_list}\\
\hat T_A\eqa (i\hat P_a,i\hat J_{ab}/2,-\hat {\bar Q}_\alpha,\hat T_r),\qquad
\lb{A_list}
\ee
for respectively translation in spacetime, Lorentz rotation, supersymmetry transformation, and Yang-Mills gauge transformation.
The operators~$\var^A T_A$ form a graded algebra whose multiplication is the commutator
\be
[\var^A_1 \hat T_A,\var^B_2 \hat T_B]=\var^B_2\var^A_1 f_{AB}{}^C\, \hat T_C\,.
\lb{structure_const_def}
\ee
On the right-hand side, $f_{AB}{}^C$ are the symmetry structure constants. The order of $\var^A_1$ and $\var^B_2$ accounts for their possible anticommutativity. 

The transformation parameters for the general local symmetry transformation are fields~$\var^A(x)$.  These parameters may not be all independent; for example, see the zero-torsion (a.k.a.~``conventional'') constraint\rf{zero_torsion_constraint} of supergravity, e.g., Ref.\ct{Freedman_Proeyen_book}. The transformation operator~$\hat U$ in\rf{F_transformation_into} equals
\be
\hat U = \exp\lf[-i\int d^3x~\var^A(\xv)\,\hat\cH_A(\xv)\rt],
\lb{U_into}
\ee
eq.\rf{U_general_local}. 

The Hamiltonian of a diffeomorphism-symmetric theory [eq.\rf{H_from_HA}] is a linear combination of the symmetry generators~$\hat\cH_A(\xv)$ from the right-hand side of\rf{U_into}. Appendix~\ref{apx_fermions} proves that in a theory with locally symmetric dynamics the wave function satisfies the constraints $\nbrk{\hat\cH_A\psi =0}$, eq.\rf{secondary_constraint_A}.

We denote the operators of bosonic fields by~$\hat f(\xv)$ and of fermionic fields by~$\hat\chi(\xv)$. We label \emph{fermionic mode occupation numbers}, each either~0 or~1, by~$\nf$. A pure state of bosonic and fermionic fields is thus specified by a wave function~$\psi(f,\nf)$.

We write fermionic fields, carrying spinor indices, in the 4-component Dirac form. We do not fix a representation of the Dirac matrices~$\gamma^a$. 
The Dirac conjugate of a fermionic field~$\chi$ is $\nbrk{\bar\chi\equiv \chi^\dagger i\gamma^0}$. For the supersymmetry directions of the symmetry transformations\rfs{F_transformation_into}{A_list}, 
\be
\var^\alpha T_\alpha = \bar\xi Q = \bar Q \xi\,
\ee
where~$\xi$ and~$Q$ are anticommuting Majorana spinors in the 4-component Dirac notation.

For the Fourier transformation of a function $\psi(q)$,  $\nbrk{q\in \Rset^N}$, we introduce a shorthand notation
\be
\dbar^N\!p \equiv \fr{d^N\!p}{(2\pi)^N}\,.
\lb{dbar_def}
\ee
Then
\be                           
\psi(q)\eqa \int \dbar^N\!p\ e^{i p\cdot q}\, \psi(p)\, \\   
\psi(p)\eqa \int d^N\!q\ e^{-i p\cdot q}\, \psi(q)\, 
\lb{FT_standard}
\ee
and
\be
\int \dbar^N\!p\  e^{i p\cdot(q-q')} = \delta^{(N)}(q-q'),
\lb{int_dbar_p}
\ee
where $\delta^{(N)}(q)$ is the $N$-dimensional Dirac delta function.
Likewise we define
\be
\deltabar^{(N)}(p) \equiv (2\pi)^N\delta^{(N)}(p),
\ee
for which
\be                   
\int \dbar^N\!p' \, f(p') \ \deltabar^{(N)}(p'-p) = f(p)  \nn
\ee
and
\be
\int d^N\!q\ e^{i (p'-p)\cdot q} = \deltabar^{(N)}(p'-p). \nn
\ee

Throughout the paper, $q$ and $p$ denote ``dynamical variables,'' which describe the physical degrees of freedom in the configuration and momentum representations respectively.  The dynamical variables~$q$ should be distinguished from spacetime coordinates $x$ or spatial coordinates $\xv$ of field operators, e.g., of~$\nbrk{\hat\phi(x)\equiv\, \hat\phi(t,\xv)}$. Similarly, the canonical dynamical momenta~$p$ should be distinguished from the spatial wavevector~$\kv$ for the Fourier modes of field operators, e.g., in
\be
\hat\phi(\xv) = \int \dbar^3 k\, e^{i \kv\cdot \xv}\, \hat\phi(\kv).
\ee

We use function notations for functions of a finite number of variables and for functionals (functions of functions). The quantum-mechanical wave function is denoted by~$\psi(f)$ and called ``wave function" even when its argument~$f$ ranges over a set of field configurations~$\{f(\xv)\}$. Since we may approximate a continuous function~$f(\xv)$ by its discretizations~$\{f(\xv_n)\}$ on a series of progressively refined grids, a functional~$\psi[f(\xv)]$ can be regarded as the limit of regular functions~$\psi(q)$ of an increasingly large number of variables~$\nbrk{q^n\equiv f(\xv_n)}$. 

Accordingly, $\int\!df$ denotes both the ordinary (e.g.\ Riemann) and functional integral with a measure~$df$. The Dirac delta function~$\delta(f)$ is defined for any---discrete, continuous, or function---argument~$f$ by
\be
\int df'\,\delta(f-f')\,\psi(f')\equiv \psi(f)
\lb{delta_Dirac_def}
\ee
for any map $\psi(f)$.

\section{Freedom of quantum evolution}
\label{sec_first_observation}

Experiments and observations provide solid evidence that quantum principles apply to the physical world from the smallest scales probed by the particle colliders to macroscopic and even cosmological distances. The validity of quantum description at the accessible microscopic scales is verified, for example, by the precision measurement of scale ``running'' of the renormalized couplings and other parameters of the Standard Model due to quantum radiative corrections. At larger scales experiments  confirm accurate and detailed predictions of atomic and condensed matter physics, relying on quantum description of the electrons and electromagnetic field. On macroscopic scales quantum mechanics has also been tested with precision in quantum optics, quantum networks and communication, and other experiments (e.g., recently, with massive mechanical oscillators\ct{entang_massive_Natur_18}). Even for the largest observable cosmological distances quantum entanglement and the superposition principle are strongly favored with the successful predictions of the inflationary paradigm, explaining the observed cosmological inhomogeneities as amplified vacuum quantum fluctuations.

This section highlights a peculiar logical consequence of the cornerstone superposition principle of quantum physics on the accessible scales. This observation will guide us toward identifying in \scts{sec_structure}-\ref{sec_why_the} realistic dynamical quantum fields as emergent phenomena in the generic set of almost arbitrary static entities.

Let us consider quantum degrees of freedom that are described by commuting (bosonic) field operators~$\hat f^\iota(\xv)$. Their discrete label~$\iota$ denotes a field type or/and specifies the component of a field multiplet (e.g., spin or isospin projection). The continuous label~$\xv$ of the field operators belongs to a 3-dimensional manifold. It can be physically interpreted as the spatial coordinates in some coordinate frame of the physical points associated with the respective field degrees of freedoms. 

Since the field operators~$\hat f^\iota(\xv)$ commute, we can expand any pure quantum state~$|\psi\rangle$ over their simultaneous eigenstates~$|f\rangle$, for which $\nbrk{\hat f^\iota(\xv)|f\rangle=f^\iota(\xv)|f\rangle}$, as
\be
|\psi\rangle= \int df\,\psi(f)\,|f\rangle\,.
\lb{psi_decomposition_f}
\ee
Here $\nbrk{df\equiv\prod_{\iota,\xv}[df^\iota(\xv)]}$ and $\nbrk{\psi(f)\equiv\psi[f^\iota(\xv)]}$. Normalizing the eigenstates~$|f\rangle$ to the delta function\rf{delta_Dirac_def},
\be
\langle f'|f\rangle = \delta(f'-f)\,,
\lb{f_state_normalization}
\ee
we arrive at representing the state $|\psi\rangle$ by its wave function~$\psi(f)$. By\rfs{psi_decomposition_f}{f_state_normalization}, the Hermitian product of quantum states is
\be
\langle \psi_1|\psi_2\rangle = \int df\,\psi^*_1(f)\,\psi_2(f)\,.
\lb{scalar_product_standard}
\ee

We define the operators of canonical momenta fields $\hat\pi_\iota(\xv)$ as the operators that are canonically conjugate to the local fields~$\hat f^\iota(\xv)$:
\be
\hat\pi_\iota(\xv)\,\psi(f) \equiv -i\,\frac{\delta}{\delta f^\iota(\xv)}\,\psi(f)\,.
\lb{pi_def}
\ee 
The operators $\hat f^\iota(\xv)$ and~$\hat\pi_\iota(\xv)$ obey the canonical commutation relations
\be
[\hat f^\iota(\xv),\hat\pi_\kappa(\yv)]= i\delta^\iota_\kappa \delta^{(3)}(\xv-\yv)\,,
\lb{fpi_commutator}
\ee
where $\delta^\iota_\kappa$ and $\delta^{(3)}(\xv)$ are respectively the Kronecker symbol and Dirac delta function. 

We now consider an arbitrary analytic function~$H'$ of the operators~$\{\hat f^\iota(\xv)\}$ and~$\{\hat\pi_\iota(\xv)\}$:
\be                           
H'(\hat f,\hat\pi)\eqa  \sum_n \int d^3x_1\,d^3x_2\dots d^3y_1\,d^3y_2\dots\,\times \lb{H_prime_general}\\*
&\times& K^{\kappa_1\kappa_2\dots}_{n\,\iota_1\iota_2\dots}(\xv_1,\xv_2,\dots;\yv_1,\yv_2,\dots)\,\times \nn\\*
&\times& \hat f^{\iota_1}(\xv_1)\hat f^{\iota_2}(\xv_2)\dots\hat\pi_{\kappa_1}(\yv_1)\hat\pi_{\kappa_2}(\yv_2)\dots~.  \phantom{\int}\nn
\ee
To avoid ambiguities, let us use only functions~$H'(\hat f,\hat\pi)$ with a finite number of terms in the sum\rf{H_prime_general} and with a finite number of operators~$\hat f^\iota$ and~$\hat\pi_{\kappa}$ in each of the terms. Also let us apply only the kernels~$K_n$ that yield well-defined convergent integrals in $H'(\hat f,\hat\pi)\,\psi$.

For a real~$\Delta t$, let
\be
\hat U \equiv \exp(-i\hat H'\Delta t)\,,
\lb{U_from_Hprime}
\ee
where
\be
\hat H'\equiv H'(\hat f,\hat\pi)
\ee 
of\rf{H_prime_general}. Further, let
\be             
 \ba{rcl}
\hat f'{}^\iota(\Delta t,\xv)&\equiv& \hat U^{-1}\hat f^\iota(\xv)\,\hat U\,,\\
\hat\pi'_\iota(\Delta t,\xv)&\equiv& \hat U^{-1}\hat \pi_\iota(\xv)\,\hat U\,.
 \ea
\lb{operators_new}
\ee
The similarity transformation\rf{operators_new} is canonical, i.e., it preserves the commutation relations\rf{fpi_commutator}.

Take $\hat H'$ to be Hermitian: $\nbrk{\hat H'{}^\dagger=\hat H'}$. The hermiticity imposes some straightforward constraints on the kernels~$K_n$ in\rf{H_prime_general}, but the remaining freedom for choosing these kernels is still vast. The corresponding operator~$\hat U$ of\rf{U_from_Hprime} is unitary: $\nbrk{\hat U^{-1}=\hat U^\dagger}$. Hence for any function $\nbrk{O(\hat f,\hat\pi)}$ and any state~$|\psi\rangle$ we have
\be                 
\langle\psi|\,O(\hat f',\hat\pi')\,|\psi\rangle \eqa \langle\psi|\,\hat U^{-1}\,O(\hat f,\hat\pi)\,\hat U\,|\psi\rangle\,= \quad\nn\\*
\eqa \langle\psi'|\,O(\hat f,\hat\pi)\,|\psi'\rangle
\ee
with
\be
|\psi'\rangle = \hat U\,|\psi\rangle\,.
\lb{evolution_Schrodinger}
\ee

When~$\hat H'$ coincides with the Hamiltonian~$\hat H$ of a physical field system then in the Schrodinger picture of quantum mechanics the system that is initially in a state~$|\psi\rangle$ evolves over the time span~$\Delta t$ into the state\rf{evolution_Schrodinger}. We will however be interested in $\nbrk{\hat H'\neq\hat H}$.

For the arguments of the rest of this section, including its important conclusion, we may also take some of the operators~$\hat f^\iota(\xv)$ and~$\hat\pi_\iota(\xv)$ to be fermionic. The commutator in the canonical  eq.\rf{fpi_commutator} should then be replaced by the usual commutator/anticommutator~$[.,.\}$ that accounts for the grading of its arguments. With fermions, we specify a quantum state by a wave function~$\psi(f,\nf)$, where  $f$~stands for the independent bosonic fields and $\nf$~for the occupation numbers, $0$ or~$1$ each, of dynamical modes of the fermionic fields (\sct{sec_fermions}).

\subsection{Freedom of evolution in Heisenberg picture}
\lb{subsec_freedom_Heisenberg}

In the Heisenberg picture $|\psi\rangle$~is static. Physical evolution is manifested as the change of the operators for observables, e.g., the energy-momentum tensor, density of a current, a field-strength tensor, or their averages over a finite region. An observable that at a time~$t$ corresponds to an operator 
\be
\hat \Op= \Op(\hat f,\hat\pi)\,,
\ee
where~$\Op$ is an analytic function of the displayed arguments, at a new time $\nbrk{t+\Delta t}$ becomes
\be
\hat \Op'_{\Delta t}=\hat U^{-1}\,\Op(\hat f,\hat\pi)\,\hat U=\Op(\hat f',\hat\pi')\,,
\lb{Op_def}
\ee
where $\hat f'$ and~$\hat\pi'$ are given by\rf{operators_new} with $\hat H'$~set to the physical Hamiltonian~$\hat H$.
The system's state at the new time is the same Heisenberg state~$|\psi\rangle$ that is now regarded as a linear combination of the eigenstates of the operator for the evolved observable~$\hat \Op'_{\Delta t}$. 

We now make our first key observation. Let the Hermitian operator~$\hat H'$ in\rf{H_prime_general} \emph{differ} from
the physical Hamiltonian~$\hat H$. Then the operator~$\hat \Op'_{\Delta t}$ in\rf{Op_def},  obtained by~$\hat U$ of\rf{U_from_Hprime} with~$\nbrk{\hat H'\neq\hat H}$, is still a Hermitian operator. The transformations $\nbrk{\hat \Op \to \hat \Op'_{\Delta t}}$ of various physical operators~$\hat \Op$ form a continuous one-dimensional group of similarity transformations, parameterized by $\nbrk{\Delta t\in\Rset}$. For several observables~$\hat \Op^\iota$, the respective~$\hat \Op'{}^\iota_{\!\!\Delta t}$ satisfy the same commutation/anticommutation relations as the original~$\hat \Op^\iota$ do. 

Most of the arbitrary ``alternative Hamiltonians''~$\hat H'$ do not possess pointer states, stable to decoherence under the evolution by~$\hat H'$. Yet for a typical physical system we can contemplate many $\nbrk{\hat H'\neq\hat H}$ with pointer states that smoothly continue the evolution of the earlier pointer states of the system. For example, consider a Hamiltonian where some particle masses and couplings or the cosmological constant change from their current values adiabatically, so that short-scale degrees of freedom retain the ground state. The projections of~$|\psi\rangle$ on the evolving pointer states of the alternative Hamiltonian~$\hat H'$ then develop into the Everett branches of evolution with~$\hat H'$. If the quantum superposition principle is an exact law of nature then these branches should be as real for their inhabitants, evolving from our current state by the alternative Hamiltonian, as a future branch of the evolution with the unchanged Hamiltonian will be for its inhabitants. Hence it should be more likely for us after any current moment to experience the evolution described by one of the alternative Hamiltonians compatible with our well-being, rather than by the unchanged past one.

This conclusion stands even if one insists that the wave function of a physical system should not be considered independently of a specific Hamiltonian~$\hat H$. Indeed, for many $\nbrk{\hat H'\neq\hat H}$ we can arrange a measurement of the projections of~$|\psi\rangle$ on the eigenstates of the operators~$\hat \Op'{}^\iota_{\!\!\!\Delta t}$, obtained by transforming the observables~$\hat \Op^\iota$ with the alternative~$\hat H'$. Let~$|1'\rangle$ and~$|2'\rangle$ be such $\hat \Op'{}^\iota_{\!\!\!\Delta t}$~eigenstates with different eigenvalues. Suppose, for simplicity, $\nbrk{|\psi\rangle=|1'\rangle +|2'\rangle}$. During the suggested measurement the joint state of a studied subsystem~$|\psi\rangle$ and of the rest of the system ($|\ext\rangle$, representing the detector, observers, and environment) changes as
\be
|\psi\rangle|\ext\rangle= \lf(\vphantom{\hat i}|1'\rangle +|2'\rangle\rt)\!|\ext\rangle \to |1'\rangle|\ext_{1'}\rangle +|2'\rangle|\ext_{2'}\rangle\,.
\nn
\ee
After a typical measurement, the states $|\ext_{1'}\rangle$ and~$|\ext_{2'}\rangle$ for the macroscopic environment decohere.
Subsequent observations of the studied subsystem for the observers, who are part of~$|\ext\rangle$, then become confined either only to the branch~$\nbrk{|1'\rangle|\ext_{1'}\rangle}$ or only to the branch $\nbrk{|2'\rangle|\ext_{2'}\rangle}$.
This is identical to evolution of the subsystem~$|\psi\rangle$ with the \emph{alternative} Hamiltonian~$\hat H'$. 

The above assumes that the detector and observer retain their integrity and continue to operate in their respective roles when they themselves evolve by the modified Hamiltonian~$\hat H'$. It is a reasonable assumption for some modifications of the Hamiltonian, e.g., for moderate and smooth variation of the cosmological constant that yet exceeds the current observational constraints on its time dependence by orders of magnitude.

Why, despite the presented arguments, the experiments consistently indicate quantum evolution by unchanged fundamental physical laws, constant in time and space?
Importantly, we cannot resolve this paradox by simply assuming a multitude of worlds with all the conceivable Hamiltonians\ct{Tegmark_mathematical_universe,Linde_foundational_conference} and suggesting that only the world with our Hamiltonian is life-friendly. The question concerns our own world. Why of all the choices for quantum evolution that should \emph{materialize here and now}, do we witness only spacetime-independent dynamics? Why do its local laws avoid the variations that would be readily detectable while compatible with non-disrupted continuation of our lives?

\subsection{Repeat in Schrodinger picture}
\label{subsec_Schrodinger}

We now reformulate the above considerations in the Schrodinger picture. Let at a time~$t$ the system be in a pure state with a wave function~$\psi(f)$. By the Born rule, then~$|\psi(f)|^2$ is the probability density in the configuration space~$\{f\}$. Let a function~$\psi'(f')$ be the result of invertible convolution of~$\psi(f)$ with some kernel~$U(f',f)$:
\be
\psi'(f')= \int df~U(f',f)\,\psi(f).
\lb{evolution_convolution}
\ee
Transformation\rf{evolution_convolution} is unitary with respect to the Hermitian product\rf{scalar_product_standard},
\be
\int df'\,|\psi'(f')|^2 = \int df\,|\psi(f)|^2,
\lb{wave_function_unitarity}
\ee
whenever
\be
\int df'\,U^*(f',f_1)\,U(f',f_2)= \delta(f_1-f_2)\,.
\lb{convolution_unitarity}
\ee

Consider unitary evolution\rf{evolution_Schrodinger} of a state~$|\psi\rangle$ with $\nbrk{\hat U=\exp[-iH'(\hat f,\hat\pi)\,dt]}$.  	The Feynman path integral lets us present\rf{evolution_Schrodinger} as convolution transformation\rf{evolution_convolution} of the state's wave function $\nbrk{\psi(f)=\langle f|\psi\rangle}$.  For an infinitesimal~$dt$
\be
\psi'(f')= \int df\,\dbar\pi\, e^{i[(f'-f)\cdot\pi-H'(f,\pi)\,dt]}\,\psi(f)\,,
\lb{evolution_path_integral}
\ee
where
\be
f{\cdot}\pi\equiv \int d^3x\,\sum_\iota f^\iota(\xv)\,\pi_\iota(\xv)
\ee
and the momentum measure~$\dbar\pi$ for a finite number of dynamical variables  is defined by\rf{dbar_def}. For a wave function (functional) whose argument $\nbrk{f=\{f^\iota(\xv)\}}$ is a function, the measure~$\dbar\pi$ is defined by the condition
\be
\int \dbar\pi\, e^{i(f'-f)\cdot\pi}= \delta(f'-f)\,,
\ee
generalizing eq.\rf{int_dbar_p}. We can verify equivalence of the Schrodinger transformation\rf{evolution_Schrodinger} and of the convolution\rf{evolution_path_integral} by noting that the evolving wave function in both cases satisfies the same first-order differential equation
\be
\frac{\partial}{\partial t}\,\psi'= -iH'(f,-i\fr{\delta}{\delta f})\,\psi'
\ee
with the same initial condition $\nbrk{\psi'|_{dt=0}=\psi}$.\footnote{
	$\nbrk{\hat H'\equiv H'(\hat f,\hat\pi)}$ may contain products of non-commuting operators~$\hat f$ and~$\hat\pi$. Then for the equivalence of eqs.\rf{evolution_Schrodinger} and\rf{evolution_path_integral} we order these products in the Hamiltonian operator\rf{U_from_Hprime} so that all the momenta $\hat\pi$ stand to the right, as in\rf{H_prime_general}.
}
Wave function evolution over a finite time span~$\Delta t$ follows by repeating the infinitesimal transformation\rf{evolution_path_integral} $\Delta t/dt$~times. The limit $\nbrk{dt\to0}$ then yields the respective convolution kernel~$U(f',f)$ in the standard Feynman path integral form.

We now reformulate the ``key observation" of the previous subsection~\ref{subsec_freedom_Heisenberg} in the Schrodinger picture. Let $\psi$ be the wave function of a physical system at a time~$t$. Consider its equivalent presentation~$\psi'$ obtained by any convolution\rf{evolution_convolution} that satisfies the unitarity condition\rf{convolution_unitarity}. Any real Hamiltonian function~$H'(f,\pi)$ generates a continuous group of such convolutions through\rf{evolution_path_integral}.

Let $\nbrk{\hat \Op=\Op(\hat f,\hat\pi)}$ be a Hermitian operator for an observable. At the current time~$t$ the probability of finding various values of this observable equals the squared norm of the projections of~$\psi$ on the eigenstates of~$\hat \Op$. Associate~$\psi'$ of\rf{evolution_path_integral} and the squares of its projections on the eigenstates of the same (not evolved) operator~$\hat \Op$ with a different time moment~$t'$. Given the existence of~$\psi$, its transform\rf{evolution_convolution} also exists as the presentation of the same physical state in another basis of the quantum Hilbert space. This presentation of the state is its Schrodinger-picture wave function that has evolved from~$t$ to~$t'$ by the Hamiltonian~$\hat H'$.

\section{Presentations of smoothed discrete distribution can form wave functions}
\label{sec_structure}

\begin{figure*}[t]
\centering
\includegraphics[width=0.7\textwidth]{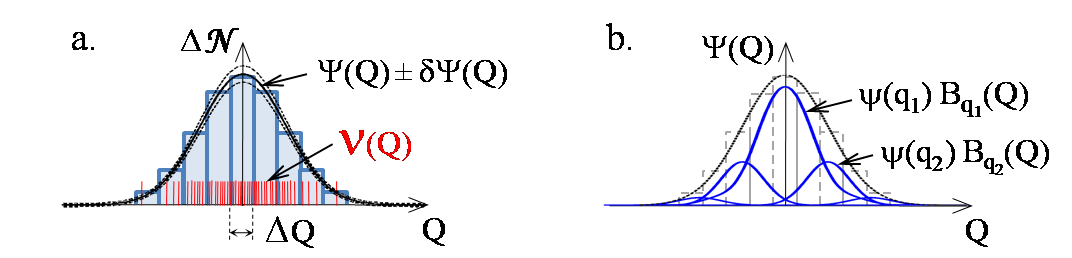}
\caption{
  (a).~The red vertical lines illustrate a one-dimensional discrete distribution~$\dist(Q)$ of a property that can be quantified by a real value~$Q$.  The distribution~$\dist(Q)$ is binned over intervals of equal width~$\Delta Q$. The resulting blue-bar histogram is fitted with a smooth function~$\Psi(Q)$. (b).~View $\Psi(Q)$  as a superposition of linearly independent, smooth functions~$B_q(Q)$ that are multiplied by adjustable fitting coefficients~$\psi(q)$, eq.\rf{psi_general}.  }
\label{fig_1d}
\end{figure*}

The paper will show, starting from this section, that certain presentations of the distribution of the generic properties of static entities of practically any nature compose the evolving wave functions of  phenomenologically viable physical worlds. The dynamics and initial conditions for some of these worlds match those for our observed universe, which therefore possibly has a similar fundamental origin. These wave functions, existing materially, determine the objective frequentist probabilities of alternate outcomes of quantum processes in the emergent worlds. By studying instead of an abstract, axiomatically defined wave function the concrete entities with the identical properties and a specific and simple implementation, we will be able to track down answers to several long-standing fundamental questions, including all of those listed in \sct{subsec_questions}. We also determine how, despite the arguments of \sct{sec_first_observation}, the Lagrangian density for observed quantum evolution can remain constant in spacetime. 	

\subsection{Alternate smooth presentations}
\label{subsec_smooth_pres}

Consider a large finite set of $\Nobj$~entities $E_a$, $\nbrk{a=1,2,\dots,\Nobj}$, of any nature that permits characterizing each of the entities by a large number of independent real properties. Pick a finite number~$N$ of their generic independent real properties~$Q^n$, $\nbrk{n=1,2,\dots,N}$. Let  $\nbrk{Q^n_a\in\Rset}$ be the value that the $n$-th property takes for the $a$-th entity.

As one of many examples, we may think of various quantifiable characteristics of the individual planets in the visible universe. We should however remember that while this or another large collection of known physical objects can indeed contain emergent evolving systems whose internal dynamical laws coincide with those of our physical world, the latter is unlikely to emerge from its own objects. Nor do we expect the fundamental entities~$E_a$ possibly behind our physical world to belong to any similar world with space and time. Instead, it appears more general for that fundamental structure to be self-contained and unrelated to any external spacetime or dynamical system.  Correspondingly, the typical fundamental entities~$E_a$ do not resemble familiar physical objects. They are visualized more appropriately as general subsets of the set of all the entities that materially exist (regardless of what it may mean). Each of these subsets possesses many independent or quasi-independent properties~$Q^n$, evaluated for the~$a$'s subset (entity~$E_a$) to~$\nbrk{Q^n_a\in \Rset}$.

Let~$\dist(Q)$ be the density of the distribution of the properties $\nbrk{Q\equiv (Q^1,\dots,Q^N)}$ over the $\Nobj$~basic entities,  enumerated by $\nbrk{a=1,2,\dots,\Nobj}$:
\be
\dist(Q)=\sum_{a=1}^\Nobj\delta(Q-Q_a)\,.
\lb{nu_q_def}
\ee
The number of the entities whose properties fall in a range $\nbrk{\Delta Q\equiv (\Delta Q^1,\dots,\Delta Q^N)}$ is
\be
\Delta \Nobj = \int_{\Delta Q} dQ \ \dist(Q)\,,
\lb{Nobj_binned}
\ee
where $\nbrk{dQ\equiv\prod_{n=1}^N dQ^n}$. Choose a resolution~$\Delta Q$ such that every region of the size~$\Delta Q$, centered at an arbitrary point~$Q_0$, contains many of the basic entities, i.e., for it $\nbrk{\Delta \Nobj\gg1}$. In addition, let the relative change in~$\Delta \Nobj$ for such adjacent regions be insignificant. Then at the resolution~$\Delta Q$ we can approximate the exact distribution density~$\dist(Q)$ of\rf{nu_q_def} by a smooth real function~$\Psi(Q)$. It means that for any window function~$W(Q)$ with characteristic width no smaller than~$\Delta Q$ and for any $Q$'s~value~$Q_0$
\be
\int\! dQ~W(Q-Q_0)\,\dist(Q)\simeq \int\! dQ~W(Q-Q_0)\,\Psi(Q)\,.~~
\lb{fit_def}
\ee

As illustrated in \fig{fig_1d}, to obtain~$\Psi(Q)$ we can, first, bin~$\dist(Q)$ over cells small in comparison to the width of~$W(Q)$ yet sufficiently large for every cell to contain many basic entities. Second, we fit the binned (i.e., coarse-grained) basic distribution with a superposition of smooth linearly-independent basis functions~$B_q(Q,t)$. Here, $t$~marks various choices of a basis of functions, whereas the parameter $\nbrk{q\equiv (q^1,\dots, q^N)}$ distinguishes the member functions of a given basis~$\{B_q(Q)\}$. Thus we match
\be
\dist(Q) \to \Psi(Q)= \int\! dq~\psi(q,t)\,B_q(Q,t)
\lb{psi_general}
\ee
so that\rf{fit_def}~holds. Importantly, the basis functions $B_q(Q,t)$~are not required to be localized around any value of~$Q$.

The alternative choices~$t$ of the basis of smooth functions~$\{B_q(Q)\}_t$ form a many-dimensional manifold. A 	certain continuous family of basis changes $\nbrk{\{B_q\}_t\to \{B_q'\}_{t'}}$ will produce evolution transformation of the Schrodinger wave function $\nbrk{\psi(q,t)\to \psi(q,t')}$, representing various temporal instances of an emergent physical field system. Some of other continuous changes of the expansion basis reflect the freedom of choosing a constant-time spatial slice of four-dimensional spacetime and choosing a gauge for the locally symmetric evolving fields. 
Association of the linear coefficients~$\psi(q,t)$ in the expansion\rf{psi_general} with something physically tangible may become intuitively natural if one considers the special case where~$B_q(Q,t)$ are the harmonic waves of specific frequencies (as illustrated by \fig{fig_outline}.b).  Possible immediate concerns about associating the linear coefficients~$\psi(q,t)$ in the expansion\rf{psi_general} with the wave function of a physical system are likely to be the following. 

First, the observed physical evolution proceeds by specific dynamical laws, constant in space and time. Why should dynamical laws restrict the change of an arbitrary basis for decomposing the fitting function? The existence of emergent wave functions of quantum fields with specific dynamical laws is demonstrated through \scts{sec_evolution}-\ref{sec_why_the}.

Second, the norm of a wave function should determine the physical probability. We will find that for the suggested association this is indeed the case. Thus, unlike all the past formulations of quantum mechanics, the probability of a macroscopic outcome of a quantum process in the generically emergent systems is not postulated but follows from the first principles. 

Third, $\psi(q,t)$ is real while the physical wave function is complex. Yet we will see soon, in \sct{subsec_complex}, that the emergent quantum states to develop from the real $\psi(q,t)$ are naturally described by a \emph{complex} wave function.

\subsection{Uncertainty of smoothing yields probability-related norm of states}
\label{subsec_norm}

Consider the linear space spanned by linear combinations of the functions~$B_q(Q)$ from a current basis in expansion\rf{psi_general}. Let these linear combinations,
\be
\Psi_i(Q)= \int\! dq~\psi_i(q)\,B_q(Q),
\lb{Psi_i_def}
\ee
have arbitrary real or complex coefficients~$\psi_i(q)$. Correspondingly, $\Psi_i(Q)$~may be real or complex. The linear combinations\rf{Psi_i_def} of particular physical significance are those that will represent dynamically isolated Everett's branches of the global wave function of an emergent physical system.

A norm on a linear space is equivalent to a Hermitian product~$\langle1|2\rangle$ for any members $|1\rangle$ and~$|2\rangle$ of the linear space. One might suggest $\nbrk{\int dQ\,\Psi_1^*(Q)\,\Psi_2(Q)}$ as a ``natural'' candidate for the Hermitian product of the linear combinations\rf{Psi_i_def}. However, this is unmotivated and, moreover, unacceptable because the quantity $\nbrk{\int\! dQ\,\Psi_1^*\Psi_2}$ depends on the subjective choice of the arbitrary coordinates~$Q$. It thus cannot specify the objective physical probability, unrelated to our description of the system.
Since the quantum-mechanical Hermitian product has unequivocal physical meaning---it determines the probability of observing a particular branch of the wave function after a measurement---to understand the emergence of the physical Hermitian product, let us understand the emergence of the probability.

Let us bin the discrete fundamental distribution~$\dist(Q)$ over coordinate cells of a size~$\Delta Q$ chosen as described just below eq.\rf{Nobj_binned}. Without limiting the generality, let all the bins have the same size~$\Delta Q$. Then, by\rf{Nobj_binned} and\rf{fit_def}, the number of the basic entities~$\Delta \Nobj_b$ in a bin~$b$ fluctuates over the bins about a smoothly changing value
\be
\langle \Delta \Nobj_b \rangle= \langle\Psi(Q_b)\rangle\,\Delta Q
\lb{Delta_bar_N}
\ee
with a variance $\sigma^2(\Delta \Nobj_b)$. 

For a finite set of the fundamental entities~$E_a$, the fitting coefficients~$\psi(q,t)$ in\rf{psi_general} and the respective smooth fitting function~$\Psi(Q)$ have an intrinsic ambiguity. 
We can require the fitting function to minimize some artificially chosen statistics. A simple and convenient choice is $\chi^2$~statistics,
\be
\chi^2 = \sum_b \frac{[\Delta \Nobj_b-\Psi(Q_b)\,\Delta Q]}{\sigma^2(\Delta \Nobj_b)}^2.
\lb{chi2_def}
\ee
For a different choice of the bins or statistics the best fit somewhat differs. Consider only binning choices for which in every bin~$\nbrk{\Delta \Nobj\gg1}$ but~$\Delta Q$ is much smaller than the characteristic variation scales of~$\langle\Psi(Q)\rangle$. Within these limits on admissible binning, a fit~$\Psi(Q)$ is either rejected at a statistically significant level for every binning and sensible statistics or is accepted for all of them. Thus the notion of~$\Psi(Q)$ fitting---up to some unavoidable uncertainty---the generic discrete distribution~$\dist(Q)$ is objective.

In terms of the variance density
\be
v(Q)\equiv \frac{\sigma^2(\Delta \Nobj)}{\Delta Q}\,,
\lb{v_q_def}
\ee
the~$\chi^2$ statistics\rf{chi2_def} equals
\be
\chi^2= \sum_b \fr{\Delta Q}{v(Q_b)}\lf[\fr{\Delta \Nobj_b}{\Delta Q}-\Psi(Q_b)\rt]^2.
\lb{chi_2_v_sum}
\ee
We assume from now on that~$\Delta \Nobj_b$ for adjacent bins are uncorrelated for any possible suitable binning; hence, locally~$v(Q)$ of\rf{v_q_def} is independent of the bin size~$\Delta Q$. Under these conditions, $\Delta \Nobj_b$~is determined, at least locally, by the Poisson process with the expectation value\rf{Delta_bar_N}.

Consider a prospective smooth fitting function 
\be
\Psi(Q)=\langle\Psi(Q)\rangle+\delta\Psi(Q)\,,
\ee 
where $\delta \Psi(Q)$, similarly to $\langle\Psi(Q)\rangle$, varies negligibly over~$Q$ in any region of the size~$\Delta Q$. By $\langle\Psi(Q)\rangle$ definition\rf{Delta_bar_N},
\be
\langle\Delta \Nobj_b/\Delta Q\rangle = \langle\Psi(Q_b)\rangle.
\lb{Delta_Nobj_b_avr}
\ee
Averaging the terms on the right-hand side of\rf{chi_2_v_sum} over nearby regions over which~$v(Q)$ changes negligibly and applying\rf{Delta_Nobj_b_avr} yields:
\be
\chi^2(\Psi)= \chi^2(\langle\Psi\rangle)+\sum_b \fr{\Delta Q}{v(Q_b)}\lf[\delta \Psi(Q_b)\rt]^2.
\lb{chi2_variation}
\ee
For the smooth~$\delta\Psi(Q)$ and~$v(Q)$, we replace the sum over the bins~$b$ by an integral and arrive finally at
\be
\delta\chi^2(\delta\Psi)\equiv \chi^2(\Psi) - \chi^2(\langle\Psi\rangle)
\simeq \int \fr{dQ}{v(Q)}\lf[\delta \Psi(Q)\rt]^2.~
\lb{delta_chi2}
\ee
This simple formula relates the fitting coefficients~$\psi(q,t)$ in\rf{psi_general} to the probability-connected wave function of an emergent physical system as follows. 

Suppose that the physical system that has emerged can be artificially separated into a simple studied subsystem with relatively few degrees of freedom and the remaining ``environment'' with many degrees of freedom. Consider ``evolution'' of~$\psi(q,t)$ due to continuously changing the basis~$\{B_q(Q)\}_t$ in decomposition\rf{psi_general}. 

Let this evolution split~$\psi(q,t)$ in, for simplicity, two terms:
\be
\psi(q,t)=\psi_1(q,t)+\psi_2(q,t)\,. 
\lb{split_coefficients}
\ee
They represent two decoherent Everett's branches of~$\psi(q,t)$
under the evolution, induced by the basis change.
The terms $\psi_1$ and~$\psi_2$ in\rf{split_coefficients} describe different states of the entire emergent system, including the studied subsystem and its environment. These terms decohere because of the interaction of the numerous degrees of freedom of the environment.

After $R$~repetitions of the process that splits the wave function as described, we arrive at a superposition of $2^R$~outcomes:
\be
\psi=\psi_{11\dots1}+\psi_{21\dots1}+\dots+\psi_{22\dots2}\,.
\lb{splits}
\ee
Let the splits be caused by a quantum measurement, e.g., determination of a projection of an election's spin. Correspondingly, in\rf{splits} an individual term~$\psi_{r_1r_2\dots r_R}$, with $\nbrk{r_i\in\{1,2\}}$, represents the measured subsystem and its environment, including the observer, in the branch with the specific sequence of the measured results~$(r_1,r_2,\dots, r_R)$. 

The alternate, decoherent branches of quantum evolution are typically mutually orthogonal in the Hilbert space of the quantum system. Indeed, first, a typical quantum measurement splits the wave function along different non-degenerate, hence orthogonal, eigenstates of a Hermitian operator of the measured observable. For example, a measurement of the vertical projection of the election spin yields the orthogonal spin-up and spin-down states. Second, after the  split and subsequent decoherence, the typical formed ``einselected''\ct{Zurek_03}, i.e., stable to further splitting due to subsystem-environment interaction, branches of future evolution are usually\ct{Zurek_81,Zurek_03} the orthogonal eigenstates of the subsystem-environment interaction Hamiltonian, which is Hermitian.

The split of the fitting coefficients\rf{split_coefficients} respectively splits the overall fitting function\rf{psi_general}: $\nbrk{\Psi(Q)=\Psi_1(Q)+\Psi_2(Q)}$. Consider the parameterization-independent\footnote{
	Let us prove that\rf{scalar_product_rho} does not depend on the choice of the variables~$Q$.
	A change of the variables $\nbrk{Q\to Q'(Q)}$ gives
	$$
	dQ'=J\,dQ\,
	$$
	where $\nbrk{J\equiv|\partial Q'/\partial Q|}$ 
	is the Jacobian of the coordinate transformation. 
	Then, 
	since $\nbrk{\langle d\Nobj\rangle=\Psi\,dQ=\Psi'\,dQ'}$,
	$$
	\Psi'=J^{-1}\Psi\,.
	$$
	The variance density~$v(Q)$ transforms as
	$$
	v'=\frac{d\sigma^2}{dQ'}=J^{-1}v\,.
	$$
	Thus the right-hand side of\rf{scalar_product_rho} 
	is manifestly parameterization-independent.
	}
scalar product
\be
\langle1|2\rangle= \int \frac{dQ}{v(Q)}\,\Psi_1(Q)\,\Psi_2(Q)\,.
\lb{scalar_product_rho}
\ee
Let the terms~$\Psi_1$ and~$\Psi_2$ represent the Everett branches for evolution by a Hamiltonian that is orthogonal under this scalar product (Hermitian under the related Hermitian product, to be identified in \sct{subsec_Hermitian_product}). Then $\Psi_1$ and~$\Psi_2$ are orthogonal under this product: $\nbrk{\langle1|2\rangle=0}$.
Then, denoting the initial (represented by~$\Psi$) state with~$|0\rangle$, we have $\nbrk{\langle 0|0\rangle=\langle 1|1\rangle+\langle 2|2\rangle}$. More generally, for repeated measurements and multiple mutually orthogonal outcomes~$\Psi_i$,
\be
\langle 0|0\rangle= \sum_{{\rm outcomes}~i} \langle i|i\rangle\,.
\lb{norm_sum_outcomes}
\ee
 
\Sct{subsec_mixed_states} will introduce more general realizations of emergent mixed quantum states. They can be described by a density matrix~$\rho(q,q',t)$. For them, eq.\rf{norm_sum_outcomes} generalizes to 
\be
\rho= \sum_{i} \rho_i\,,
\lb{rho_split_preview}
\ee 
eq.\rf{split_rho}.
Decoherence ensures that the off-diagonal terms~$\rho_{ij}$, $\nbrk{i\neq j}$, in the pointer state basis rapidly decay and vanish\ct{Zurek_03}. For clarity, let us for now focus on pure states, represented by wave functions. Generalization of the obtained results to mixed states is straightforward, \sct{subsec_mixed_states}.

Let the evolution of~$\psi(q,t)$ of\rf{psi_general} split it and respectively split the overall smooth fitting function~$\Psi(Q)$ into a sum of decoherent Everett's branches:
\be
\Psi=\sum_i\Psi_i\,.
\lb{Psi_split}
\ee
Consider one of the branches,~$\Psi_i$.
Depending on~$\delta\chi^2$ that results from setting the variation~$\delta\Psi(Q)$ in\rf{delta_chi2} to~$\Psi_i(Q)$, we can distinguish two qualitatively different situations: 
\ben
\item[a.]
When~$\delta\chi^2(\Psi_i)$ of\rf{delta_chi2} is large,
removal of the term~$\Psi_i$ from the sum over the alternate decoherent branches\rf{Psi_split} degrades the confidence level of the overall fit, changing from~$\Psi$ to~$\nbrk{\Psi-\Psi_i}$, by an order of~$100\%$,
\een
vs.
\ben
\item[b.]
When~$\delta\chi^2(\Psi_i)$ is small,
removal of the branch~$\Psi_i$ does not qualitatively affect the fit's confidence level. 
\een
We define (up to a factor of the order of unity) a \emph{statistical significance threshold}~$\delta\chi^2_{\rm min}$ as the borderline between these situations.

The squared norm~$\langle i|i\rangle$ of the branch~$\Psi_i$ with respect to the scalar product\rf{scalar_product_rho} equals $\delta\chi^2(\Psi_i)$ of\rf{delta_chi2}:
\be
\langle i|i\rangle=\delta\chi^2(\Psi_i)\,.
\lb{i_squared_norm}
\ee
Suppose that the squared norm\rf{i_squared_norm} of~$\Psi_i$ has fallen below the introduced threshold~$\delta\chi^2_{\rm min}$ of unambiguous determination of the fitting function~$\Psi$:
\be
\langle i|i\rangle \,\lesssim\ \delta\chi^2_{\rm min}\,.
\lb{norm_unit}
\ee
Then~$\Psi_i$ no longer represents an objectively existing branch of the system's evolution. A quantum state~$|i\rangle$ may be discussed mathematically, but it has no objective representation through the underlying fundamental entities. 

It may be helpful to rephrase the above as follows. We cannot describe the \emph{discrete} basic distribution~$\dist(Q)$ by a \emph{smooth} approximation~$\Psi(Q)$ beyond certain precision. Once we require a fit~$\Psi(Q)$ to be too accurate, it can no longer be acceptably smooth because the underlying discrete distribution~$\dist(Q)$ is not. 

Thus every physically meaningful term in\rf{norm_sum_outcomes} should exceed the positive threshold~$\delta\chi^2_{\rm min}$. Consequently, in the emergent physical system the number of the physically existing Everett branches for the alternate outcomes of a quantum process is finite. This will let us \emph{calculate} the frequentist probability of the outcomes instead of postulating it. It will be the subject of \sct{sec_probabilities}.

As shown, the condition $\nbrk{\langle i|i\rangle \gg \delta\chi^2_{\rm min}}$ is necessary for the branch~$\Psi_i(Q)$ to be an objectively existing component of the smoothed generic discrete distribution~$\dist(Q)$. It yet remains insufficient for the physical existence of an object described by the wave function~$\Psi_i(Q)$, or by its equivalent presentations~$\psi_i(q)$. We will see this in \sct{subsec_implications}. Most of our discussion will nonetheless neglect other requirements to objective existence of the objects represented by the emergent wave function. Thus for most of the present paper we operationally regard as physical all the emergent Everett branches whose squared norm well exceeds~$\chi^2_{\rm min}$ and whose evolution is unambiguous and regular.
 
For arbitrary branches~$|1\rangle$~and~$|2\rangle$ of an emergent wave function, the scalar product\rf{scalar_product_rho} in terms of the fitting coefficients~$\psi(q,t)$ of\rf{psi_general} at a fixed~$t$ is
\be
\langle1|2\rangle= \int dq\,dq'\,\psi_1(q)\,M(q,q')\,\psi_2(q')\,,
\lb{scalar_product_gen_f_psi}
\ee
where
\be
M(q,q') \eqa \int \fr{dQ}{v(Q)}\,B_q(Q)\,B_{q'}(Q)\,.
\lb{M_scalar_product_gen}
\ee
So far,~$\psi(q)$ is real because the fitting function\rf{psi_general} is real. The next section will show that the standard complex wave function and the Hermitian product arise for the emergent quantum systems automatically.

The canonical form for the scalar product,
\be
\langle1|2\rangle= \int dq~\psi_1(q)\,\psi_2(q)\,,
\lb{scalar_product_psi_real}
\ee
follows whenever in\rf{scalar_product_gen_f_psi}
\be
M(q,q') = \delta(q-q')\,.
\ee
For example, the scalar product\rfs{scalar_product_gen_f_psi}{M_scalar_product_gen} takes the canonical form\rf{scalar_product_psi_real} for
\be
B_q(Q)=v^{1/2}(Q)\,\delta(Q-q).
\lb{B_canonical}
\ee
Any orthogonal transformation of this basis produces another basis~$\{B_q(Q,t)\}$ where the scalar product\rf{scalar_product_gen_f_psi} also has the canonical form\rf{scalar_product_psi_real}.

\section{Emergent evolving wave function and field operators}
\label{sec_evolution}

We continue to explore emergent quantum systems whose evolving wave functions consist of alternate presentations~$\psi(q)$ of the smoothed (coarse-grained) generic discrete distribution~$\dist(Q)$. This section will show that the emergent systems automatically possess a \emph{complex} wave function, local field operators, and probability-related Hermitian product for the wave function's dynamically isolated Everett branches, becoming quantum states. The section will also present a simple example of quantum field evolution constructed from alternate presentations of the static distribution. This example cannot describe any physical world by the reasons stated at the end of the section. Yet based on it, subsequent \scts{sec_gauge_fields}-\ref{sec_why_the} will identify physically viable emergent quantum fields.

\subsection{Emergence of a complex wave function}
\label{subsec_complex}

Consider a presentation~$\psi(q)$ of the smoothed $\dist(Q)$ for which the scalar product\rf{scalar_product_rho} has the canonical form\rf{scalar_product_psi_real}. Expand this~$\psi(q)$ over the irreducible representations of the abelian group of uniform translation of its variables~$\nbrk{q= (q_1,\dots,q_N)}$:
\be
q\to q' = q-\Delta q\,,
\lb{q_shift}
\ee
where~$\nbrk{\Delta q\equiv (\Delta q_1,\dots,\Delta q_N)}$ is constant (independent of~$q$). 
The group\rf{q_shift} corresponds to the various shifts of the coordinates~$q$ for the configuration space of the emergent system to be identified with~$\psi(q)$.

The irreducible representations of the group\rf{q_shift} are two-dimensional real (one-dimensional complex) spaces of linear combinations of basis functions $\nbrk{\{\cos(q{\cdot}p),\,\sin(q{\cdot}p)\}}$ for any fixed $\nbrk{p=(p_1,\dots,p_N)}$. The expansion over them is the Fourier transformation:
\be            
\psi(q)\eqa  \int\!\dbar p\lf(\cos(q{\cdot}p)\,,\, -\sin(q{\cdot}p)\vphantom{\hat I}\rt)
	\lf(\ba{c}\psi^r(p)\\ \psi^i(p)\ea\rt)= \nn\\*
\eqa \int\!\dbar p\,\re\lf[e^{iq{\cdot}p}\,\psi(p)\rt],
\lb{psi_FT}
\ee
where
\be
i \equiv \lf(\ba{cr}0&-1\\1&0\ea\rt), & & q{\cdot}p \equiv \sum_n q^n p_n\,,\quad
\lb{i_def}\\
\psi(p) &\!\equiv\! & \lf(\ba{c}\psi^r(p)\\ \psi^i(p)\ea\rt),
\lb{psi_p}
\ee
and the real-part function~``$\re$'' in\rf{psi_FT} selects the upper component of its two-component argument. The two-component function\rf{psi_p} is another representation of the smoothed generic distribution. This representation is dual to the representation~$\psi(q)$. Under a shift\rf{q_shift} of the variables~$q$, the values of~$\psi(p)$ transform as
\be
\psi(p)\to e^{i\Delta q{\cdot}p}\,\psi(p)\,.
\ee

The nature of the dual, ``momentum'' representation~$\psi(p)$ as the amplitudes of the ``waves that compose''~$\psi(q)$, eq.\rf{psi_FT}, for the emergent wave function is analogous to the nature of the representation~$\psi(q)$ itself: the amplitudes of the ``waves that compose'' the generic distribution of arbitrary quantities [eq.\rf{psi_general} or \fig{fig_outline}]. Quantum physical evolution, discussed starting from \sct{subsec_operators}, intermixes the configuration and momentum representations of the evolving wave function. It is therefore reasonable that the emergent wave function in both representations has the same fundamental nature.

We may as well consider the representation that is dual to the dual representation\rf{psi_p}. To this end, we expand an arbitrary complex~$\psi(p)$ over the waves that transform irreducibly under shifts of its coordinates~$p$:
\be
p\to p' = p-\Delta p\,,
\lb{p_shift}
\ee
where~$\nbrk{\Delta p\equiv (\Delta p_1,\dots,\Delta p_N)}$ is constant.
This expansion,
\be
\psi(p)= \int d q\,e^{-iq{\cdot}p}\,\psi(q)\,
\lb{psi_p_FT_inv}
\ee
where now
\be
\psi(q)= \lf(\ba{c}\psi^r(q)\\ \psi^i(q)\ea\rt),
\lb{psi_q_complex}
\ee
leads naturally to a complex emergent wave function~$\psi(q)$. (In a mathematician's view, this is a manifestation of the Pontryagin duality for the irreducible representations of an abelian group.)

In a physicist's view, for a real~$\psi(q)$, eq.\rf{psi_FT} defines only the even (to reflection $\nbrk{p\to -p}$) component of~$\psi^r(p)$ and the odd component of~$\psi^i(p)$. 
Odd contributions to~$\psi^r(p)$ or even contributions to~$\psi^i(p)$ do not change the left-hand side of\rf{psi_FT}.
However, the wave function of a physical, observationally accessible state will be given \emph{not} by the overall wave function~$\psi(q)$ but by separate terms of its splitting, e.g.\rf{splits}, into the Everett branches, which decohere during its evolution.
The overall sum\rf{splits} for $\nbrk{\psi(p)\equiv \psi^r(p)+i\psi^i(p)}$ in the momentum representation\rf{psi_FT} of the real~$\psi(q)$ satisfies the condition $\nbrk{\psi(p)=\psi^*(-p)}$. This condition does not however apply to a term~$\psi_i$ that describes an individual decoherent Everett's branch, e.g., a single term on the right-hand side of\rf{splits}. For it, we need the general complex~$\psi_i(p)$. In $q$-space such a physical state is represented by a two-component or, equivalently, a \emph{complex} function~$\psi_i(q)$, related to~$\psi_i(p)$ by\rf{psi_p_FT_inv}:
\be
\psi_i(q)= \int\!\dbar p~e^{iq{\cdot}p}\,\psi_i(p)\in \Cset\,.
\lb{psi_FT_complex}
\ee

Thus we necessarily arrive at representing the individual decoherent branches of quantum evolution by a complex wave function even when the basic elements that produced the evolving system do not possess a complex structure. The two key ingredients for the emergence of the complex  representation are: 
\ben
\item[(i)] mixing of the configuration and momentum representations of the evolving wave function upon its evolution; \item[(ii)] decoherence and the resulting isolation of its individual Everett's branches~$\psi_i$. 
\een

To complete the proof of automatic emergence of complex structure that is required for quantum mechanics, let us show that the linear transformation of the real~$\psi(q)$ in\rf{psi_general} due to a physically relevant change of the basis~$\{B_q\}$ is linear on the \emph{complex} space\rf{psi_FT_complex}. (As seen next, this statement is not as trivial as it might appear superficially.) 

Let us drop the subscript~``$i$'' when referring to an emergent quantum state that is represented by a single Everett's branch\rf{psi_FT_complex}. Let a complex~$\psi(q)$ of\rf{psi_FT_complex} transform into
\be
\psi' =	\hat A\psi \,,
\ee
standing for
\be
\lf(\ba{c}\psi'{}^r \\ \psi'{}^i\ea\rt)=
	\lf(\ba{cc}\hat A_r^r & \hat A_i^r \\ \hat A_r^i & \hat A_i^i\ea\rt)\lf(\ba{c}\psi^r \\ \psi^i\ea\rt).
\ee
Here $\psi^{r,i}$,~$\psi'{}^{r,i}$ are real and the operators~$\hat A^\iota_\kappa$ are linear. Complex linearity,
\be
\hat A\,(c_1\psi_1+c_2\psi_2)= c_1\hat A\psi_1+c_2\hat A\psi_2
\ee
for any $\nbrk{c_1,\,c_2\in\Cset}$, holds if and only if the operator~$\hat A$ commutes with the matrix~$i$ from\rf{i_def}:
\be
\hat A\,i =i\,\hat A\,.
\ee 
Hence for the complex linearity of operators it is necessary and sufficient that
\be
\hat A_r^r= \hat A_i^i\,,\quad \hat A_i^r= -\hat A_r^i\,.
\ee

All the configuration-space coordinate and momentum operators
\be
\hat q^n=q^n\quad\mbox{and}\quad \hat p_n=-i\partial/\partial q^n
\lb{pq_n_def}
\ee
satisfy this condition. Therefore, they are linear on the complex space of the wave functions\rf{psi_FT_complex}. So is complex-linear any Hamiltonian that is a multinomial or analytic function of these operators. Thus evolution transformation generated by such a Hamiltonian is complex-linear.

\subsection{Hermitian product}
\label{subsec_Hermitian_product}

We now identify the unique probability-related Hermitian product of the complex wave functions of emergent states\rf{psi_FT_complex}. 

Consider a  smoothed~$\dist(Q)$'s representation~$\psi(q)$ for which the scalar product\rf{scalar_product_rho} has the canonical form\rf{scalar_product_psi_real}.
Then, by Fourier-decomposing the real~$\psi(q)$ of\rf{psi_general} with\rf{psi_FT}, 
\be                           
\langle\psi|\psi\rangle \eqa \! \int dq\,\psi^2(q)=\int\dbar p\,\lf\{[\psi^r(p)]^2 + [\psi^i(p)]^2\rt\}=\nn\\*
\eqa  \int\dbar p\,\lf|\psi(p)\rt|^2\,.
\ee
By \sct{sec_structure}, $\langle\psi|\psi\rangle$ quantifies the capacity of a branch (state)~$\psi_i$ to withstand future splits before it stops being an objectively existing smooth constituent of the underlying discrete structure. \Sct{sec_probabilities} will prove that, as a result, $\langle\psi_i|\psi_i\rangle$~determines the frequentist probability for an intrinsic observer in the emergent system to follow this branch. 
Likewise, the complex wave function~$\psi(q)$ of an emergent physical state\rf{psi_FT_complex} satisfies:
\be
\langle\psi|\psi\rangle= \int\dbar p\,\lf|\psi(p)\rt|^2=
\int\,dq\,|\psi(q)|^2\,.
\lb{norm_standard}
\ee 

Look for a probability-related product of the complex prospective wave functions\rf{psi_FT_complex} in the general bilinear form
\be
\langle\psi_1|\psi_2\rangle = (\psi_1^r,\psi_1^i)
	\lf(\ba{cc}\hat M_{rr} & \hat M_{ri} \\ \hat M_{ir} & \hat M_{ii}\ea\rt)\lf(\ba{c}\psi^r_2 \\ \psi^i_2\ea\rt).
\lb{product_general_matrix}
\ee
Require that the product\rf{product_general_matrix}
\ben
\item[(i)]
 reduces to the physical measure\rf{norm_standard} for $\nbrk{\psi_1=\psi_2}$;
\item[(ii)]
is linear in its second argument on the \emph{complex} linear space\rf{psi_q_complex}:
\be
\langle\psi_1|c\,\psi_2\rangle = c\,\langle \psi_1|\psi_2 \rangle 
\qquad \forall c\in\Cset\,.
\ee
\een
It is straightforward to establish that these two requirements uniquely determine the operators $\hat M_{\iota\kappa}$ in\rf{product_general_matrix} so that\rf{product_general_matrix} becomes the canonical Hermitian product:
\be
\langle\psi_1|\psi_2\rangle= \int dq\,\psi^*_1(q)\,\psi_2(q)\,.
\lb{Hermitian_product_canonical}
\ee

\subsection{Normal wave function}
\label{subsec_psi_0}

The considerations presented next suggest that---regardless of the nature of the underlying fundamental entities---the smoothed distribution~$\Psi(Q)$ of their generic properties~$Q$ takes in appropriate variables the generic Gaussian form. (However, the corresponding Gaussian wave function~$\psi\gaus(q)$
in\rf{psi_0} below is generally \emph{not the initial wave function} of the short-scale modes that appear from the Planck scale during inflation or during black hole evaporation. The initial wave function of these modes is determined as described in \sct{subsec_Planck_scale}.)

Consider a huge collection of different objects any of which can be characterized by many independent quantifiable properties. Most properties~$Q^p$ of practical relevance that characterize objects familiar to us are \emph{not} distributed normally. However, the general linear combinations~$Q^n$ of many independent properties~$Q^p$ are Gaussian, at least, for the numerous cases that meet the sufficiency conditions of the central limit theorem of probability theory. For such randomly selected $N$~uncorrelated ``generic properties'' $\nbrk{\{Q^n\}\equiv Q}$, it is straightforward to find a coordinate basis in which their smoothed distribution~$\Psi$ has the generic Gaussian (normal) form
\be
\Psi\gaus(Q)= A^2e^{-Q^2}.
\lb{rho_0}
\ee
Here $\nbrk{Q^2\equiv\sum_{n=1}^N (Q^n)^2}$ and a positive parameter~$A$ sets normalization of the distribution. 

\begin{figure}[t]
\centering
\includegraphics[width=0.3\textwidth]{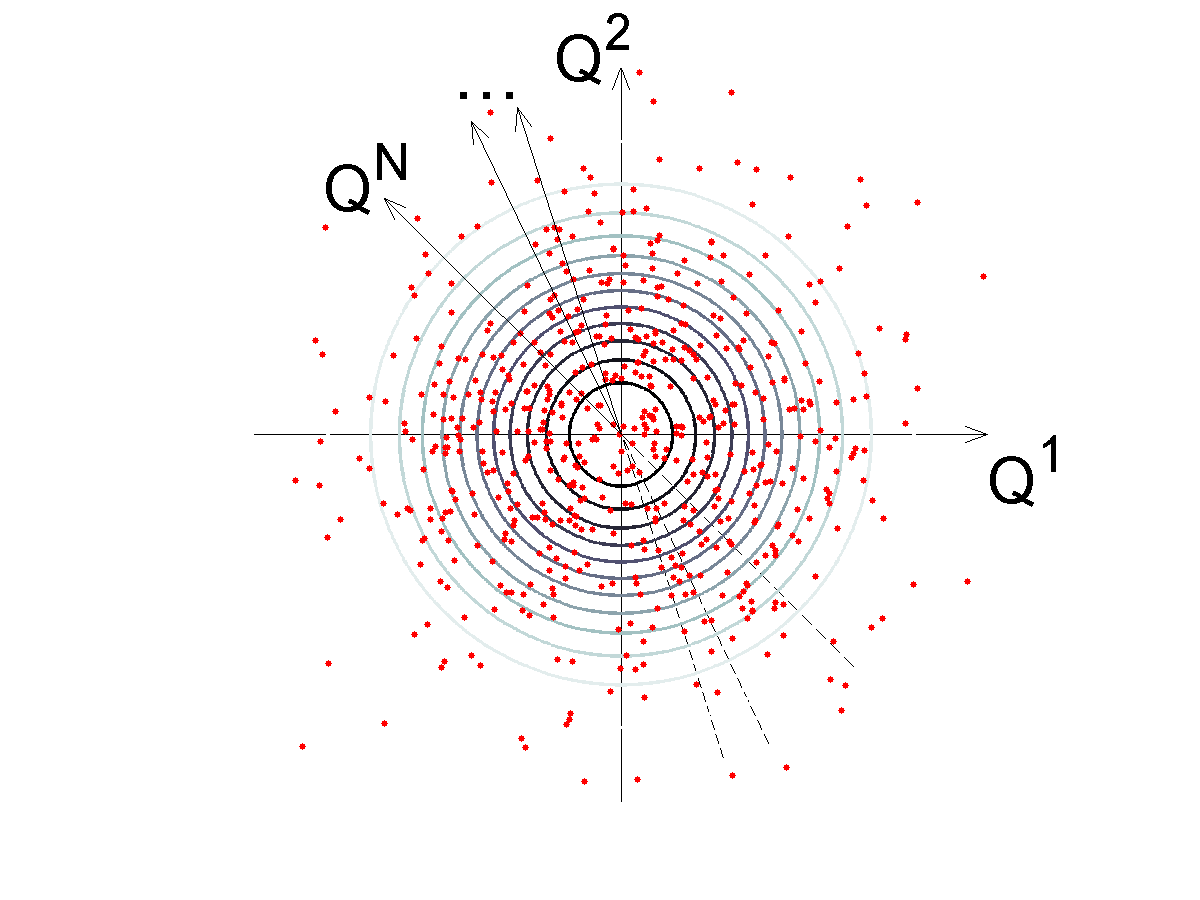}
\caption{
Visualization of the generic distribution of~$N$ arbitrary properties $\nbrk{(Q^1,Q^2,\dots,Q^N)}$. We depict its projection on the plane~$\nbrk{(Q^1,Q^2)}$. The concentric circles are isovalue curves of its fit by the generic Gaussian function\rf{rho_0}.} 
\label{fig_A}
\end{figure}

\Fig{fig_A} visualizes the exact discrete distribution~$\dist(Q)$ of some $N$~generic properties~$Q^n$ by the red dots. The concentric circles in \fig{fig_A} depict isovalue surfaces of their smoothed distribution~$\Psi(Q)$.

When~$\dist(Q)$ is determined by a local Poisson process, described immediately below\rf{chi_2_v_sum}, the variance density\rf{v_q_def} equals:
\be
v(Q)\equiv \frac{d\sigma^2}{dQ~}= \frac{\langle d\Nobj\rangle}{dQ}\simeq \Psi(Q)\,,
\ee
using\rf{Delta_bar_N} and replacing~$\langle\Psi(Q)\rangle$ by its approximation~$\Psi(Q)$.
The presentation~$\psi(q)$ of~$\Psi\gaus(Q)$ from\rf{rho_0} for the canonical basis\rf{B_canonical} is therefore also Gaussian:
\be
\psi\gaus(q)= A\,e^{-q^2/2}\,, 
\lb{psi_0}
\ee
with $q^2\equiv \sum_n (q^n)^2$.

The normal wave function\rf{psi_0} has two suggestive properties. First, it \emph{would} describe the initial ground state (the Bunch-Davies vacuum\ct{BunchDavies78}) of the field modes during inflation if the fields did not interact. Second, it is invariant under the Fourier transformation\rf{FT_standard}. Moreover, it is invariant under the broader, continuous group of \emph{fractional} Fourier transformation [eqs.\rfs{psi_rotation}{H_rotation} below], arising naturally in the considered structure as seen next.

\subsection{Evolving field operators}
\label{subsec_operators}

We start from a single degree of freedom~$q\in \Rset$ with a wave function~$\psi(q)$. Its configuration- and momentum-space representations~$\psi(q)$ and~$\psi(p)$ are continuously connected by a Lie group of transformations
\be
\psi \to \psi' = \hat U\psi\,, \quad \hat U =\exp(-i\hat H\varphi)\,
\lb{psi_rotation}
\ee
with
\be
\hat H \eqa  \fr12\lf(\hat p^2 + \hat q^2 -1\rt)\!,
\lb{H_rotation}\\*
\hat q\, \psi(q) \eqva  q\, \psi(q)\,,  \quad  \hat p\, \psi(q) \equiv -i\frac{\pd}{\pd q}\, \psi(q)\,.
\ee
An infinitesimal Schrodinger transformation\rf{psi_rotation} by $\nbrk{\varphi=d\varphi}$ is equivalent to the following Heisenberg transformation of the operators $\hat q$ and $\hat p$:
\be
 \ba{lll}
\hat q &\to &  \hat U^{-1} \hat q\, \hat U= \hat q + i[\hat H,\hat q]\,d\varphi= \hat q + \hat p\, d\varphi\,, \quad\quad \\
\hat p &\to & \hat U^{-1} \hat p\, \hat U=\hat p + i[\hat H,\hat p]\,d\varphi= \hat p - \hat q\, d \varphi \,.
 \ea
\ee
Then for a finite transformation parameter~$\varphi$
\be
	\ba{lll}
\hat q &\to & \hat q\,\cos \varphi + \hat p \,\sin \varphi \,,  \\
\hat p &\to & -\hat q\,\sin \varphi + \hat p\, \cos \varphi \,.
	\ea
\ee
This Lie group of transformations includes the Fourier transformation ($\varphi =\pi/2$) and the inverse Fourier transformation ($\varphi =-\pi/2$). Since the Hamiltonian\rf{H_rotation} annihilates the Gaussian wave function~$\psi\gaus$ of\rf{psi_0}, the group of transformations\rf{psi_rotation}, generated by this Hamiltonian, leaves~$\psi\gaus$ invariant.

The prospective emergent wave functions of bosonic fields that will be constructed in this and the next two sections are not objectively existing physical systems yet. The end of the current section will show why. Emergent wave functions of actual physical systems with specific dynamics (of locally supersymmetric fields) will be discussed in \sct{sec_emergent_SUSY}.

The canonical generic ``emergent wave function''~$\psi\gaus$ from\rf{psi_0} is identical to the wave function of the vacuum state of a free scalar field in a finite region of space constructed as follows. For simplicity, we set the space~$\{\xv\}$ to a cube $\nbrk{0\leq x,y,z\leq L}$ with coordinate volume
\be
V=L^3. 
\ee
We impose the periodic boundary conditions on its sides, i.e., construct a quantum field in a spatial region with torus topology. 

Consider the following Hermitian linear combinations of the basic operators $\hat q^n$ and $\hat p_n$ from\rf{pq_n_def}:
\be                 
 \ba{r}
\hat\phi(\xv) \equiv {\displaystyle \fr1{V^{1/2}}\sum_{\mv}\fr1{\sqrt{2\omega_{\mv}}}}\,[\hat q^c_\mv \cos (\kv_\mv{\cdot}\xv) \,+~~\\
   +\ \hat q^s_\mv \sin (\kv_\mv{\cdot}\xv) ]
 \ea
\lb{phi_standing_harmonics}
\ee
where $\hat q^c_\mv$ and $\hat q^s_\mv$ are various operators~$\hat q^n$ of the arguments~$q^n$ of the ``emergent wave function''\rf{psi_0},
\be
\kv_\mv\equiv \frac{2\pi}{L}\,\mv\,,
\lb{kv_m}
\ee
and $\nbrk{\mv=(m_1,m_2,m_3)}$ runs\footnote{
	The terms with $\nbrk{\mv=(0,0,0)\equiv\bm0}$ require special treatment, especially for a massless field. 
	Yet for a large volume~$V$ observable effects of those terms are negligible. 
	We thus simply disregard the $\nbrk{\mv=\bm0}$ terms and remove them from the sum\rf{phi_standing_harmonics}.
}
over the sets of three integers with $\nbrk{|m_{1,2,3}|<M/2}$. We will specify the positive real numbers~$\omega_{\mv}$ later in\rf{omega_kv}.
In this and the next section, until promoting the spacetime metric to a dynamical degree of freedom in \sct{sec_gravity}, $\nbrk{{\bm a}{\cdot}{\bm b}}\equiv\sum_i a_ib_i$ for any spatial vectors~${\bm a}$ and~${\bm b}$. 
	
Set 
\be
\omega_{-\mv}=\omega_{\mv}\,.
\lb{omega_reflection}
\ee
This lets us symmetrize $\hat q^c_\mv$ and antisymmetrize $\hat q^s_\mv$ in $\nbrk{\mv\to -\mv}$ without changing the sum in\rf{phi_standing_harmonics}. We therefore require that
\be
\hat q^c_{-\mv} = \hat q^c_\mv\,,\quad \hat q^s_{-\mv} = -\hat q^s_\mv\,.
\lb{q_reflection}
\ee

Eq.\rf{phi_standing_harmonics} describes a change of coordinates in the configuration space from~$\{q^{c,s}_\mv\}$ to~$\{\phi(\xv_{\nv})\}$. It is convenient to take
\be
\xv_{\nv}= \frac{L}{M}\,\nv\,,
\lb{x_n_def}
\ee
$\nbrk{\nv=(n_1,n_2,n_3)}$ with the integers $\nbrk{0\leq n_{1,2,3}<M}$. 

Also consider a Hermitian momentum field operator
\be            
 \ba{r}
\hat\pi(\xv) \equiv {\displaystyle \fr1{V^{1/2}}\sum_{\mv}\sqrt{\fr{\omega_{\mv}}{2}}}\,[\hat p^c_\mv \cos (\kv_\mv{\cdot}\xv)\, - ~~\\
-\,\hat p^s_\mv \sin (\kv_\mv{\cdot}\xv)]\,, 
 \ea
\lb{pi_standing_harmonics}
\ee
where
\be
\hat p^{c,s}_\mv \equiv -i\pd/\pd q^{c,s}_\mv
\lb{p_cs_def}
\ee 
are the operators canonically conjugate to~$\hat q^{c,s}_\mv$. 
By\rf{q_reflection},
\be
\hat p^c_{-\mv} = \hat p^c_\mv\,,\qquad \hat p^s_{-\mv} = -\hat p^s_\mv\,.
\lb{p_reflection}
\ee
For~$\xv_\nv$ of\rf{x_n_def}, 
\be
[\hat\phi(\xv_\nv),\hat\pi(\xv_{\nv'})]= \frac{i}{V}\,\delta_{\nv\nv'}\,,
\lb{phipi_n_commutators1}
\ee
and for any spatial points~$\xv$ and~$\xv'$,
\be
[\hat\phi(\xv),\hat\phi(\xv')]= [\hat\pi(\xv),\hat\pi(\xv')] = 0\,.
\ee

Let $\nbrk{\hat F(\xv)\equiv F(\hat \phi(\xv),\hat \pi(\xv),\xv)}$ be an arbitrary operator field that is constructed from the elementary field operators~$\hat \phi(\xv)$ and~$\hat \pi(\xv)$ of\rf{phi_standing_harmonics} and\rf{pi_standing_harmonics}. Define
\be
\int_V d^3x~\hat F(\xv)\equiv \frac{V}{M^3}\sum_\nv \hat F(\xv_\nv)\,,
\lb{int_n}
\ee
where the sum has~$M^3$ terms.
Then\rf{phipi_n_commutators1}  is equivalent to the canonical field commutator
\be
[\hat\phi(\xv),\hat\pi(\xv')]= i\delta^{(3)}(\xv-\xv')\,.
\lb{phipi_n_commutators}
\ee
The Dirac delta function for the integral\rf{int_n} is defined by the general prescription\rf{delta_Dirac_def}.

Consider a Hamiltonian\footnote{
	One of the factors~$1/2$ in\rf{H_standing_waves} corrects for counting every cosine and sine mode twice, at $\mv$ and~$-\mv$. 
}
\be
\hat H = \fr12\sum_\mv \frac{\omega_{\mv}}{2}\sum_{\iota=c,s}\lf[\lf(\hat p_\mv^\iota\rt)^2 + \lf(\hat q_\mv^\iota\rt)^2 -1\rt] .
\lb{H_standing_waves}
\ee
Evolution\rf{psi_rotation} that is generated by this Hamiltonian does not change the Gaussian wave function~$\psi\gaus$ of\rf{psi_0}, which is also its ground state.

Introduce annihilation operators
\be
\hat a_\mv\equiv \fr1{\sqrt2}\lf(\hat a^c_\mv + i\hat a^s_\mv\rt)
\lb{annihilation_ops_def}
\ee
where (for $\nbrk{j=c,s}$)
\be
\hat a_\mv^j\equiv \fr1{\sqrt2}\lf(\hat q^j_\mv\! + i \hat p^j_\mv\rt).
\ee
For the operators\rf{annihilation_ops_def},
\be
 \ba{l}
[\hat a_{\mv}^{\vphantom{\dagger}},\hat a^\dagger_{\mv'}] = \delta_{\mv\mv'}\,,\\
{}[\hat a_{\mv},\hat a_{\mv'}]= [\hat a^\dagger_{\mv},\hat a^\dagger_{\mv'}] = 0\,,
 \ea
\lb{a_commutators}
\ee
and
\be
\hat a_\mv \psi\gaus = 0\,.
\ee

Given the $\nbrk{\mv\to -\mv}$ symmetry properties of $\omega_{\mv}$, $\hat q^{c,s}_\mv$ and~$\hat p^{c,s}_\mv$  [eqs.\rft{omega_reflection} {q_reflection}{p_reflection}],  the Hamiltonian\rf{H_standing_waves} equals
\be
\hat H = \sum_\mv \omega_{\mv}^{\vphantom{\dagger}} \hat a^\dagger_{\mv} \hat a_\mv^{\vphantom{\dagger}} \,.
\lb{H_free}
\ee
It generates evolution of the modes of a free field. In particular, in the Heisenberg picture, the mode annihilation operators then evolve as
\be
\hat a_\mv(t)\equiv e^{i\hat Ht} \hat a_\mv e^{-i\hat Ht}=\hat a_\mv e^{-i\omega_{\mv}t}\,.
\lb{a_evolving}
\ee
With the aforementioned $\nbrk{\mv\to -\mv}$ symmetry properties, the local field\rf{phi_standing_harmonics} and its conjugate momentum field\rf{pi_standing_harmonics} at the current time equal:
\be
\hat\phi(\xv) \eqa \fr1{V^{1/2}}\sum_{\mv}\fr1{\sqrt{2\omega_{\mv}}}\lf(\hat a_\mv e^{i\kv_\mv\cdot\xv}\!\!+\hat a^\dagger_\mv e^{-i\kv_\mv\cdot\xv}\rt),   
\lb{phi_discrete_exp}
\\
\hat\pi(\xv) \eqa \fr{-i}{V^{1/2}}\sum_{\mv}\sqrt{\fr{\omega_{\mv}}{2}}\lf(\hat a_\mv e^{i\kv_\mv\cdot\xv}\!\!-\hat a^\dagger_\mv e^{-i\kv_\mv\cdot\xv}\rt).\qquad\quad
\ee
Under the evolution\rf{a_evolving}, these operators at the general time become:
\be
\hat\phi(x) \eqa  \int\!\dbar^3k\, \fr1{\sqrt{2\omega_{\kv}}}\lf(\hat a_\kv e^{ik_\mu x^\mu}\!\!+\hat a^\dagger_\kv e^{-ik_\mu x^\mu}\rt),   
\lb{phi_free}\\
\hat\pi(x) \eqa  \int\!\dbar^3k\, \sqrt{\fr{\omega_{\kv}}{2}}\,(-i)\lf(\hat a_\kv e^{ik_\mu x^\mu}\!\!-\hat a^\dagger_\kv e^{-ik_\mu x^\mu}\rt).   \qquad\quad
\lb{pi_free}
\ee
Here $\nbrk{x\equiv(t,\xv)}$ and $\nbrk{k_\mu\equiv (-\omega_{\kv},\kv)}$. We also took the continuous limit for the field modes ($\nbrk{M\to\infty}$ at $\nbrk{L=\const}$) and  introduced a continuous operator field
\be
\hat a_\kv \equiv V^{1/2} \hat a_{\mv(\kv)}\,,
\lb{a_continuous}
\ee
with $\nbrk{\mv(\kv)\equiv \mathop{\rm round}[\kv L/(2\pi)]}$, rounded to the nearest integer. 
For the continuous annihilation operators\rf{a_continuous}
\be
 \ba{l}
[\hat a_\kv^{\vphantom{\dagger}},\hat a_{\kv'}^\dagger] = \deltabar^{(3)}(\kv-\kv')\,,\\
{}[\hat a_\kv,\hat a_{\kv'}] = [\hat a_\kv^\dagger,\hat a_{\kv'}^\dagger] = 0\,.
 \ea
\ee

The fields\rfs{phi_free}{pi_free} satisfy
\be
\pd_t \hat\phi= \hat\pi\,.
\lb{eom_free1}
\ee
The equation for~$\pd_t\hat\pi$ likewise becomes local in~$\xv$ if we set the frequency parameters  $\nbrk{\omega_{\kv}\equiv\omega_{\mv(\kv)}}$ to 
\be
\omega_{\kv}^2 = k^2+\mu^2\,,
\lb{omega_kv}
\ee
where $\nbrk{k^2\equiv k^2_x + k^2_y + k^2_z}$ and $\mu$~is a real constant. Then
\be
\pd_t \hat\pi= (\nabla^2+\mu^2)\,\hat\phi\,,
\lb{eom_free2}
\ee
where $\nbrk{\nabla^2\equiv \pd^2_x + \pd^2_y + \pd^2_z}$. The corresponding Hamiltonian\rf{H_free} can be expressed as the spatial integral of local Hamiltonian density,
\be
\hat H = \int d^3x\,\fr12\lf[\hat\pi^2+({\bm\pd}\hat\phi)^2+\mu^2\hat\phi^2\rt].
\lb{Hamiltonian_density_free}
\ee

Importantly, the field operators~$\hat\phi(x)$ and~$\hat\pi(x)$ evolve by local equations not only for the Hamiltonian above but for any Hamiltonian given by spatial integral of any, possibly spacetime-dependent, local Hamiltonian density,
\be
\hat H= \int d^3x\,\cH(\hat\phi(x),\hat\pi(x),x)\,.
\lb{Hamiltonian_local}
\ee
However, the Gaussian function~$\psi\gaus$ is generally \emph{no longer} the ground state of the arbitrary local Hamiltonian\rf{Hamiltonian_local}. 

Add to the non-interacting Hamiltonian density in\rf{Hamiltonian_density_free} other Hermitian local operators $\nbrk{\alpha(t)\,\Op(\hat\phi(x),\hat\pi(x))}$ with time-dependent $c$-number coefficients~$\alpha(t)$. Let the coefficients start from zero and grow adiabatically to one over a characteristic timescale~$T$. Under the evolution of the initial state~$\psi\gaus$ from\rf{psi_0} by this Hamiltonian, the modes with frequency $\nbrk{\omega\gg T^{-1}}$ remain in the ground state of the adiabatically changing Hamiltonian. This demonstrates that requiring the field dynamics to be local and the high-frequency modes of the field to be in the ground state is not sufficient for fixing the Hamiltonian.

The free scalar quantum field~$\hat\phi$ with the wave function~$\nbrk{\psi(\phi)\equiv\psi(q[\phi])}$---constructed from the generic distribution~$\dist(Q)$ as described---is thus not an objectively existing emergent quantum system. Its evolution by the Hamiltonian\rfs{Hamiltonian_local}{Hamiltonian_density_free} does not stand out from nearby paths of its continuous evolution by arbitrarily modified Hamiltonians, local or non-local. As argued in \sct{subsubsec_basic}, such systems are \emph{in principle} unsuitable for biological evolution and natural development of internal intelligent observers. 

Below we identify other ubiquitously present emergent quantum field systems whose dynamical laws are fixed. These systems also possess such features of our observed physical world as gauge and gravitational interactions, and they contain bosonic as well as fermionic fields.

\section{Gauge fields}
\label{sec_gauge_fields} 

The observed physical dynamics is highly symmetric. Its equations are invariant under the vast groups of local gauge and diffeomorphism transformations. Correspondingly, we will identify emergent quantum fields with gauge- and diffeomorphism-symmetric dynamics. In this section we will discuss gauge symmetry. The next \sct{sec_gravity} will addresses diffeomorphism symmetry. These symmetries do not yet fix the dynamics of an emergent system, i.e., they do not resolve the fundamental problem pointed out in \sct{sec_first_observation}. However, further \sct{sec_why_the} will demonstrate that the dynamics of emergent fields with larger local symmetries, in particular, with local supersymmetry, is fixed.

\subsection{Locally symmetric dynamics entails symmetric wave function}
\label{subsec_wf_constant} 

Consider an axiomatically introduced quantum field theory whose dynamics is invariant under a group of  local transformations of the fields. As seen next, the wave function~$\psi(f,t)$ of the fields is then invariant under the restriction of these transformations to the wave function's field arguments, i.e., to the field configurations~$f(\xv)$ at a fixed time~$t$. 
To understand this intuitively, consider the equation for quantum evolution $\nbrk{\psi(f,t)\to \psi(f',t')}$ of bosonic fields in the convolution form\rf{evolution_convolution}. Let us for the moment not worry about defining the integration over~$df$ in\rf{evolution_convolution} for locally symmetric fields---we will prove the eventual result rigorously later.  

Consider a local symmetry transformation that changes some of the fields~$f'(\xv)$ at the time~$t'$ but does not affect the fields~$f(\xv)$ at the time~$t$. Being a symmetry of the dynamics, the transformation preserves the convolution kernel~$U(f',f)$ in\rf{evolution_convolution}. Hence $\psi'(f')$ on the left-hand side of\rf{evolution_convolution} is invariant under this transformation of~$f'$. The invariance of the wave function under such a transformation at the general time~$t'$ is achieved through initial-value constraints on the wave function and through the symmetry of the dynamical equations, preserving the constraints.

We demonstrate the wave function's invariance for abelian gauge symmetry explicitly in the next subsection~\ref{subsec_abelian}, eq.\rf{secondary_constraint_gauge_abelian}. Appendix~\ref{apx_gauge} shows it for any non-abelian gauge symmetry, eq\rf{delta_gauge_psi_is_0}. \Sct{sec_gravity} addresses diffeomorphism symmetry, eqs.\rfs{dpsi_diff}{Hamiltonian_momentum_constraints}. Finally, Appendix~\ref{apx_fermions} proves this for the general local (super)\,symmetry, possibly mixing bosonic and fermionic fields, eqs.\rfs{delta_psi_evol_A}{secondary_constraint_A}.

\subsection{Wave function of abelian gauge fields}
\label{subsec_abelian} 

We now use alternate presentations of the smoothed generic static distribution~$\dist(Q)$ to construct prospective emergent wave functions of fields with gauge symmetry. This section discusses \emph{abelian} gauge symmetry and flat spacetime. Appendix~\ref{apx_gauge} extends the construction to renormalizable theories with non-abelian gauge symmetry and arbitrary metric in 3+1 dimensional spacetime.

First, consider formal (introduced axiomatically) quantum abelian gauge field theory.
Let $\nbrk{f\equiv\{f^\iota(x)\}}$ stand for all its fields and $\nbrk{\pi\equiv\{\pi_\iota(x)\}}$~denote their canonically conjugate momenta. The Legendre transformation
\be
\cL(\dot f,f)= \lf.\lf[\dot f^\iota\pi_\iota - \cH(f,\pi)\rt]\rt|_{\partial\cH/\partial\pi=\dot f}\,
\ee
matches a local Hamiltonian\rf{Hamiltonian_local} to a local action $\nbrk{S = \int d^4x\,\cL}$.
Let the Hamiltonian be at most quadratic in the momenta fields. Then we can explicitly take the integral over~$\dbar\pi$ in\rf{evolution_path_integral} to obtain
\be
\psi'(f')= \int[df]\,e^{idS(\dot f,f)}\ \psi(f)\,.
\lb{path_integral_action}
\ee
Here
\be
dS = dt\!\int\! d^3x~\cL(\dot f(x),f(x))\,,
\ee
$\nbrk{\dot f=(f'-f)/dt}$, and the Feynman path-integral measure~$[df]$ absorbs the constants from integration in\rf{evolution_path_integral} over the independent momenta $\nbrk{\pi_{\iota\nv}\equiv \pi_\iota(\xv_\nv)}$ [cf.\ eq.\rf{int_n}].
The action~$S(\dot f,f)$ specifies the kernel~$U(f',f)$ for evolution\rf{evolution_convolution} and involves only local fields but not their momenta. Hence, the action is the quantity of choice for establishing whether or not a local transformation is a symmetry of the fields' dynamics.

Consider a renormalizable Lagrangian density\footnote{
   We do not treat the complex conjugate field~$\phi^*$ as formally independent 
   from~$\phi$. Hence in\rf{action_gauge_abelian} the kinetic term for~$\phi$
   has the factor~$1/2$.
}
\be
\cL = \frac1{4e^2}\,F_{\mu\nu}F^{\mu\nu}
		-\fr12\,(D^\mu\phi)^*D_\mu\phi - V(|\phi|^2)\,,\quad
\lb{action_gauge_abelian}
\ee
where
\be
F_{\mu\nu}\eqa \partial_\mu A_\nu-\partial_\nu A_\mu\,,
\lb{F_abelian}\\
D_\mu \eqa \partial_\mu-iA_\mu\,,
\ee
$\nbrk{\phi=(\phi^1,\phi^2)}$ is a complex scalar field, $A_\mu$~is a real vector field, and $e$~is a non-zero coupling constant. This Lagrangian density is invariant under the gauge transformation
\be
A_\mu &\to& A_\mu + \pd_\mu\varphi\,
\lb{A_gauge_transformation}\\
\phi   &\to&  e^{i\varphi}\phi\,
\lb{phi_gauge_transformation}
\ee
for any~$\varphi(x)$.

The standard procedure, reviewed for the general gauge symmetry and metric in \apx~\ref{apx_gauge}, matches the Lagrangian density\rf{action_gauge_abelian} to the Hamiltonian density
\be
\hat \cH = \cH_N(\hat\pi^i,\hat\pi_\iota,\hat F_{ij},D_i\hat\phi,|\phi|^2)+
	\,A_0{}\hat\cH_\varphi\,.
\lb{H_gauge_abelian}
\ee
This $\hat \cH$~is a function of the displayed local fields and their canonically conjugate momenta operators
\be
\hat\pi^i(\xv) \!&\equiv&\! -i\frac{\delta}{\delta A_i(\xv)}\,,
\lb{pi_i_def}\\
\hat\pi_\iota(\xv) \!&\equiv&\! -i\frac{\delta}{\delta\phi^\iota(\xv)}\,,\quad
\iota\in\{1,2\}.
\ee
In the last term of\rf{H_gauge_abelian},
\be
\hat\cH_\varphi(\xv)\equiv -i\frac{\delta}{\delta\varphi(\xv)}
\equiv \!\sum_{f=\phi^\iota\!,A_i}\frac{\delta f}{\delta \varphi}\,\hat\pi_f
= \hat j^0-\pd_i\hat\pi^i\,\quad
\lb{varepsilon_gauge_abelian}
\ee
where
\be
\hat j^0 = \frac{\pd \hat\phi^\iota}{\pd\varphi}\, \hat\pi_\iota = i^\iota_\kappa\hat\phi^\kappa\hat\pi_\iota = \hat\phi^1\hat\pi_2-\hat\phi^2\hat\pi_1\,
\lb{j_phi}
\ee
is the matter charge density. Here and below $i^\iota_\kappa$ is an element of the 2$\times$2
matrix~$i$ of\rf{i_def}. 

Let us show that $\hat\cH_\varphi(\xv)$ of\rf{varepsilon_gauge_abelian} generates the gauge transformations of the theory operators. Let $\hat \Op$ be an arbitrary operator that is composed of the elementary local fields $\nbrk{(\hat\phi^\iota,\hat A_i)}$ and their conjugate momenta fields~$\nbrk{(\hat\pi_\iota,\hat\pi^i)}$ at a common time. The operator~$\hat\Op$ varies under an infinitesimal gauge transformation as
\be
\delta_\varphi \hat\Op = \int\! d^3x~\varphi(\xv)\,i\,[\hat\cH_\varphi(\xv),\hat \Op]\,.
\lb{operator_gauge_change}
\ee
Its gauge variation is thus produced by similarity transformation\rf{Op_def} with
\be
\hat U \eqa \,\exp\lf[-i\int d^3x~\varphi(\xv)\,\hat\cH_\varphi(\xv)\rt]=
\lb{U_gauge}\\
\eqa \,\exp\lf[-\!\int d^3x~\varphi(\xv)\,\fr{\delta}{\delta\varphi(\xv)}\rt].\nn
\ee

The normal component~$\cH_N$ of the Hamiltonian density\rf{H_gauge_abelian} for the Lagrangian density\rf{action_gauge_abelian} equals 
\be
\cH_N &\!=& \fr{e}{2}^2\pi_i\pi^i+ 
	\fr{1}{4e^2}\,F^{ij}F_{ij}\,+ 
\lb{H_N_gauge}\\
	&+& \fr12\lf(|\pi|^2+|{\bm D}\phi|^2\rt) + V(|\phi|^2) \,.\nn
\ee
It is manifestly gauge-invariant:
\be
\delta_\varphi \hat\cH_N(\yv)= \int\! d^3x~\varphi(\xv)\,i\,[\hat\cH_\varphi(\xv),\hat\cH_N(\yv)]=0\,.~
\lb{dgauge_cH_N}
\ee

For a classical field system with the gauge-invariant Lagrangian density\rf{action_gauge_abelian} the conjugate to~$A_0$ momentum is identically zero:
\be
\pi^0 = \frac{\pd \cL}{\pd (\pd_0 A_0)} = 0\,,
\lb{class_pi_0}
\ee
because~$\cL$ of\rf{action_gauge_abelian} does not contain~$\partial_0 A_0$, 
In quantum theory  eq.\rf{class_pi_0} corresponds to the \emph{primary constraint}
\be
\hat\pi^0(\xv)\,\psi\equiv -i\frac{\delta}{\delta A_0(\xv)}\,\psi = 0\,.
\lb{primary_constraint_gauge_abelian}
\ee
By the last equation, the wave function of this gauge-symmetric system does not depend on the non-dynamical field~$A_0$.
In other words, $A_0$~is not an argument of the system's wave function.

The wave function should remain independent of~$A_0$ after an infinitesimal step of its evolution, $\nbrk{d\psi=-idt\hat H\psi}$. With the Hamiltonian density\rf{H_gauge_abelian}, this is the case only when the wave function satisfies the \emph{secondary constraint}
\be
\hat\cH_\varphi(\xv)\,\psi \equiv -i\frac{\delta}{\delta\varphi(\xv)}\,\psi
= 0\,
\lb{secondary_constraint_gauge_abelian}
\ee
[the first identity recalls\rf{varepsilon_gauge_abelian}]. The secondary constraint\rf{secondary_constraint_gauge_abelian} demands that the physical wave function does not depend on the gauge degree of freedom~$\varphi$, i.e., on the gauge choice. We observed in the previous subsection~\ref{subsec_wf_constant} that this is the general property of the wave functions of systems with locally symmetric dynamics.

Since the action~$S$ for the gauge-invariant Lagrangian density\rf{action_gauge_abelian} does not involve~$\pd_0 A_0$, we have
\be
\frac{\delta S}{\delta A_0(x)} = \frac{\delta L}{\delta A_0(\xv)} = -\,\frac{\delta H}{\delta A_0(\xv)} = -\,\cH_\varphi\,
\ee
[the last equality above follows from\rf{H_gauge_abelian}].
The secondary constraint\rf{secondary_constraint_gauge_abelian} hence corresponds in the Lagrangian formulation to a requirement
\be
\frac{\delta \hat S}{\delta A_0(x)}\,\psi = 0\,.
\lb{constraint_action}
\ee

\subsection{Emergent quantum gauge fields}
\label{subsec_gauge_dofs} 

\Sct{sec_evolution} discussed emergent real scalar field operators~$\hat\phi(x)$ and the canonically conjugate momenta field operators~$\hat\pi(x)$. An emergent \emph{complex} scalar field~$\hat\phi(x)$ is straightforwardly composed of two real scalar fields~$\hat\phi^\iota(x)$, $\nbrk{\iota\in\{1,2\}}$. 
To construct~$\hat\phi^\iota(x)$ and the canonically conjugate~$\hat\pi_\iota(x)$, substitute in\rf{phi_standing_harmonics} each of the operators~$\hat q_\mv$ (standing for either~$\hat q^c_\mv$ or~$\hat q^s_\mv$) by two operators~$\hat q^\iota_\mv$ of some independent arguments~$q^n$ of a presentation $\nbrk{\psi(q)\equiv \psi(q^1,\dots,q^N)}$ of the generic static distribution, eq.\rf{psi_general}. Then proceed with the rest of the construction as in \sct{subsec_operators}. 

For the pair of the real fields that compose the complex field we may extend the Hamiltonian density\rf{Hamiltonian_density_free} to
\be
\hat\cH = \sum_{\iota=1,2} \fr12\lf[\hat\pi_\iota^2+({\bm\pd}\hat\phi^\iota)^2+\mu^2(\hat\phi^\iota)^2\rt],
\lb{Hamiltonian_density_free_many}
\ee
assigning equal mass~$\mu$ to both components~$\phi^\iota$.
We can as well consider another path of evolution that is generated by a Hamiltonian given by a spatial integral of a local function of the fields, $\nbrk{H =\int d^3x\ \cH(\hat\pi_\iota(x),\hat\phi^\iota(x),x)}$.	For it, as noted at the end of \sct{subsec_operators}, the dynamics remains local. 

As a possible modification of the Hamiltonian density\rf{Hamiltonian_density_free_many}, consider
\be
\hat\cH \eqa \sum_\iota \fr12\lf[\hat\pi_\iota^2+({\bm D}\hat\phi^\iota)^2+\mu^2(\hat\phi^\iota)^2\rt]\equiv
\nn \\
&\equiv & \fr12\lf(|\hat\pi|^2+|{\bm D}\hat\phi|^2+\mu^2|\hat\phi|^2\rt) \lb{Hamiltonian_density_gauged}
\ee
where
\be
{\bm D}\hat\phi^\iota \equiv {\bm\pd}\hat\phi^\iota - i^\iota_\kappa {\bm A}(x)\,\hat\phi^\kappa\,.
\lb{D_def}
\ee
To obtain an emergent gauge-symmetric field system with the Hamiltonian density\rfd{H_gauge_abelian}{H_N_gauge}, identify the amplitudes of  the \emph{transverse} modes of the arbitrary vector field~$\Av(x)$ in\rf{D_def} with additional---independent of~$q^\iota_\mv$---coordinates~$q^n$ of the presentation~$\psi(q)$. 

Specifically, promote $\Av(x)$ in\rf{D_def} to a vector operator field
\be
\hat\Av=\hat\Av^T+{\bm\pd}\varphi\,
\lb{A_VS_decomposition}
\ee
where $\varphi(x)$ is an arbitrary real function on spacetime. At some reference time~$t$, set
\be                   
\hat\Av^T(\xv) = \fr{e}{V^{1/2}}\sum_{\mv,\lambda}\fr{{\bm \epsilon}_{\mv\lambda}}{\sqrt{2|\kv_\mv|}}\lf[\hat q^{c\lambda}_\mv \cos (\kv_\mv\pdot\xv) \, +~~\rt. \nn\\
\lf. +\ \hat q^{s\lambda}_\mv \sin (\kv_\mv\pdot\xv) \rt] ~~
\lb{A_standing_harmonics}
\ee
with $\nbrk{\lambda\in\{1,2\}}$, $\nbrk{{\bm \epsilon}_{\mv\lambda}\pdot\,{\bm \epsilon}_{\mv\lambda'}=\delta_{\lambda\lambda'}}$, and $\nbrk{\kv_\mv\pdot\,{\bm \epsilon}_{\mv\lambda}= 0}$, giving $\nbrk{{\bm\pd}\pdot\hat\Av^T = 0}$. The amplitudes~$\hat q^{c\lambda}_\mv$ and $\hat q^{s\lambda}_\mv$ of the transverse modes are identified with the operators~$\hat q^n$ of independent basic coordinates~$q^n$ of the presentation~$\psi(q)$.
Now any value of~$\nbrk{q= (q^1,\dots, q^N)}$ specifies a realization of the complex scalar and transverse vector fields~$(\phi(\xv),\Av^T(\xv))$. 
Finally, for any~$\varphi(\xv)$ in\rf{A_VS_decomposition}, set 
\be
\psi(\phi,\Av)\equiv \psi(e^{-i\varphi}\phi,\Av^T)\equiv \psi(q(e^{-i\varphi}\phi,\Av^T))\,.
\lb{psi_gauge_construct}
\ee

Consider evolution of the constructed gauge-invariant prospective wave function\rf{psi_gauge_construct} with the Hamiltonian density\rfd{H_gauge_abelian}{H_N_gauge}. We can follow this gauge-symmetric evolution using any~$\varphi(x)$. The Hamiltonian density\rf{H_gauge_abelian} corresponds to the Hamiltonian
\be
\hat H = \hat H_N+\hat H_\varphi
\lb{Hprime_A_0}
\ee
with a gauge-invariant~$\hat H_N$ and
\be
\hat H_\varphi=\int d^3x\,A_0 \hat \cH_\varphi\,.
\lb{H_varphi}
\ee

The constructed wave function\rf{psi_gauge_construct} obeys the secondary constraint\rf{secondary_constraint_gauge_abelian}. Hence, in the Schrodinger picture, the Hamiltonian\rf{Hprime_A_0} generates a gauge-independent change of this wave function:
\be
d\psi = -idt\hat H\psi = -idt\hat H_N\psi\,.
\ee
By\rf{dgauge_cH_N}, the evolving wave function continues to obey the secondary constraint\rf{secondary_constraint_gauge_abelian}. Hence, it continues to be invariant under the gauge transformation of its argument fields.

Eq.~(\ref{varepsilon_gauge_abelian}) shows that the operator $\nbrk{-idt\hat H_\varphi}$, with~$\hat H_\varphi$ from\rf{H_varphi}, is the operator of the gauge variation by
\be
\delta\varphi = dt\,A_0\,.
\lb{phase_evolution}
\ee
While, as shown above, gauge transformation does not affect the wave function, it generally changes field operators. The evolution of the Heisenberg operators~$\hat\phi(x)$ and~$\hat A_i(x)$ generated by the Hamiltonian\rf{Hprime_A_0} hence depends on the temporal component,~$A_0(x)$, of the gauge connection field~$A_\mu$:
\be
\pd_t \hat\phi \eqa i[\hat H,\hat\phi]= i[\hat H_N,\hat\phi]+iA_0\hat\phi\,,\\
\lb{phi_change}
\pd_t \hat A_i \eqa i[\hat H,\hat A_i]= i[\hat H_N,\hat A_i]+ \partial_i A_0\,.
\ee
Equivalently, in the covariant form,
\be
D_0\hat\phi^\iota \eqa i[\hat H_N,\hat\phi^\iota]\,,\\
\hat F_{0i} \eqa i[\hat H_N,\hat A_i]\,.
\ee

The described construction of locally interacting quantum fields could fail at small length scales. Without regularization of the field dynamics in the ultraviolet limit, the contribution of short-wavelength modes to the Hamiltonian of a typical renormalizable interacting theory for $\nbrk{3+1}$ dimensional spacetime diverges. 
Later \sct{sec_Planck_scale} however shows that diffeomorphism-symmetric, gravitationally interacting emergent quantum fields, whose dynamics can also be gauge-symmetric, are naturally regularized in the ultraviolet limit. Only a finite number of their modes requires basic representation by independent dimensions of the fundamental distribution~$\dist(Q)$. The same \sct{sec_Planck_scale} identifies a regular physical emergent wave function of these fields.

Appendix~\ref{apx_Hamiltonian} derives the Hamiltonian for the general locally Lorentz-invariant, renormalizable, local action of bosonic fields with arbitrary gauge symmetry, abelian or not. This Hamiltonian describes quantum evolution of non-abelian gauge fields \emph{without} introducing the Faddeev-Popov \emph{ghosts}\ct{FP_ghosts}. Evolution by it preserves the primary and secondary constraints on the wave function,  eqs.\rf{psi_A0_independence} and\rf{secondary_constraint_gauge} respectively. 
The variables~$q$ of a presentation~$\psi(q)$ of the smoothed generic distribution~$\dist(Q)$ can be mapped to the corresponding matter fields~$\phi^\iota$ and gauge fields~$A_i^r$ modulus their joint gauge transformation similarly to the presented mapping for abelian gauge theory in flat spacetime.

\section{General covariance and quantum gravity}
\label{sec_gravity} 

We proceed to identifying emergent quantum fields whose dynamics has diffeomorphism symmetry. 
Any field transformation, whether being a symmetry of the dynamics or not, to another description of a quantum field system changes its Heisenberg field operators~$\hat f(x)$ as
\be
\hat f'(x) = \hat U^{-1}\hat f(x)\, \hat U\,.
\lb{fU_diffeomorphism}
\ee
The change $\nbrk{\delta \hat f(x)\equiv\hat f'(x)-\hat f(x)}$ of a scalar, vector, or higher-tensor field under
\emph{diffeomorphism} transformation by an infinitesimal 4-vector parameter field~$\veps^\mu(x)$ is the field's Lie derivative along~$\veps^\mu(x)$. In particular, 
\be
 \ba{l}
\delta \hat\phi = L_\veps \hat\phi = \veps^\lambda \hat\phi_{,\lambda}\,\\
\delta \hat A_\mu = L_\veps \hat A_\mu = \veps^\lambda \hat A_{\mu,\lambda}+\veps^\lambda_{,\mu} \hat A_\lambda^{}\,\\
\delta \hat g_{\mu\nu} = L_\veps \hat g_{\mu\nu} = \veps^\lambda \hat g_{\mu\nu,\lambda}^{}+ \veps^\lambda_{,\mu}\hat g_{\lambda\nu}^{}+\veps^\lambda_{,\nu}\hat g_{\mu\lambda}^{}\,.
 \ea
\lb{diffeomorphism_transformation_def}
\ee
The transformed fields~$\hat f'(x)$ describe the physical system in a new spacetime coordinate frame that is displaced relative to the old coordinates by $\nbrk{\Delta x^\mu(x)=\veps^\mu(x)}$.  The transformed scalar, vector, and higher-tensor fields are related to the original fields as:
\be
\hat\phi'(x')= \hat\phi(x),\quad  \hat A_\mu'(x')= \frac{\pd x^\nu}{\pd x'{}^\mu}\,\hat A_\nu(x),\quad \dots\quad 
\lb{fpf}
\ee
with
\be
x'= x - \veps\,.
\lb{xpx}
\ee

\subsection{Hamiltonian for generally covariant fields}
\label{subsec_gravity_A}

To identify the quantum systems with diffeomorphism-symmetric dynamics, we evoke arguments presented by Dirac\ct{Dirac} for classical general relativity. We use the more modern ADM\ct{ADM_59,ADM_62_republished} notation\rf{gADM}, and we remember that in quantum theory observables are described by operators.

Consider a prospective observable whose operator~$\hat\eta$  is a function of dynamical fields~$\hat f^\iota(\xv)$ at a common time~$x^0$:
\be
\hat\eta \equiv \eta(\hat f) \equiv \eta[\hat f^\iota(\xv)]\,.
\lb{hat_eta_def}
\ee
This observable may or may not be localized at one spatial point~$\xv$. Diffeomorphism transformation by an infinitesimal vector~$\veps^\mu(x)$ changes its operator~$\hat\eta$ by
\be
\delta\hat\eta = \int d^3x~\veps^\mu(\xv)\,\hat v_\mu(\xv),
\lb{delta_eta}
\ee
where in the linear order the operators~$\hat v_\mu$ do not depend on the displacement~$\veps^\mu$. 

We require that the transformation operator~$\hat U$ from\rf{fU_diffeomorphism} preserves the canonical form of the Hermitian product\rf{Hermitian_product_canonical}. Then $\hat U$~is unitary. For an infinitesimal~$\veps^\mu$, it has the form
\be
\hat U= \exp\lf[-i\int d^3x~\veps^\mu(\xv)\,\hat\cH_\mu(\xv)\rt],
\lb{U_eps_explicit}
\ee
where~$\hat\cH_\mu$ are Hermitian operators that are independent of~$\veps^\mu$. Transformation of~$\eta(\hat f)$ under\rfd{fU_diffeomorphism}{U_eps_explicit} produces the variation\rf{delta_eta} with
\be
\frac{\delta \hat\eta}{\delta \veps^\mu(\xv)}= \hat v_\mu(\xv)=i[\hat\cH_\mu(\xv),\hat \eta]\,.
\lb{eta_variation}
\ee

Hamiltonian density\rf{H_gauge_abelian} or\rf{Hamiltonian_density_gauge} of \emph{gauge} theories allowed us to specify ``gauge-normal'' evolution with $\nbrk{A_0=0}$. Similarly, let
any configuration of emergent \emph{diffeomorphism-symmetric} fields allow us to specify unit\footnote{
    Fields with conformally invariant dynamics do not have a unique unit length for the normal displacement. 
    Nevertheless, the dynamics of our further interest is not conformally invariant. In particular, 
    it is characterized by a physical Planck scale.
} 
temporal displacement~$n^\mu(\xv)$ that is normal to the current-time hypersurface. Akin to gauge theories, evolution along the normal displacement~$n^\mu(\xv)$ (corresponding to the synchronous coordinates) can be transformed into equivalent evolution along an arbitrary displacement~$\veps^\mu(\xv)$ (corresponding to the general coordinates) as follows.

Decompose the displacement vector~$\veps^\mu(\xv)$ into the components that are normal and tangential to the current-time hypersurface:
\be
\veps^\mu= [Cn^\mu+(0,\bm{C})]\,dt\,.
\lb{eps_foliation}
\ee
Here the vector in the brackets has arbitrary normalization. Extend the spatial coordinates~$\{\xv\}$ to other times $\nbrk{t\neq x^0}$ along the vector field~$\veps^\mu(\xv)$. Set the coordinate time at the displaced spatial hypersurface as $\nbrk{t\equiv x^0 + dt}$ with~$dt$ from\rf{eps_foliation}. In the introduced spacetime coordinates the displacement vector\rf{eps_foliation} equals
\be
\veps^\mu(\xv) = (1,\bm{0})\,dt\,.
\lb{eps_foliation_unit}
\ee

The systems where every spatial hypersurface has a normal unit vector~$n^\mu$ possess a \emph{metric tensor}.\footnote{
     For a proof, construct an orthonormal basis (vierbein) by taking the vector $e^\mu_1$ as the normal 
     to an arbitrary hypersurface that passes through the considered point, another vector~$e^\mu_2$ 
     as the normal to a hypersurface tangent to~$e^\mu_1$, and so on. 
     The metric tensor is the inverse of $\nbrk{g^{\mu\nu}=\sum_a e^\mu_a e^\nu_a}$.
}
Let the metric tensor in the constructed coordinates, for which\rf{eps_foliation_unit} applies, be parameterized by the ADM lapse~$\lapse(\xv)$ and shift~$\shift^i(\xv)$, eq.\rf{gADM}. The unit vector~$n^\mu$ that is normal to the current-time ($\nbrk{dt=0}$) hypersurface in the ADM parameterization\rf{gADM} equals:
\be
n_\mu= (-\lapse,\bm{0})\,,\quad
n^\mu= \lf(\fr1\lapse,\,-\frac{~\shift^i}{\lapse}\rt).
\lb{n_mu}
\ee 
Substitution of this expression for~$n^\mu$ into\rf{eps_foliation} reveals that the displacement $\veps^\mu$ from\rf{eps_foliation} becomes\rf{eps_foliation_unit} when $\nbrk{C=\lapse}$ and $\nbrk{C^i=\shift^i}$. With these values of~$C$ and~$\bm{C}$,  eq.\rf{eps_foliation} reads
\be
\veps^\mu= [\lapse n^\mu+(0,\shift^i)]\,dt \,.
\lb{eps_foliation_normal}
\ee
Then by eqs.~(\ref{delta_eta},\ref{eps_foliation_normal},\ref{eta_variation})
\be
\delta\hat\eta = i[\hat H,\hat\eta]\,dt
\lb{delta_eta_equation}
\ee
with
\be
\hat H \eqa  \int d^3x\,\lf(\lapse\,\hat\cH_\lapse + \shift^i\hat\cH_i\rt)
\lb{Na_def0}
\ee
where 
\be
\hat\cH_\lapse\equiv n^\mu\hat\cH_\mu\,. 
\lb{cH_N_def}
\ee
Define
\be
\shift^a\equiv(\lapse,\shift^i) 
\quad\mbox{and}\quad 
\hat\cH_a\equiv(\hat\cH_\lapse,\hat\cH_i)\, .
\lb{Na_def}
\ee
Then\rf{Na_def0} becomes
\be
\hat H= \int d^3x\ \shift^a\hat\cH_a\,.
\lb{Hamiltonian_covariant}
\ee
Neither~$\hat\cH_\mu$ nor~$n^\mu$ depend on the displacement field\rf{eps_foliation_normal}. Hence $\hat\cH_a$ of\rfd{Na_def}{cH_N_def} are also independent of~$\shift^a$. In other words, $\hat\cH_a$~are composed of only fields that specify the current physical state but not our choice of extending the coordinates to future times.

\subsection{Emergent generally covariant systems}
\label{subsec_emergent_gravity}

Diffeomorphism invariance could be achieved by introducing affine connections---the direct counterpart of gauge connections---and promoting them to additional dynamical degrees of freedom. This would result in the Einstein-Cartan theory of gravity\ct{Cartan_23}. Yet we can reduce the number of independent gravitational degrees of freedom by imposing the Einstein equivalence principle. It restricts the connections~$\Gamma^\lambda_{\mu\nu}$ to be metric-compatible ($\nbrk{g_{\mu\nu;\lambda}=0}$) and torsion-free ($\nbrk{T_{\mu\nu}^\lambda\equiv \Gamma_{\mu\nu}^\lambda-\Gamma_{\nu\mu}^\lambda\!= 0}$). 
Then the standard machinery of general relativity lets us construct the symmetric Christoffel symbols~$\Gamma^\lambda_{\mu\nu}$ from the metric~$g_{\mu\nu}$. The Christoffel symbols, metric, and integer-spin matter fields can be combined into a diffeomorphism-invariant action, yielding the desired Hamiltonian. This procedure fails to incorporate covariantly half-integer spins. They will be handled separately in \scts{sec_fermions} and~\ref{sec_emergent_SUSY}. 

We start by constructing emergent quantum field systems with
\emph{spatial} diffeomorphism symmetry, whose displacement vectors are by definition limited to~$\nbrk{(0,{\bm \veps}(x))}$. Akin to gauge transformations, spatial diffeomorphism transformations mix dynamical fields only within the spatial slices of constant time. 

For the integer-spin fields, we implement spatial diffeomorphism symmetry by introducing new, gravitational, degrees of freedom. They are described by a $3$-tensor field operator~$\hat\gamma_{ij}(\xv)$ that corresponds to the spatial metric in the emergent system. We incorporate these degrees of freedom into a prospective emergent wave function similarly to the gauge degrees of freedom~$\hat A_i$ in \sct{sec_gauge_fields}. 

Namely, consider various evolution paths of the matter fields, here~$\nbrk{(\hat\phi,\hat A_i)}$. Some of the paths are generated by a family of matter Hamiltonians $H^m(\hat\pi_\phi,\hat\pi_{A_i},\hat\phi,\hat A_i,\gamma_{ij})$ where the last argument,~$\gamma_{ij}(x)$, is a parameter field and $H^m(\pi_\phi,\pi_{A_i},\phi,A_i,\gamma_{ij})$ corresponds to a diffeomorphism-invariant action. (Appendix~\ref{apx_Hamiltonian} gives these~$H^m$ explicitly.)

Then, similarly to the gauge symmetry, map the dynamical components (identified below) of~$\gamma_{ij}(\xv)$ to yet unused independent coordinates~$q^n$ of the representations~$\psi(q)$ of the smoothed basic distribution~$\dist(Q)$. These independent coordinates thus become the gravitational degrees of freedom of an emergent quantum system.

The wave function at any time~$t$ is independent of the lapse and shift~$\shift^a$. Indeed, the \emph{same} wave function  specifies the current state for the subsequent evolution with \emph{any conceivable}~$\shift^a$ in the arbitrary displacement\rf{eps_foliation_normal}.  

The wave function in diffeomorphism-symmetric theory evolves over an infinitesimal time interval~$dt$ by a Hamiltonian of the general form\rf{Hamiltonian_covariant}:
\be
d\psi= -idt\hat H\psi = -idt\int d^3x\ \shift^a\hat\cH_a\psi\,.
\lb{dpsi_diff}
\ee
The wave function at the new time, $\nbrk{t+dt}$, should again be independent of~$\shift^a$. By\rf{dpsi_diff}, this requires
\be
\hat\cH_a \psi = 0\,.
\lb{Hamiltonian_momentum_constraints}
\ee
These are the Hamiltonian and momentum constraints of canonical quantum gravity\ct{DeWitt}.

The \emph{momentum constraints}, $\nbrk{\hat\cH_i\psi=0}$, express constancy of the wave function~$\psi(\phi,A_i,\gamma_{ij})$ on the orbits of spatial diffeomorphism transformations of its argument fields,\ct{DeWitt,LapchRubakov} or eq.\rf{spatial_Lie_generator} in \apx~\ref{apx_Hamiltonian}. 
These transformations change fields by their Lie derivatives along the \emph{spatial} displacements~${\bm \veps}(x)$, i.e., by the 3-dimensional equivalents of eqs.\rf{diffeomorphism_transformation_def}.  

The constancy of~$\psi(\phi,A_i,\gamma_{ij})$ under the spatial diffeomorphism displacements  lets us gauge away 3~of the~6 independent components of the symmetric tensor~$\gamma_{ij}$ as follows. A spatial diffeomorphism transformation changes the 3-metric tensor as
\be
\delta_{\bm\veps} \gamma_{ij} = L_{\bm\veps} \gamma_{ij} = \veps_{(i|j)}\,,
\lb{diffeomorphism_transformation_3metric}
\ee
where $\nbrk{\veps_i=\gamma_{ij}\veps^j}$, the vertical bar~``$\,|\,$'' indicates covariant derivative based on the spatial metric~$\gamma_{ij}$, and parentheses denote symmetrization. We can find 3~functions~$\veps_i(\xv)$ such that transformation\rf{diffeomorphism_transformation_3metric} brings 3 components of~$\gamma_{ij}$ to some standard form, amounting to a gauge condition. As noted in the previous paragraph, the spatial diffeomorphism transformation with these (or other) parameters~$\veps_i(\xv)$ preserves the wave function~$\psi$.

Define a \emph{scale factor} field~$a(\xv)$ by
\be
\gamma_{ij}= a^2 h_{ij}\,, \quad \det h_{ij}= 1\,.
\lb{a_def}
\ee
The \emph{Hamiltonian constraint}, $\nbrk{\hat\cH_\lapse\psi=0}$, relates the values of~$\psi$ at different~$a(\xv)$ [cf.\  the Wheeler-DeWitt equation\rf{WDW_with_matter} below]. This lets us regard~$a(\xv)$ as another arbitrary transformation parameter, analogous to $\varphi(\xv)$ and~$\veps^i(\xv)$. Thus only two components of~$\gamma_{ij}$ remain independent dynamical variables (gravitons). We map them to the basic coordinates~$q$ similarly to\rf{A_standing_harmonics}.

\subsection{Recovering canonical quantum gravity}
\label{subsec_canonical_gravity}

Formulation of generally covariant dynamical laws is much easier in the Lagrangian formalism. Consider a gauge- and diffeomorphism-invariant local action 
\be                   
S \equiv \int d^4x \lf[\frac{\sqrt{-g}}2\,R - \!\!
	\sum_{\rm simple\atop subgr.}\frac{\sqrt{-g}}{4e^2}\,F_{\mu\nu}^{r}F^{r\,\mu\nu}\,+\rt.\quad\nn\\
	\lf.\vphantom{\frac{\sqrt{-g}}2} +\,\cL^\phi(D_\mu\phi,\phi,g_{\mu\nu})\rt].\quad
\lb{action_full}
\ee
In it, $\nbrk{m^2_P\equiv(8\pi G)^{-1}}\equiv 1$, $\nbrk{g\equiv\det g_{\mu\nu}}$, $R$~is the Riemann curvature scalar, and $e$~are the gauge couplings (generally different for different simple subgroups of the overall gauge group). Also, $\nbrk{S^\phi \equiv \int d^4x\,\cL^\phi}$ is required to be gauge- and diffeomorphism-invariant. We suppose that the matter part of the action contains only renormalizable terms because only they are generically expected at sub-Planckian energy\ct{Polchinski_renorm_84}. For the same reason, we describe dynamics of the metric by Lagrangian density that is proportional to~$R$ -- the local invariant of the lowest energy dimension constructed of the metric tensor and its derivatives. These considerations still allow coupling of~$R$ to scalar fields. We disregard such coupling based on the experimental bounds on it. (For some of the latest bounds and references to earlier constraints see, e.g.,\ct{omega_BD_constr_13,omega_BD_constr_14}.)

The Hamiltonian for the gravitational part of\rf{action_full}, the Hilbert action
\be
S^g = \int d^4x \,\frac{\sqrt{-g}}2\,R \,,
\lb{action_Hilbert}
\ee
is\ct{Dirac,DeWitt}
\be
H^g = \int d^3x \ \shift^a \cH_a^g(\pi^{ij},\gamma_{ij})\,.
\lb{H_g}
\ee
Here, $\pi^{ij}$ are the operators of the momenta conjugate to the 3-metric~$\gamma_{ij}$,
\be
\pi^{ij}(\xv)= -i\,\frac{\delta}{\delta \gamma_{ij}(\xv)}\,,
\lb{pi_ij_def}
\ee
and $\nbrk{\cH_a^g=(\cH^{g\lapse},\gamma_{ij}\cH^{gj})}$ with\ct{Dirac,DeWitt}
\be
\cH^{g\lapse} \eqa  G_{AB}\,\pi^A\pi^B - \fr12\,\sqrt\gamma\ {}^{(3)}\!R\,,
\lb{HgN}\\
\cH^{gi} \eqa  -2\pi^{ij}{}_{|j}\,.
\lb{Hgi}
\ee
In\rf{HgN}: the indices $A$ and $B$ run over all the pairs~$ij$,
\be
G_{ij\,kl}\equiv \fr1{\sqrt\gamma}\lf(\gamma_{ik}\gamma_{jl}+\gamma_{il}\gamma_{jk}-\gamma_{ij}\gamma_{kl}\rt),
\ee
${}^{(3)}\!R$ is the Riemann curvature scalar for the 3-metric~$\gamma_{ij}$, and $\nbrk{\gamma\equiv \det\gamma_{ij}}$.
For cleaner formulas, we no longer place hats over quantum operators.

Appendix~\ref{apx_Hamiltonian} explicitly shows that the full Hamiltonian that corresponds to the entire action\rf{action_full} indeed has the form
\be
H = \int d^3x \, N^a \cH_a\,,
\lb{H_covariant_general}
\ee
anticipated in \sct{subsec_gravity_A}.
Here
\be
\cH_a= \cH_a^g +\cH_a^A +\cH_a^\phi
\ee
are local functions of: the fields $\nbrk{f=(\phi^\iota,A^r_i,\gamma_{ij})}$, their spatial derivatives, and conjugate momenta fields.

Appendix~\ref{apx_Hamiltonian} also demonstrates that
\be
e^{-i \int d^3x \, N^i \cH_i}\,\psi(f)=\psi(f-L_\Nv f) \,.
\lb{psi_spatial_displacement}
\ee
The respective similarity transformation\rf{fU_diffeomorphism} with $\nbrk{\veps^\mu=(0,\bm{\shift})}$ transports the field operators~$\hat f(\xv)$ in space by\rf{diffeomorphism_transformation_def}. Since the constructed wave function~$\psi$ satisfies the momentum constraints $\nbrk{\cH_i\psi=0}$,  eq.\rf{psi_spatial_displacement} shows that $\psi$~is invariant under the spatial displacement of the fields:
\be
\psi(f)=\psi(f-L_\Nv f) \,.
\lb{constancy_spatial_orbits}
\ee

By the Hamiltonian constraint, $\nbrk{\cH_\lapse\psi=0}$, the total Hamiltonian\rf{H_covariant_general} for the general~$\lapse(\xv)$ also annihilates~$\psi$. While the Schrodinger equation would then produce time-independent~$\psi$, the Hamiltonian constraint itself specifies physical evolution\ct{DeWitt}. The role of evolution time is now formally\footnote{
	Connecting the evolution of the quantum-gravitational wave function~$\psi$ in~$a(\xv)$ to the evolution of quantum fields in the quasiclassical time is by no means straightforward. This is described in Refs.\ct{DeWitt,Gerlach,LapchRubakov} and further in \sct{sec_physical_world}.
} 
taken by the metric scale factor~$a(\xv)$ of\rf{a_def}.

Indeed, \apx~A of Ref.\ct{DeWitt} proves that
\be
G_{AB}\,\pi^A\pi^B= -\,G_{AB}\,\fr{\delta^2}{\delta \gamma_A\, \delta \gamma_B}\,
\ee
in the gravitational Hamiltonian density\rf{HgN} is a hyperbolic Laplacian operator with signature $\nbrk{(-+++++)}$. The scale factor~$a(\xv)$ corresponds to the ``timelike" coordinate of the six coordinates~$\gamma_A(\xv)$ at every~$\xv$. Namely, from\ct{DeWitt},
\be
G_{AB}\,\pi^A\pi^B= \fr1{24a}\,\fr{\delta^2}{\delta a^2}-\,G'_{A'\!B'}\,\fr{\delta^2}{\delta \zeta_{A'}\delta \zeta_{B'}}\,
\ee
where~$G'_{A'\!B'}$ is positive definite and~$\zeta_{A'}$ are five independent coordinates that parameterize~$h_{ij}$ of\rf{a_def}. The Hamiltonian constraint is thus a hyperbolic equation. Explicitly, it is 
\be        
\lf(\fr1{24a}\,\fr{\delta^2}{\delta a^2}-\,G'_{A'\!B'}\,\fr{\delta^2}{\delta \zeta_{A'}\delta \zeta_{B'}} 
	- \fr12\,\sqrt\gamma\ {}^{(3)}\!R \,+\rt. \qquad\quad\quad \nn\\
\lf.\vphantom{\fr{\delta^2}{\delta a^2}} +\,\cH_\lapse^A +\cH_\lapse^\phi\rt)\psi = 0\,,\quad\quad
\lb{WDW_with_matter}
\ee
which is the Wheeler-DeWitt equation of canonical quantum gravity with matter fields.

\section{Emergence of quasiclassical spacetime and of the Schrodinger equation}
\label{sec_physical_world}

This section will describe the emergence of quasiclassical spacetime. It will then derive the Schrodinger equation for the particle and metric fields in it. This description, consolidating and generalizing results of Refs.\ct{Gerlach,LapchRubakov,Kim_95}, applies to any wave function that satisfies the Wheeler-DeWitt equation with matter\rf{WDW_with_matter}, including the studied generically  emergent field wave functions.

\subsection{Emergence of quasiclassical background}
\label{subsec_emergent_classical}

As reviewed at the end of the previous section, a suitable evolution parameter for solutions of the Wheeler-DeWitt equation\rf{WDW_with_matter} is the scale-factor field~$a(\xv)$, defined by\rf{a_def}. Write solutions of\rf{WDW_with_matter} symbolically as~$\psi(f)$, where now $\nbrk{f\equiv(\phi,A_i,h_{ij},a)\equiv(\phi,A_i,\gamma_{ij})}$. Given initial $\psi$ and~$\delta\psi/\delta a(\xv)$ for all~$(\phi,A_i,h_{ij})$ at some $\nbrk{a(\xv)= a_0(\xv)}$, the hyperbolic Wheeler-DeWitt equation\rf{WDW_with_matter} uniquely determines~$\psi(f)$ for other configurations~$a(\xv)$.

Let us split the fields~$f(\xv)$ into two terms 
\be
f^a(\xv) = \bar f^a(\xv) + \pert{f}^a(\xv)\,
\lb{f_bar_tilde_split}
\ee
as follows. Let~$\bar f$ be composed of the field modes whose spatial wavelength at a reference configuration~$a_0(\xv)$ of the evolution parameter~$a(\xv)$ is larger than a certain fixed value. Let $\pert{f}$ be composed of the remaining, short-wavelength, modes.
We will specify the borderline wavelength later.
The variables $\bar f$ will apply to quasiclassical degrees of freedom for macroscopic scales. Their energy and De-Broyle frequency will typically exceed the Planck energy by orders of magnitude.

Consider solutions~$\psi(\bar f,\pert{f})$ of\rf{WDW_with_matter} that describe quasiclassical evolution of the macroscopic degrees of freedom~$\bar f$.
Start with the quasiclassical ansatz
\be
\psi(\bar f,\pert{f}) = A(\bar f)\,e^{iS(\bar f)}\,\pert{\psi}(\bar f,\pert{f})\,
\lb{psi_quasiclassical}
\ee
where $A(\bar f)$~and $S(\bar f)$ are real, but~$\pert{\psi}(\bar f,\pert{f})$ is generally complex.
The wave function\rf{psi_quasiclassical} is quasiclassical in~$\bar f$ when the phase rate of change~$\delta S/\delta\bar f$ and the prefactor~$A\pert{\psi}$ vary negligibly in~$\bar f$ over the increments $\Delta\bar f$~for which $\nbrk{\Delta S\sim1}$.

Set $S(\bar f)$ and~$A(\bar f)$ in\rf{psi_quasiclassical} to be respectively the phase and amplitude of the ``background'' wave function
\be
\bar\psi(\bar f)=A(\bar f)\,e^{iS(\bar f)}\,
\lb{psi_bar}
\ee
that by definition obeys
\be
\bar \cH_a(\bar f,\bar\pi)\, \bar\psi(\bar f) = 0\,.
\lb{constraint_background}
\ee
Here~$\bar \cH_a(\bar f,\bar\pi)$ is obtained from $\nbrk{\cH_a(\bar f+\pert{f},\bar \pi,\pert{\pi})}$ by dropping all the terms that involve the variables~$\pert{f}$ or their conjugate momenta~$\pert{\pi}$, i.e., by setting~$\pert{f}$ and~$\pert{\pi}$ to zero.
For the leading order of quasiclassical expansion, replace the operators $\nbrk{{\bar\pi}=-i\delta/\delta\bar f(\xv)}$ in\rf{constraint_background} by c-numbers
\be
\bar\pi = \fr{\delta S}{\delta\bar f(x)}\,.
\ee
This yields the Hamilton-Jacobi equations for~$S(\bar f)$, Ref.\ct{DeWitt,Gerlach}:
\be
\bar \cH_a(\bar f,\bar\pi) \equiv
\bar \cH_a\!\lf(\bar f,\frac{\delta S}{\delta\bar f}\rt) = 0\,.
\lb{Hamilton_Jacobi}
\ee

Gerlach\ct{Gerlach} showed with admirable care how for general relativity without matter the Hamilton-Jacobi equations\rf{Hamilton_Jacobi} lead to classical evolution of the dynamical fields.
Later Kim\ct{Kim_95} stressed that for consistent quasiclassical description of inflating universe the ``background" Hamiltonian components~$\bar\cH_a$ must include not only the long-wavelength metric degrees of freedom~$\bar\gamma_{ij}$ but also, at least, the long-wavelength part of the field~$\phi$ that drives the inflation. Omission of the inflaton field from~$\bar\cH_a$ would lead to \emph{non-oscillating} solution\ct{Kim_95} for the background wave function\rf{psi_bar}. Then the quasiclassical requirement of slow variation of the prefactor~$A(\bar f)$ in\rf{psi_bar} could not be fulfilled.

\begin{figure}[t]
\centering
\includegraphics[width=0.15\textwidth]{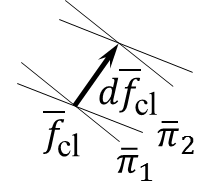}
\caption{
An interval $[\bar f\cl,\bar f\cl+d\bar f\cl]$ of a classical trajectory~$\bar f\cl(x)$ is formed by constructive interference of quasiclassical solutions  $\nbrk{\psi_{\bar\pi}(\bar f)\propto e^{i\bar\pi\cdot d\bar f}}$ of\rf{constraint_background}---waves of~$\psi$ in the configuration space---with various momenta~$\bar\pi$ in a range $\nbrk{[\bar\pi\cl-\delta\bar\pi,\,\bar\pi\cl+\delta\bar\pi]}$.
The thin lines depict constant-phase surfaces of the waves  with momenta~$\bar\pi_1$ and~$\bar\pi_2$ from that range. The waves add up along the classical path~$d\bar f\cl$, depicted by the thick arrow, that fulfills condition\rf{extremum_unconstrained}.}
\label{fig_interference}
\end{figure}

It is straightforward to generalize Gerlach's arguments\ct{Gerlach} to $\bar f$ and $\bar\cH_a$ that incorporate, besides the metric, the matter fields. An observed trajectory of the system\rf{constraint_background} is a continuous set of the field configurations~$\bar f\cl(\xv)$ that are points of constructive interference of solutions of eqs.\rf{constraint_background}. Consider the solutions of\rf{constraint_background} that are waves of the form
\be
\psi_{\bar\pi}(\bar f_0+d\bar f)\propto e^{i\bar\pi\cdot d\bar f}\,,
\lb{psi_pi}
\ee
where
\be
\bar\pi\cdot d\bar f\equiv \int d^3x\,\sum_\iota \bar\pi_\iota(\xv)\, d\bar f^\iota(\xv)\,.
\ee
Consider waves\rf{psi_pi} whose wavenumbers~$\bar\pi$ belong to a narrow range $\nbrk{[\bar\pi\cl-\delta\bar\pi,\,\bar\pi\cl+\delta\bar\pi]}$, centered at some~$\bar\pi\cl$. These waves interfere constructively along an interval~$d\bar f\cl$ of a trajectory in the configuration space when\ct{Gerlach}
\be
\delta\bar\pi\cdot d\bar f\cl= 0\,.
\lb{extremum_unconstrained}
\ee
This condition is illustrated by \fig{fig_interference}.

In the quasiclassical limit, the phase $\nbrk{\bar\pi\cdot d\bar f}$ of every solution\rf{psi_pi} obeys the Hamilton-Jacobi equations\rf{Hamilton_Jacobi}. Following\ct{Gerlach}, we use the Lagrange multiplies to find the extremum\rf{extremum_unconstrained} under the constraints\rf{Hamilton_Jacobi} for every~$\xv$. Introducing the Lagrange multiplies $dt\,N^a(\xv)$ of variations of\rf{Hamilton_Jacobi}, we thus require for all infinitesimal~$\delta\bar\pi$ the following variation to vanish:
\be
\delta\bar\pi\cdot d\bar f\cl - \int d^3x\,dt\,N^a \delta\bar\cH_a(\bar f\cl,\bar\pi) =0\,.
\lb{extremum_Lagrange_mult}
\ee
With
\be
\dot{\bar f}\cl\equiv d\bar f\cl/dt= \partial_t\bar f\cl\,,
\ee
eq.\rf{extremum_Lagrange_mult} becomes:
\be
dt\!\int\! d^3x\ \delta\bar\pi \lf(\dot{\bar f}\cl - N^a \frac{\partial \bar\cH_a}{\partial\bar\pi\cl}\rt) = 0\,.
\lb{extremum_constrained}
\ee
Condition\rf{extremum_constrained} yields half of the Hamilton equations of classical motion:
\be
\partial_t\bar f\cl(x) = N^a \frac{\partial \bar\cH_a}{\partial\bar\pi\cl} = \frac{\delta \bar H(\bar f\cl,\bar\pi\cl,\shift^a)}{\delta\bar\pi\cl(\xv)}\,,
\lb{Hamilton_equations_f}
\ee
with
\be
\bar H = \int d^3x\,N^a \bar\cH_a(\bar f,\bar\pi)\,.
\lb{bar_H_def}
\ee
The other half of the classical Hamilton equations,
\be
\partial_t\bar \pi\cl(x) = -\,\frac{\delta \bar H(\bar f\cl,\bar\pi\cl,\shift^a)}{\delta\bar f\cl(\xv)}\,,
\lb{Hamilton_equations_pi}
\ee
then also follows as shown in\ct{Gerlach}. The classical equations of motion\rfs{Hamilton_equations_f}{Hamilton_equations_pi} should be complemented by the constraints\rf{Hamilton_Jacobi}:
\be
\lf.\fr{\delta\bar H}{\delta\shift^a(\xv)}\rt|_{\bar f,\bar\pi=\bar f\cl,\bar\pi\cl}\!\!\!\!= \bar\cH_a(\bar f\cl,\bar\pi\cl)=0\,.
\lb{constraints_Hamiltonian_cl}
\ee

The lapse and shift~$N^a(x)$ [cf.\rfs{Na_def}{Hamiltonian_covariant}] can be chosen arbitrarily. Their choice, however, affects $\bar f\cl(x)$ and $\bar\pi\cl(x)$ that result from integrating the dynamical equations\rfs{Hamilton_equations_f}{Hamilton_equations_pi}. For any~$N^a(x)$, the obtained configurations~$\bar f\cl(t,\xv)$, considered as functions of~$\xv$ for various spatial hypersurfaces with fixed~$t$, are points of constructive interference of the waves\rf{psi_pi}. 
The set of all such configurations~$\{\bar f\cl(\xv)\}$ can be parameterized\ct{Gerlach} by Tomonaga's ``many-fingered time" $\sigma(\xv)$\ct{Tomonaga} as $\nbrk{\{\bar f\cl(\sigma(\xv),\xv)\}}$. This set may be interpreted physically as the configurations of the classically evolving fields~$\bar f\cl(x)$ on all the various spatial slices $\nbrk{(t =\sigma(\xv),\xv)}$ of the physical spacetime.

\subsection{Evolution of emergent quantum fields' modes in the quasiclassical background}
\label{subsec_tilde_Hamiltonian}

\subsubsection{Hamiltonian formulation}

Let us return to the full diffeomorphism-symmetric quantum field system. Its wave function~$\psi(f)$ satisfies $\nbrk{\hat\cH_a\psi(f)=0}$, eq.\rf{Hamiltonian_momentum_constraints}. We separated the fields~$f$  into large-wavelength and short-wavelength components\rf{f_bar_tilde_split}. In the respective~$\psi(\bar f,\pert{f})$, we now confine the quasiclassical, large-scale degrees of freedom~$\bar f$ to the trajectories of constructive interference~$\bar f\cl(x)$, shown by the previous subsection to be classical trajectories. Let us demonstrate that then the system's quantum, small-scale degrees of freedom~$\pert{f}$ have a wave function that evolves by the Schrodinger equation $\nbrk{i\partial_t\pert{\psi}(\pert{f},t) = \pert{H}\pert{\psi}(\pert{f},t)}$ with 
\be
\pert{H}=\int d^3x\,N^a\cH_a-\bar H.
\lb{Htilde_def}
\ee 

Lapchinsky and Rubakov explored solutions of the Wheeler-DeWitt equation with matter\rf{WDW_with_matter} in\ct{LapchRubakov}. They confined the gravitational degrees of freedom to the classical configurations and treated all the matter fields quantum-mechanically. They showed that in the leading order of quasiclassical expansion the prefactor~$\pert{\psi}(\pert{f},\bar f\cl(t))$ of the ansatz\rf{psi_quasiclassical} for such solutions does satisfy the Schrodinger equation.

However, Kim later demonstrated\ct{Kim_95} that the conditions for quasiclassicality necessarily fail when in an expanding universe only gravity is treated quasiclassically. We therefore extend the arguments of Lapchinsky and Rubakov to the situation where the quasiclassical variables~$\bar f$ are the long-wavelength components of the metric \emph{as well as  of matter} fields. 

Start from the Hamiltonian constraint for the full wave function\rf{psi_quasiclassical}:
\be
(\bar H + \pert{H})\lf[A(\bar f)\,e^{iS(\bar f)}\,\pert{\psi}(\bar f,\pert{f})\rt]=0\,.
\lb{full_quasiclassical_equation}
\ee
Of the various derivatives over~$\bar f$ that appear from $\nbrk{\bar\pi = -i\delta/\delta\bar f}$ in~$\bar H(\bar f,\bar\pi)$ acting on the expression in the brackets, retain only the first derivatives of $S$~and~$\pert{\psi}$.\footnote{
	The derivatives of~$A(\bar f)$ on the left-hand side of\rf{full_quasiclassical_equation} are canceled by the terms that appear at the higher orders of quasiclassical expansion of eq.\rf{constraint_background} for the background~$\nbrk{\bar\psi(\bar f)\equiv A\,e^{iS}}$. Hence $A(\bar f)$~derivatives do not affect the short-scale wave function~$\pert{\psi}$ in\rf{full_quasiclassical_equation} at the considered leading order.
}
This gives
\be
\bar H(\bar f,\frac{\delta S}{\delta\bar f})\,\pert{\psi} -i
\!\int\! d^3x\,\frac{\delta \bar H}{\delta\bar\pi_\iota(\xv)}\,\frac{\delta \pert{\psi}}{\delta\bar f^\iota(\xv)}+ \pert{H}\pert{\psi} = 0\,.~~
\lb{qu_expanded}
\ee
The first term of the last equation vanishes by\rf{Hamilton_Jacobi}. If we confine~$\bar f$ to the classical solutions~$\bar f\cl$ then\rf{Hamilton_equations_f} lets us replace $\delta \bar H/\delta\bar\pi_\iota(\xv)$ in the second term of\rf{qu_expanded} by~$\partial_t\bar f\cl^\iota$. Then\rf{qu_expanded} becomes
\be
-i\int d^3x\,(\partial_t\bar f\cl^\iota)\,\frac{\delta \pert{\psi}(\pert{f},\bar f\cl)}{\delta\bar f\cl^\iota(\xv)}+ \pert{H}\pert{\psi} = 0\,.
\ee
Thus
\be
i\,\frac{d}{dt}\,\pert{\psi}(\pert{f},\bar f\cl(t)) = \pert{H}\pert{\psi}\,.
\lb{Schrodinger_eq_derived}
\ee
This is the standard Schrodinger equation for the wave function
\be
\pert{\psi}(\pert{f},t)\equiv \pert{\psi}(\pert{f},\bar f\cl(t))\,.
\lb{psi_reduced}
\ee

The Hamiltonian~$\pert{H}$, defined by\rf{Htilde_def}, involves the canonically conjugate momenta of the matter fields ($\pert{\phi}$ and~$\pert{A}_i$) and of the metric~$\tilde\gamma_{ij}$. The metric momenta contribute to~$\pert{H}$ a wrong-sign kinetic energy term~$\delta^2/\delta\pert{a}^2$
from the first term in the Wheeler-DeWitt equation\rf{WDW_with_matter}.
We can avoid it, and the resulting (spurious) instabilities, by excluding the metric scale factor~$a$ from~$\pert{f}$ as follows. 

The modes of~$a(\xv)$ are not dynamical degrees of freedom because $a(\xv)$~enters the dynamical equation\rf{WDW_with_matter} as an evolution parameter. In particular, the modes of~$a(\xv)$ should not in \sct{subsec_emergent_gravity} be identified with independent dimensions~$q^n$ of a representation~$\psi(q)$ of the basic distribution~$\dist(Q)$. These modes should correspondingly be excluded from the integration measure~$df$ in the norm $\nbrk{\int df\,|\psi(f)|^2}$ for the probability of various macroscopic outcomes.
We thus can consider the wave function of the remaining fields $\nbrk{\phi=\bar\phi+\pert{\phi},\dots,h_{ij}=\bar h_{ij}+\pert{h}_{ij}}$ with $\nbrk{a=\bar a}$. I.e., we can impose a gauge condition $\nbrk{\pert{a}\equiv 0}$.
Then the arguments of the \emph{reduced wave function}\rf{psi_reduced} for the system\rf{action_full} explicitly are  $\nbrk{\pert{\psi}(\pert{\phi},\pert{A}_i,\pert{h}_{ij},t)}$. This reduced wave function does not involve~$\pert{a}$. Its evolution equation thus does not contain the spurious kinetic energy term~$\delta^2/\delta\pert{a}^2$ of the wrong sign.

Addition to the background Hamiltonian~$\bar H$ of the term~$\pert{H}$ for the short-wavelengths modes generally displaces the trajectory of constructive interference of the quasiclassical modes away from the classical background solution~$\bar f\cl(t)$. Indeed, the classical force~$-\delta\bar H/\delta\bar f$ in the Hamilton equation\rf{Hamilton_equations_pi} for~$\partial_t\bar\pi\cl$ should receive additional contribution~$-\delta\pert{H}/\delta\bar f$, described by a quantum operator. In contrast, the left-hand side of\rf{Hamilton_equations_pi} is a classical expression,~$\partial_t\bar \pi\cl$. This indicates that the approximations leading to\rf{Hamilton_equations_pi} are too crude to properly account for the quantum backreaction~$-\delta\pert{H}/\delta\bar f$.

One can solve the complete equation $\nbrk{(\bar H + \pert{H}) \,\psi(\bar f,\pert{f})=0}$ consistently by expanding~$\psi(\bar f,\pert{f})$ over the eigenstates of~$\pert{H}$\ct{Kiefer_88} or another convenient complete set of states\ct{Kim_95} as
\be
\psi(\bar f,\pert{f})= \sum_n c_n(\bar f)\,\psi_n(\bar f,\pert{f})\,.
\ee
The resulting formalisms \ct{Kiefer_88,Kim_95} are more involved than just a single Schrodinger equation for~$\pert{\psi}(\pert{f},t)$. These approaches analyze multiple branches~$\{n\}$ of background evolution. The backreaction~$-\delta\pert{H}/\delta\bar f$ is specific in each of the branches but is branch-dependent.

\subsubsection{Lagrangian formulation}

The Lagrangian formulation involves the fields~$f$ and their spacetime derivatives~$\partial_\mu f$ but not their canonical momenta~$\pi$. It hence manifestly reveals local symmetries of the field dynamics. The classical trajectories~$\bar f\cl$\rfs{Hamilton_equations_f}{Hamilton_equations_pi} of the quasiclassical macroscopic variables~$\bar f$ extremize the ``background" action $\nbrk{\bar S \equiv \int dt\,L(\dot{\bar f},\bar f,\shift^a)}$:
\be
\lf.\frac{\delta \bar S}{\delta\bar f(x)\vphantom{\dot{\bar f}}}\rt|_{\bar f=\bar f\cl}\!\!=0\,.
\lb{Euler_Lagrange}
\ee
The background solutions also satisfy the constraints
\be
\lf.\frac{\delta \bar S}{\delta\shift^a(x)}\rt|_{\bar f=\bar f\cl}\!\!\!=\lf.\frac{\delta  L(\dot{\bar f},\bar f,\shift^a)}{\delta\shift^a(\xv)}\rt|_{\bar f=\bar f\cl}\!\!\!=0\,.
\lb{constraints_Lagrangian_bar}
\ee
They are equivalents of the constraints\rf{constraints_Hamiltonian_cl} of the Hamiltonian formulation.

By\rf{ginvADM}, $\shift^a$~can be traded for the components~$g^{0\mu}$ of the inverse metric tensor. Hence, for the action of general relativity, the constraints\rf{constraints_Lagrangian_bar} give 4 of the 10 independent components of the Einstein equations for the classical background fields:
\be
\bar G_{0\mu}-\bar T_{0\mu}=\frac{2}{\sqrt{-\bar g}}\,\fr{\delta \bar S}{\delta g^{0\mu}(x)}= 0\,,
\lb{Einstein_0mu_bar}
\ee
where $\nbrk{G_{\mu\nu}\equiv R_{\mu\nu}-\fr12\,g_{\mu\nu}R}$, $R_{\mu\nu}$ is the Riemann tensor, and~$T_{\mu\nu}$ is the matter energy-momentum tensor.  

The full system, including its quantum degrees of freedom~$\pert{f}$, is described by a diffeomorphism-invariant action  $\nbrk{S=\int dt\,L}$ with a local Lagrangian
\be
L(\dot f,f,\shift^a)=\int d^3x\ \cL\,.
\ee 
By the secondary constraints\rf{Hamiltonian_momentum_constraints},
\be
\lf.\fr{\delta L}{\delta\shift^a(\xv)}\rt|_{f=\hat f}\!\psi=
-\lf.\fr{\delta H}{\delta\shift^a(\xv)}\rt|_{(f,\pi)=(\hat f,\hat\pi)}\!\!\!\!\psi=0\,.
\lb{constraints_Lagrangian}
\ee
For the studied general relativistic action\rf{action_full}, these equations reduce, similarly to\rf{Einstein_0mu_bar}, to
\be
(\hat G_{0\mu}-\hat T_{0\mu})\psi=0\,.
\ee

The overall wave function~$\psi(\bar f,\pert{f})$ is a superposition of numerous decohered Everett's branches. Its branches represent various realizations of classical background~$\bar f\cl(x)$. We may treat $\bar f\cl(x)$ of any individual Everett's branch as external classical fields. 
In that branch we can describe the physical world by a pair $\nbrk{(\bar f\cl,\pert{\psi})}$. Its $\bar f\cl(x)$~evolves by the classical Euler-Lagrange eqs.\rf{Euler_Lagrange} for the Hamiltonian~$\bar H$. 
(It can be corrected for the backreaction of quantum degrees of freedom as, e.g., suggested in\ct{Kim_95}.) The reduced wave function $\pert{\psi}(\pert{f}(\xv),t)$ evolves by the Hamiltonian~$\pert{H}$. Both $\bar H$ and~$\pert{H}$ correspond to the same diffeomorphism-invariant local action~$\nbrk{S=\int dt\,L}$.

\section{Fermions and local supersymmetry}
\label{sec_fermions}

\subsection{Emergent fermionic quantum fields}
\label{subsec_fermions}

Low-resolution presentations of the generic distribution~$\dist(Q)$ from \sct{sec_structure} can also be identified with wave functions of systems that include fermions. Consider a wave function~$\nbrk{\psi(f,\nf)}$ for bosonic fields $\nbrk{\hat f\equiv\{\hat f^\iota(\xv)\}}$ and for $M$~modes of fermionic fields $\nbrk{\hat\chi\equiv\{\hat\chi^\alpha(\xv)\}}$. The second argument of~$\psi(f,\nf)$ refers to the occupation numbers of the fermionic modes:
\be
\nf\equiv(\nf_1,\nf_2,\dots,\nf_M)\,, \quad \nf_m\in\{0,1\}\, .
\lb{nf_def}
\ee
The wave function~$\nbrk{\psi(f,\nf)}$ is thus composed of $2^M$~bosonic wave functions---one for every sequence of $0$'s and~$1$'s in\rf{nf_def}.

Such a wave function arises from dividing the considered earlier smooth representations~$\psi(q)$ of the generic distribution into $2^M$~classes. For example, we may label every basic entity~$E_a$, depicted by a dot in \fig{fig_A}, by $M$~bits $\nbrk{(\nf_1,\dots,\nf_M)}$. Then the wave function~$\psi(f,\nf)$ of the bosonic fields~$f$ for specific values of the fermionic occupation numbers~$\nf$ from\rf{nf_def} can be identified (repeating the steps in the earlier sections) with a representation~$\psi_\nf(q)$ of the smoothed distribution~$\dist_\nf(Q)$ over only the entities~$E_a$ whose labels equal~$\nf$. A more elegant and economical mechanism for splitting the smoothed representations of the generic basic distribution~$\dist(Q)$ into $2^M$~bosonic realizations will be presented in a future paper.

Define linear, anticommuting \emph{elementary fermionic operators}~$\hat\chi_m$, $\nbrk{m=1,\dots,M}$, that annihilate the component of~$\psi(f,\nf)$ with~$\nf_m=0$ and lower its component with~$\nf_m=1$ to~$\nf_m=0$, e.g., as:
\be
\hat\chi_m\!\lf(\ba{c} \psi(\dots \nf_m\!=0 \dots)\\ \psi(\dots \nf_m\!=1 \dots) \ea\!\rt)\!=\!  \lf(\!\ba{c} s_m\psi(\dots \nf_m\!=1 \dots)\\ 0 \ea\!\rt)\!.\qquad~~
\lb{chi_n_def}
\ee
The sign factor~$s_m$ on the right-hand side is
\be
s_m(\lambda)=(-1)^{\sum_{i=1}^{m-1}\nf_i}\,,
\ee
set to achieve anticommutativity of~$\hat\chi_m$.
The standard Hermitian product, derived for the emergent wave functions in \scts{sec_structure} and~\ref{subsec_Hermitian_product}, with the fermionic degrees of freedom in an appropriate basis becomes
\be
\langle\psi_1|\psi_2\rangle= \int df\sum_{\nf} \psi^*_1(f,\nf)\,\psi_2(f,\nf)\,.
\lb{Hermitian_product_fermionic}
\ee
The integral and sum on the right-hand side should for locally symmetric systems be taken only over the dynamical degrees of freedom. 

Operators~$\hat\chi_m^\dagger$, defined as the Hermitian conjugates of~$\hat\chi_m$ from\rf{chi_n_def}
for the Hermitian product\rf{Hermitian_product_fermionic}, act as
\be
\hat\chi_m^\dagger\lf(\ba{c} \psi(\dots \nf_m\!=0 \dots)\\ \psi(\dots \nf_m\!=1 \dots) \ea\rt)=    \lf(\ba{c} 0 \\ s_m(\nf)\,\psi(\dots \nf_m\!=0 \dots) \ea\rt).
\nn
\ee
The operators~$\hat\chi_m$ and~$\hat\chi_m^\dagger$ satisfy the canonical anticommutation relations
\be
 \ba{l}
\acom{\hat\chi_m^{\vphantom{\dagger}},\hat\chi_{m'}^\dagger}= \delta_{mm'}\,\\
{}\acom{\hat\chi_m,\hat\chi_{m'}}= \acom{\hat\chi_m^\dagger,\hat\chi_{m'}^\dagger}= 0\,.
 \ea
\lb{anticommutation_basic}
\ee

We identify the anticommuting operators~$\hat\chi_m$ with the annihilation operators of modes of fermionic fields. Alternatively, we could identify~$\hat\chi_m$ directly with the local fermionic field operators~$\hat\chi(\xv)$ at certain spatial points of the current-time hypersurface [for example, the points~$\xv_{\nv}$ of\rf{x_n_def}].
The mode-annihilation operators and local field operators are related by the Bogoliubov transformation 
\be
\hat\chi_{n}' = \alpha_{nm}^{} \hat\chi_m^{}+\beta_{nm}^{}\hat\chi_m^\dagger\, .
\lb{general_fermionic_transformation}
\ee
It preserves the canonical anticommutators\rf{anticommutation_basic} when the matrices~$\alpha_{nm}$ and $\beta_{nm}$ obey
\be
 \ba{l}
\alpha\alpha^\dagger+\beta\beta^\dagger=I\,\\
\alpha\beta^T+\beta\alpha^T=0\,,
 \ea
\ee
where $I$ is the identity matrix. 
The matrix~$\beta$ in\rf{general_fermionic_transformation} for the transformation from the modes to the local fields is nonzero. Hence the Fock vacuum of the modes is not annihilated by the field operators~$\hat\chi(\xv)$ and vice versa. 

For easier description of the Hamiltonian eigenstates, we identify the elementary anticommuting operators~$\hat\chi_m$ of\rf{chi_n_def} with the mode annihilation operators. Correspondingly, we introduce the local fermionic fields via their standard expansion over the modes. For example,  a spinor field in flat space equals
\be       
 \ba{r}
\hat\chi^\alpha(\xv) = {\displaystyle \fr1{V^{1/2}} \sum_{\mv}\fr1{\sqrt{2\omega_{\mv}}}}\, 
 [\hat\chi_\mv^{1s}u_s^\alpha(\kv_\mv)\, e^{i\kv_\mv\cdot\xv} +~~~\\ 
 +~\hat\chi^{2s}_\mv{}^\dagger v_s^\alpha(\kv_\mv)\, e^{-i\kv_\mv\cdot\xv}]\,. \qquad~ 
  \ea
\lb{chi_x_construct}
\ee
Here the operators~$\hat\chi_\mv^{1s}$ are independent of~$\hat\chi_\mv^{2s}$ for Dirac spinors or linearly related to them for Weyl and Majorana spinors. The index~$s$ stands for either a spin projection~($\pm \fr12$) or chirality~($L$, $R$). The canonical anticommutators
\be
 \ba{l}
\acom{\hat\chi^\alpha(\xv),\hat\chi^\dagger_\beta(\xv')}= \delta^\alpha_\beta\,\delta^{(3)}(\xv-\xv')\,\\
{}\acom{\hat\chi^\alpha(\xv),\hat\chi^\beta(\xv')}= \acom{\hat\chi^\dagger_\alpha(\xv),\hat\chi^\dagger_\beta(\xv')}= 0\,
 \ea
\lb{ferm_fields_anticom}
\ee
should be understood, similarly to the bosonic canonical commutators\rf{phipi_n_commutators}, as the continuous limit of the analogous anticommutators for discrete points~$\xv_\nv$.

We construct the Rarita-Schwinger spin-$3/2$ fermionic field operator~$\hat\chi^\alpha_\mu(\xv)$ for gravitino via its similar expansion over modes with specific helicities. We identify the annihilation and creation operators for its dynamical (helicity~$\pm 3/2$) modes with independent elementary anticommuting operators $\hat\chi_m$ and~$\hat\chi_m^\dagger$. We will identify the other (helicity~$\pm 1/2$) gravitino modes of the emergent locally supersymmetric systems in the next \sct{sec_emergent_SUSY}.

\subsection{Local supersymmetry}
\label{subsec_supersym}

\Scts{sec_gauge_fields}--\ref{sec_gravity} interpreted smooth presentations~$\psi(q)$ of the generic static structure as wave functions of gauge- and diffeomorphism-symmetric bosonic fields~$f$. Every value of $\psi$'s argument~$q$ was mapped to an orbit of symmetry transformation of the fields~$f$. 
This yielded the wave function~$\nbrk{\psi(f)\equiv\psi(q(f))}$. The constancy of the wave function on the symmetry orbits considerably restricted the Hamiltonian for its evolution. Nevertheless, it failed to produce dynamical laws. For example, the couplings in the Hamiltonian could still vary arbitrarily in time. 

\Sct{sec_why_the} will show that emergent systems of bosonic and fermionic fields with \textit{local supersymmetry} evolve by unchanged laws. The same generic basic structure gives rise to alternate emergent locally supersymmetric systems with different dynamical laws. Yet the laws in each of the systems are unchanged during its evolution. The laws are encoded in the map of the basic coordinates~$q$ to (the dynamical modes of) the field configurations~$f(\xv)$, \sct{subsec_fixing}. 

Since supersymmetry transformation interchanges bosons and fermions, its variation parameters anticommute. These parameters, along with fermionic fields, in axiomatically formulated supersymmetric theories are elements of a formal Grassmann algebra. The considered here emergent fermionic fields, on the contrary,  are operators defined on the space built of materially represented $c$-number functions~$\psi(q,\nf)$. They do not require the Grassmann numbers, still handy for routine calculations, or other formal mathematical  constructions. The previous subsection~\ref{subsec_fermions} explicitly provided an implementation of suitable anticommuting fermionic operators. Further \sct{sec_emergent_SUSY} will discuss various realizations of the fermionic transformation parameters of supersymmetry.

Consider a system of bosonic and fermionic fields
\be
\hat F^\iota(\xv)  \equiv  (\hat f^\iota(\xv),\hat \chi^\iota(\xv))
\ee
whose dynamical equations are covariant under a Lie group of local similarity transformations
\be
\hat F^\iota(\xv) \to \hat F'{}^\iota(\xv) = \hat U^{-1}\hat F^\iota(\xv)\, \hat U\,.
\lb{transf_general}
\ee
For infinitesimal transformations from this group
\be
\hat U = \exp\lf[-i\int d^3x~\var^A(\xv)\,\hat\cH_A(\xv)\rt] 
\lb{U_general}
\ee
where $\var^A(\xv)$~are the transformation parameters, and $\hat\cH_A(\xv)$~are functions of the field operators~$\hat F^\iota(\xv)$ and their conjugate momenta $\hat \Pi_\iota(\xv)$, satisfying the commutation/anticommutation relations\rf{canonical_brackets}.
The functions $\cH_A(\hat F(\xv),\hat \Pi(\xv))$~may contain spatial derivatives of the field operators but not their temporal derivatives. 

From now on, we stop placing hats above operators. Except for wave functions, fields' indices, or spacetime coordinates, 
every quantity below can be considered as an operator that acts on wave functions.

Following Refs.\ct{Proeyen_SUSY_approach,Freedman_Proeyen_book}, we describe the degrees of freedom of a locally supersymmetric system in terms of fields from one of the two classes: covariant matter fields and connections. 

\emph{Covariant matter fields}, to be denoted
\be
\Phi^\iota= (\phi^\iota,\chi^\iota)\,,
\lb{covar_fields}
\ee 
with~$\phi^\iota$ and~$\chi^\iota$ for respectively bosonic and fermionic fields, change under the general symmetry transformation\rfs{transf_general}{U_general} so that their increment does not depend on spacetime derivatives of the transformation parameters~$\var^A(x)$\ct{Proeyen_SUSY_approach,Freedman_Proeyen_book}:
\be
\delta\Phi^\iota(x)\equiv \Phi'{}^\iota(x)-\Phi^\iota(x) = \var^A(x)\,(\tau_A\Phi)^\iota\,.
\lb{delta_Phi}
\ee
As long as transformation\rf{delta_Phi} is infinitesimal, the operators $(\tau_A\Phi)^\iota$ on its right-hand side are independent of~$\var^A$. In the Hamiltonian formulation, they are linear combinations of~$\{\Phi^\kappa\}$,  canonically conjugate~$\{\Pi_\kappa\}$, and spatial derivatives of~$\{\Phi^\kappa\}$ or~$\{\Pi_\kappa\}$. 

\emph{Connections}, to be collectively denoted~$\Omega^A_\mu$, for the discussed symmetries\rfs{e_list}{A_list} are the vierbein~$e^a_\mu$, spin connection~$\omega^{ab}_\mu$, gravitino\footnote{
    We avoid the conventional gravitino notation~$\psi_\mu$ to reserve the symbol~$\psi$ for wave functions.
}~$\chi^\alpha_\mu$, 
and Yang-Mills gauge fields~$A^r_\mu$:
\be
\Omega^A_\mu = (e^a_\mu,\omega^{ab}_\mu,\chi^\alpha_\mu,A^r_\mu).
\lb{Omega_general}
\ee
In both general relativity and supergravity the spin connection~$\omega^{ab}_\mu$ is not an independent field. It is composed of $e^a_\mu$ and~$\chi^\alpha_\mu$ as determined by the zero-torsion condition\rf{zero_torsion_constraint}, e.g., Refs.\ct{Proeyen_SUSY_approach,Freedman_Proeyen_book}. Our analysis does not depend on whether an emergent system obeys the constraint\rf{zero_torsion_constraint} or torsions are its independent degrees of freedom. Theories of either type emerge from the generic basic structure; although, due to different dimensionalities of their configuration spaces, their chances to be our world may differ drastically. 

Appendix~\ref{apx_fermions} summarizes the algebraic structure of local symmetries that include local supersymmetry. It presents formulas for transformation of the connections~$\Omega^A_\mu$\rf{Omega_general} and for covariant curvatures~$\R_{\mu\nu}^A$.

Temporal evolution is diffeomorphism transformation whose displacement parameters equal $\nbrk{\veps^\mu(\xv)=(dt,\bf 0)}$. Evolution in diffeomorphism-symmetric theory is therefore a special case of the general local symmetry transformation\rfs{transf_general}{U_general}. Let $\var^A_{\rm diff}(x,\veps)$ be the transformation parameters from\rf{U_general} that yield the diffeomorphism transformation\rfs{fpf}{xpx} with the displacement~$\veps^\mu$ [eq.\rf{delta_dif_decomposition}]
For transformation\rf{transf_general} of temporal evolution,
\be
 U = \exp(-i Hdt)\,
\lb{UH_general}
\ee
where the Hamiltonian operator~$H$ by\rf{U_general} equals
\be
H= \int d^3x~\frac{\partial\var^A_{\rm diff}}{\partial\veps^0}\,\cH_A= \int d^3x~\Omega^A_0\cH_A\,.
\lb{H_from_HA}
\ee
The last equality used the relation
\be
\Omega^A_0 = \partial\var^A_{\rm diff}/\partial\veps^0\,
\lb{Omega0}
\ee
that follows from\rf{OmegaA_mu_def}. The temporal components~$\Omega^A_0$ of the connections~$\Omega^A_\mu$ thus specify the gauge at future time. Hence $\Omega^A_0$~can be chosen at will.

Appendix~\ref{apx_supersymmetric_evolution} describes how to  construct generators~$\cH_A$ that produce a covariant local Hamiltonian\rf{H_from_HA}. The generators~$\cH_A$ can be built from a local action functional\rf{S_supersymmetric_apx} that is invariant under the desired symmetry transformations. Eq.\rf{H_A_explicit} explicitly gives~$\cH_A$ that correspond to the general locally symmetric action\rf{S_supersymmetric_apx}:
\be
\cH_A=(\tau_A\Phi)^\iota\Pi_\iota - \D_i\Pi^i_A - K_A^0\,.
\lb{cH_A_display}
\ee
Here,~$\D_i\Pi^i_A$ is the covariant divergence\rf{D_i_Pi} of the momenta fields~$\Pi^i_A$ (canonical conjugates of~$\Omega^A_i$), and $K_A^0$~is the temporal component of~$K_A^\mu$ from the Lagrangian density variation\rf{K_mu_def}. $K_A^0$~can be a function of the dynamical fields~$(\Phi^\iota,\Omega^A_i)$ and their conjugate momenta~$(\Pi_\iota,\Pi^i_A)$.

Appendix~\ref{apx_supersymmetric_evolution} demonstrates that a wave function of a locally supersymmetric system does not depend on~$\Omega^A_0$ and satisfies the secondary constraint\rf{secondary_constraint_A}:
\be
\cH_A \psi = 0\,.
\lb{secondary_constraint_display}
\ee
With~$\cH_A$ of\rf{cH_A_display}, the constraint\rf{secondary_constraint_display} is
\be
\D_i\Pi^i_A\psi=  \lf[(\tau_A\Phi)^\iota\Pi_\iota - K_A^0\rt]\psi\,. 
\lb{constraint_general}
\ee
The divergence~$\D_i\Pi^i_A$ on the left-hand side of\rf{constraint_general} may also enter $K_A^0$ on the right-hand side. Therefore, we algebraically resolve\rf{constraint_general} to obtain
\be
\D_i\Pi^i_A\psi=  \rho_A\psi\,
\lb{constraint_resolved}
\ee
where the generalized ``charge density''~$\rho_A(F,\Pi)$ no longer includes~$\D_i\Pi^i_A$. 
Eq.\rf{constraint_resolved} is our main equation for the wave function of locally supersymmetric fields.

\section{Emergent locally supersymmetric fields}
\label{sec_emergent_SUSY}  

We now identify generically emergent quantum fields that are locally supersymmetric. Their wave function exists materially. It is composed of presentations\rf{psi_general} of the smoothed generic discrete distribution as described in this section.

\subsection{Bosonic constraints and gauge modes}

Consider the local symmetries\rf{A_list}, consisting of local translation, local Lorentz, gauge symmetries, and local supersymmetry. Then the index~$A$ ranges over~$\nbrk{\{a,ab,r\}}$ for the bosonic components of the constraint\rf{constraint_resolved} and over~$\{\alpha\}$ for its fermionic components. 

The bosonic components of the constraint\rf{constraint_resolved} read
\be
-i\D_i\frac{\delta}{\delta\omega_i^A(\xv)}\,\psi\eqa  \rho_A\psi\,,
\lb{constraint_bosonic}
\ee
where
\be
\omega_i^A=(e^a_i,\omega^{ab}_i,A^r_i)
\ee
are the spatial components of the bosonic connections. 
Similarly to\rf{A_VS_decomposition} for the gauge connection~$A_i$, the general bosonic connection~$\omega_i$ decomposes into transverse and longitudinal parts:
\be
\omega_i^A=\omega_i^{TA} + \D_i \var^A\,.
\lb{omega_TA}
\ee
Here,~$\omega_i^{T}$ obeys some arbitrary gauge-fixing condition, e.g., $\nbrk{\gamma^{ij}\D_i^{}\omega_j^{T}=0}$.
An emergent wave function of bosonic fields stems from identifying the variables $\nbrk{q= (q^1,q^2,\dots)}$ of a presentation~$\psi(q)$ of the smoothed generic distribution with the dynamical variables $\nbrk{(\phi(\xv),\omega_i^{T}(\xv))}$, where $\phi$~are bosonic matter fields. This determines the wave function~$\psi(\phi,\omega_i^{T})$ of the fields in the gauge~$\nbrk{\var^A=0}$. The details repeat \sct{subsec_gauge_dofs} and are therefore omitted.  

Determine the wave function~$\psi(\phi,\omega_i)$ of the locally symmetric bosonic fields in any gauge. This wave function consists of the alternate representations of the above~$\psi(\phi,\omega_i^{T})$ that are obtained by transforming it along~$\var^A(\xv)$ as
\be
\delta\psi=\int d^3x~\delta\var^A(\xv)\,\frac{\delta\psi}{\delta\var^A(\xv)}\,
\lb{bosonic_ext_psi}
\ee 
with
\be
\frac{\delta\psi}{\delta\var^A(\xv)}=
\D_i\,\frac{\delta}{\delta\omega_i^A(\xv)}\,\psi=  i\rho_A\psi\,.
\lb{bosonic_extension}
\ee
The last line of the equations above follows from\rf{omega_TA} and\rf{constraint_bosonic}.

The gauge modes $\nbrk{\omega_i^{L}\equiv \D_i\var}$ [see\rf{omega_TA}] of the emergent system thus do not map to any underlying ``material'' variables. Rather, they are parameters of the transformation that creates~$\psi(\phi,\omega_i)$ from~$\psi(\phi,\omega_i^{T})$. This is visualized on the right side of further \fig{fig_psi_susy}.

\subsection{Fermionic constraints}
\lb{subsec_ferm_constr}

Let us discuss the fermionic directions of local symmetry transformation. To be specific, consider local supersymmetry with $\nbrk{\mathcal{N}=1}$. Its fermionic connections are the gravitino field, $\nbrk{\Omega^\alpha_\mu=\chi^\alpha_\mu}$. Being motivated by the above construction for bosons, we would like to extend the wave function~$\psi$ along the non-dynamical fermionic degrees of freedom (gravitino modes of helicities~$\pm 1/2$) by transformation. Correspondingly, we do \emph{not} map the occupation numbers of non-dynamical fermionic modes to independent bit variables~$\nf_m$\rf{nf_def} of presentations of a basic material structure. 

The momentum field~$\Pi^i_A$ on the left-hand side of the fermionic component of\rf{constraint_resolved} is the gravitino canonical momentum $\nbrk{\pi^i_\alpha=\Pi^i_\alpha}$. Its relation to the local gravitino field~$\chi^\alpha_\mu$ follows from the kinetic part of the gravitino action\ct{sugra_orig_76,Freedman_Proeyen_book}:
\be
S_{\pert{g}} =  -\int d^4x\,\frac{e}{2}\,\bar\chi_\mu \gamma^{\mu\nu\rho}D_\nu\chi_\rho \,,
\lb{S_gravitino}
\ee
where $\nbrk{e\equiv\det e_\mu^a}$, $\nbrk{\gamma^{\mu\nu\rho}\equiv \fr1{3!}\gamma^{[\mu}\gamma^\nu\gamma^{\rho ]}}$, $\nbrk{\gamma^\mu\equiv e^\mu_a\gamma^a}$, and
\be
D_\nu\chi_\rho = \partial_\nu\chi_\rho + \fr18\,\omega^{ab}_\mu[\gamma_a,\gamma_b]
\chi_\rho\,.
\nn
\ee
Variation of this action with respect to~$\partial_0^{}\chi_i^\alpha$ yields the gravitino canonical momentum
\be
\pi^i_\alpha= \frac{e}{4}\,(\bar\chi_j \gamma^0[\gamma^i,\gamma^j])_\alpha\,.
\lb{Pi_alpha_gen}
\ee

\begin{figure}[t]
\centering
\includegraphics[width=0.4\textwidth]{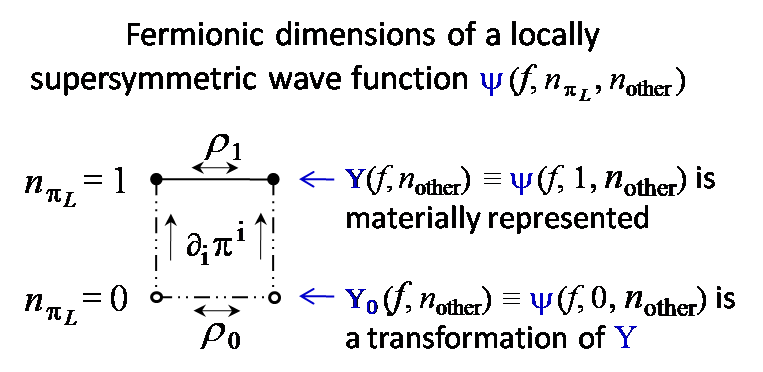}           
\caption{
Bit dimensions of the configuration space for a constrained gravitino mode ($\nf_{\pi_L}$, the vertical direction) and the other fermionic degrees of freedom ($\nf_{\rm other}$, the horizontal direction). The dependence of a locally supersymmetric wave function~$\psi(f,\nf_{\pi_L},\nf_{\rm other})$ on~$\nf_{\pi_L}$ is constrained by\rf{constraint_fermionic1}. The constraint requires that the lower component~($Y_0$) of the wave function in the basis\rf{divPi_fermionic} is a transformation\rf{psi_n_L_constr} of the upper component~($Y$).
}
\label{fig_psi_fermionic}
\end{figure}

\Fig{fig_psi_fermionic} depicts the fermionic component of the constraint\rf{constraint_resolved},
\be
\D_i\pi^i_\alpha\psi=  \rho_\alpha\psi\,.
\lb{constraint_fermionic}
\ee
For the anticommuting fermionic momenta fields, $\nbrk{(\pd_i\pi^i_\alpha)^2=0}$ at any~$\xv$ and~$\alpha$. (Let us from now on suppress the index~$\alpha$ as well.) In some basis
of the Hilbert space $\pd_i\pi^i$~is therefore a raising operator:\footnote{
    The upper-component states for this basis belong to the subspace annihilated by~$\pd_i\pi^i$. 
    The lower-component states belong to the orthogonal subspace of the Hilbert space.
}  
\be
\pd_i\pi^i = \lf(\ba{cc}0 & \lambda \\0 & 0\ea\rt),
\lb{divPi_fermionic}
\ee
where~$\lambda$ is an invertible bosonic operator. Given\rf{D_i_Pi}, we can rearrange the terms in\rf{constraint_fermionic} to obtain
\be
\pd_i\pi^i \psi=  \tilde\rho\,\psi\,
\lb{constraint_fermionic1}
\ee
where, for the representation\rf{divPi_fermionic},
\be
\tilde\rho = \lf(\ba{cc}\rho_1 & 0 \\ \rho_{01} & \rho_0\ea\rt).
\lb{rho_matrix}
\ee

Let $\nf_{\pi_L}$ be the bit coordinate of the configuration space (i.e., the fermionic occupation number) that is raised by~$\pd_i\pi^i$ of\rf{divPi_fermionic}. Consider the fields' wave function~$\psi(f,\nf_{\pi_L},\nf_{\rm other})$, where $\nbrk{f=(\phi,\omega_i)}$ are its bosonic arguments. Parameterize it in the considered representation as
\be
\psi(f,\nf_{\pi_L},\nf_{\rm other}) = \lf(\ba{c}Y(f,\nf_{\rm other}) \\Y_0(f,\nf_{\rm other})\ea\rt).
\lb{psi_n_L}
\ee
The constraint\rf{constraint_fermionic1} with~$\pd_i\pi^i$ from\rf{divPi_fermionic} and~$\tilde\rho$ from\rf{rho_matrix} amounts to
\be
\lf\{\ba{rcl} \lambda \,Y_0 \eqa \rho_1 Y\\ 
   0 \eqa \rho_{01} Y + \rho_0 Y_0.
\ea\rt.
\lb{constraint_ferm_two_conditions}
\ee

By the first of conditions\rf{constraint_ferm_two_conditions}, $Y_0$~is a transformation of the wave function's component~$Y$:
\be
Y_0= \lambda^{-1} \rho_1 \,Y,
\lb{psi_n_L_constr}
\ee 
where~$\lambda^{-1}$ is the operator inverse of the bosonic operator~$\lambda$ from\rf{divPi_fermionic}.
Thus the configuration-space bit dimensions~$\nf_{\pi_L}$, represented on the right-hand side of\rf{psi_n_L} by the vertical dimension, are non-dynamical. 
They correspond to the longitudinal modes of the gravitino momentum~$\pi^i_\alpha$.
Similarly to the longitudinal modes of bosonic gauge fields, they arise by transforming a function~$Y(f,\nf_{\rm other})$ into its equivalent presentation~$Y_0(f,\nf_{\rm other})$ by\rf{psi_n_L_constr}.  \Fig{fig_psi_fermionic} illustrates this.

The second equation of\rf{constraint_ferm_two_conditions} and eq.\rf{psi_n_L_constr} subject the remaining component~$Y$ of the wave function at every~$\xv$ to 4~constraints
\be
(\rho_{01} + \rho_0 \lambda^{-1} \rho_1)\,Y=0\,,
\ee
for the 4~values of the index~$\alpha$ of the Majorana fermionic operators in\rf{constraint_fermionic}. These constraints can be written as $\nbrk{\mathcal{P}_\alpha\psi=0}$ where the operators~$\mathcal{P}_\alpha$ is self-conjugate. They are equivalent to one Hamiltonian and three momentum constraints\rf{Hamiltonian_momentum_constraints} at every~$\xv$ with the Hermitian operators~$\cH_a(\xv)$.

Since the wave function of a locally supersymmetric system satisfies~$\nbrk{\cH_\alpha\psi=0}$ [eq.\rf{secondary_constraint_display}], it is invariant under local supersymmetry transformations:
\be
e^{-i\int d^3x\,\xi^\alpha(\xv)\cH_\alpha(\xv)}\psi =\psi\,.
\lb{SUSY_psi_sym}
\ee
This formula holds regardless of the implementation of the fermionic parameter field~$\xi^\alpha(\xv)$. Local supersymmetry transformation changes the gravitino field as
\be
\delta\chi^\alpha_i= (\D_i \xi)^\alpha
\lb{delta_chi_i}
\ee
[cf.\rf{delta_Omega_A}]. The parameter field~$\xi^\alpha(\xv)$ thus represents the longitudinal, gauge modes of gravitino.
 
It is instructive to identify the gauge, constrained, and dynamical modes of gravitino in a gauge where at the studied spacetime point $\nbrk{e^a_\mu=\delta^a_\mu}$ and the other connections vanish. Consider a small vicinity of this point inside which the terms nonlinear in connections are negligible. In that vicinity, expand the gravitino field~$\chi_i^\alpha$ over modes with specific spatial wavevector and helicity.
Focus on the modes with wavevector $\nbrk{{\bm k}=(0,0,k)}$. The gravitino's longitudinal component,~$\chi_3^\alpha$, is gauge. It can be adjusted arbitrarily by supersymmetry transformation\rf{delta_chi_i}, $\nbrk{\delta\chi_3=ik\xi}$.
This component decomposes into modes of helicity~$\pm 1/2$.
For the components $\chi_1^\alpha$ and~$\chi_2^\alpha$, the constraint\rf{constraint_fermionic} with~$\pi^i_\alpha$\rf{Pi_alpha_gen} becomes
\be
\frac{k}{4}\,\chi_j^\dagger[\gamma^3,\gamma^j] \psi=  
-\,\frac{k}{2}\,(\chi_1^\dagger \gamma^1+\chi_2^\dagger \gamma^2)\gamma^3 \psi=  \rho \psi\,. \quad
\lb{constraint_special_gauge}
\ee
It specifies the remaining modes of helicity~$\pm 1/2$, which are thus the constrained modes. Finally, the modes of $\chi_1^\alpha$ and~$\chi_2^\alpha$ of helicity~$\pm 3/2$ are physical fermionic degrees of freedom. Their annihilation operators for the emergent fields are represented by the basic anticommuting operators~$\chi_m$ of \sct{subsec_fermions}. 

\subsection{Fermionic transformation parameters}
\lb{subsec_ferm_params}

The fermionic transformation parameters~$\xi^\alpha(\xv)$ specify a considered symmetry transformation or a choice of gauge. Their values are not objective characteristics of the physical system. Correspondingly, they are not related to bit dimensions of the configuration space along which the emergent wave function is represented by an underlying material structure. In the explicit construction above, these parameters where auxiliary. They could either have an arbitrary material implementation that reproduces their relevant anticommutation properties, or be abstract elements of the Grassmann algebra without any material implementation. For the next \sct{sec_why_the}, it is helpful to clarify the admissible properties of the  parameters~$\xi^\alpha(\xv)$.

The transformation parameters of the standard supersymmetry anticommute with themselves and with the physical fermionic fields~$\chi^\alpha(\xv)$. The standard supersymmetry also postulates that the complex conjugates, as well as the Dirac and charge conjugates, of the fermionic transformation parameters~$\xi^\alpha$ anticommute with the original~$\xi^\alpha$. For example, for the Dirac conjugate, $\nbrk{\{\xi^\alpha(\xv),\bar\xi_\beta(\yv)\}=0}$. This is assumed in the powerful superspace formalism for constructing a supersymmetric action and calculating scattering amplitudes.

When $\xi$ and~$\bar\xi$ are defined to be independent anticommuting fields then, for example, the right-hand side of the supersymmetry commutator\rf{SUSY_com_standard} corresponds to translation in spacetime by
\be
\veps^a=\,\bar\xi_1\gamma^a\xi_2\,. 
\lb{veps_12}
\ee
For the anticommuting~$\bar\xi_1$ and~$\xi_2$, however, this spacetime displacement is a nilpotent operator, $\nbrk{(\veps^a)^2 =0}$, instead of an ordinary spacetime vector. 

A more natural option is to set~$\xi^\dagger$ in the Dirac conjugate
\be
\bar\xi=\xi^\dagger i\gamma^0\, 
\ee
to the actual Hermitian conjugate of~$\xi$ on an extended Hilbert space~$\psi(f,\nf,\nf_\xi)$, where the Hermitian product straightforwardly extends\rf{Hermitian_product_fermionic} to the added bit dimensions~$\nf_\xi$. Then let $\xi(\xv)$~be a non-dynamical Majorana spinor field. Upon proper normalization, it has the canonical anticommutator
\be
\{\xi(\xv), \bar \xi(\yv)\}=i\gamma^0\,\delta(\xv-\yv)\,.
\lb{zeta_anticom_modified}
\ee
[For the Majorana spinors, $\nbrk{\bar\xi=\xi^T C}$ where $C$~is the charge conjugation matrix. Hence, given\rf{zeta_anticom_modified}, $\nbrk{\{\xi^\alpha,\xi^\beta\}\neq 0}$. These field operators still square to zero: $\nbrk{(\xi^\alpha)^2=0}$.]
For a different normalization, e.g., $\nbrk{\xi^\alpha(\xv)\to \alpha(\xv)\,\xi^\alpha(\xv)}$, the anticommutator\rf{zeta_anticom_modified} rescales as
\be
\{\xi(\xv), \bar \xi(\yv)\}=|\alpha(\xv)|^2 i\gamma^0\,\delta(\xv-\yv)\,. 
\lb{zeta_anticom_a}
\ee
The conventional assumption $\nbrk{\{\xi, \bar\xi\}=0}$ is recovered in the limit~$\nbrk{\alpha\to 0}$.

Despite $\nbrk{\{\xi, \bar\xi\}\neq0}$, the supersymmetry algebra retains its standard form. Consider a small spatial region in which the local transformation of covariant operators can be approximated by a global transformation. Its parameters~$\xi$ and generators~$Q$, which are mutually anticommuting Majorana spinors, satisfy $\nbrk{\bar\xi Q=\bar Q\xi}$. Hence
\be
[\bar\xi_1 Q, \bar\xi_2 Q]\eqa
\bar\xi_{1\alpha} \{Q^\alpha, \bar Q_\beta\}\, \xi^\beta_2 - \bar Q_\beta \{\xi^\beta_2, \bar\xi_{1\alpha}\}\,Q^\alpha  \,.\qquad
\lb{SUSY_com_expanded}
\ee
The last term vanishes trivially for mutually anticommuting $\bar\xi_1$ and~$\xi_2$.
It also vanishes when $\xi_1$ and~$\xi_2$ are proportional to the same Majorana spinor~$\xi$ with $\nbrk{\{\xi, \bar\xi\}=|\alpha|^2 i\gamma^0}$, cf.\rf{zeta_anticom_a}, because $\nbrk{\bar Q\gamma^a Q=0}$ for the fermionic Majorana spinor~$Q$. This recovers the standard supersymmetry algebra:\footnote{ \lb{ftnote_susy_norm}
         We normalize the supersymmetry generators to
	$$
	\{Q, \bar Q\}=\gamma^a iP_a\,.
	$$   
   The historical normalization of this anticommutator contains an extra factor of~$2$ on the right-hand side. Without it, the algebraic structure and field transformations of supersymmetry are described by cleaner formulas. We can convert formulas from one normalization to the other by rescaling the supersymmetry generators and transformation parameters as: 
   $\nbrk{Q_{\rm old}=\sqrt2\,Q_{\rm new}}$ and \nbrk{\xi_{\rm old}=\xi_{\rm new}/\sqrt2}.
}
\be
[\bar\xi_1 Q, \bar\xi_2 Q] = \bar\xi_1 \{Q, \bar Q\}\, \xi_2 = \bar\xi_1\gamma^a\xi_2\, iP_a\,.
\lb{SUSY_com_standard}
\ee
We thus may employ the less abstract implementations of the supersymmetry transformation parameters~$\xi^\alpha(\xv)$ with anticommutation\rf{zeta_anticom_modified}.

\subsection{Map to the basic structure}
\lb{subsec_susy_map}

The generic static distribution~$\dist(Q)$ from \sct{sec_structure} thus gives rise to wave functions of locally supersymmetric fields as follows. Consider a locally supersymmetric action. An example is the action of supergravity with an arbitrary superpotential for matter fields. Any locally supersymmetric action is diffeomorphism-invariant because the local supersymmetry group includes diffeomorphism. The action can have other local symmetries, in particular, gauge symmetries. The symmetry generators~$\cH_A$ for this action equal\rf{H_A_explicit}.

\begin{figure}[t]
\centering
\includegraphics[width=0.45\textwidth]{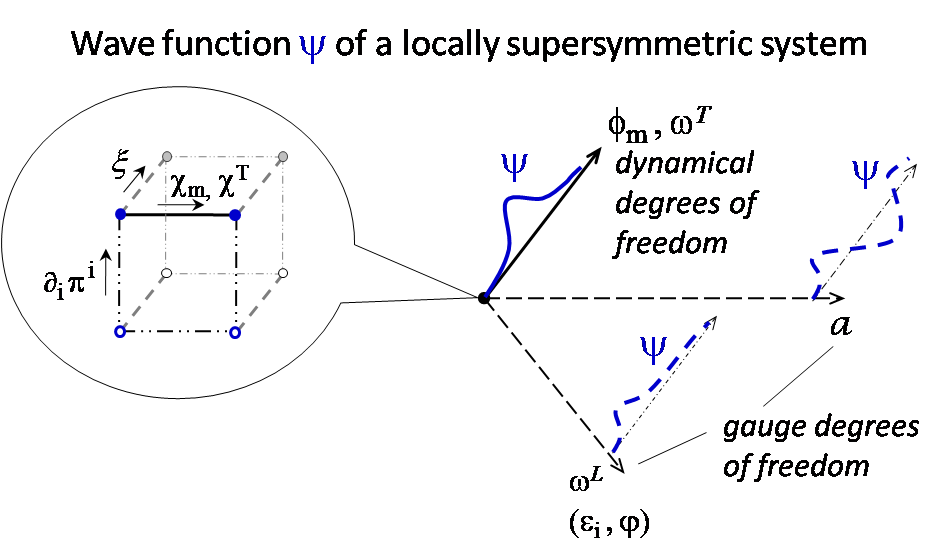}       
\caption{
The complete emergent wave function of a field system with $\nbrk{\mathcal{N}=1}$ local supersymmetry. The thick solid lines depict its configuration-space dimensions spanned by the amplitudes of the fields' \emph{dynamical} (independent, physical) modes. They are represented by the continuous (bosonic) and bit (fermionic) dimensions of a presentation~$\psi(q,\nf)$ of the smoothed generic basic distribution~$\dist(Q)$, \fig{fig_outline}. The dimensions for the fields' \emph{gauge} modes, shown with dashed lines, are composed by stacking symmetry-equivalent presentations of the wave function of the dynamical modes. 
}
\label{fig_psi_susy}
\end{figure}

Let~$\psi(q,\nf)$, with many independent continuous arguments~$\nbrk{q=\{q^n\}}$ and bit arguments $\nbrk{\nf=\{\nf_m\}}$, be an approximate, smooth in~$q$ description of a distribution of properties for the large discrete set of the basic entities~$\{E_a\}$ of \sct{subsec_smooth_pres}. An example of specifying~$\psi(q,\nf)$ was presented in \sct{subsec_fermions}. This~$\psi(q,\nf)$ can be regarded as a wave function of the dynamical modes of emergent bosonic and fermionic fields with the considered action in a fixed gauge. The amplitudes of the dynamical modes of the bosonic fields are identified with the continuous arguments~$q^n$ [e.g.,\rf{phi_standing_harmonics}]. The annihilation operators of the dynamical modes of the fermions are identified with the operators that lower independent~$\nf_m$ [e.g.,\rf{chi_x_construct}].
We depict these, physical dimensions of the configuration space of the emergent system in \fig{fig_psi_susy} with thick solid lines.

The transformation\rfs{bosonic_ext_psi}{bosonic_extension} extends this wave function to other gauges along the bosonic gauge degrees of freedom~$\var^A(\xv)$. The fields' gauge components $\var^A(\xv)$~are variation parameters of this transformation. Thus the wave function of all the bosonic (matter and connection) fields $\nbrk{f=(\phi,\omega_i)}$ consists of the original presentation~$\psi(q,\nf)$ for the chosen gauge and its alternative, equivalent presentations for the other gauges.

The transformation\rf{psi_n_L_constr} extends the wave function along the fermionic \emph{constrained} modes as described in \sct{subsec_ferm_constr}.
This specifies the wave function for all the modes of the field theory except for the fermionic gauge modes, $\nbrk{\delta\chi^\alpha_i= (\D_i \xi)^\alpha}$.

Since supersymmetry transformation\rf{SUSY_psi_sym} does not change the wave function, the fermionic gauge degrees of freedom,~$\xi^\alpha(\xv)$, can be represented by any fermionic variables with the properties described in the previous subsection~\ref{subsec_ferm_params}. They may be abstract or be operators that act on a material structure. In particular, they may be matched to the presentation provided by the emergent physical objects in the gauge specified by these~$\xi^\alpha$.  (The emergent physical objects and background fields can transform in other gauges into something equivalent yet rather dissimilar to their familiar physical picture.)  \fig{fig_psi_susy} depicts the resulting full emergent wave function of locally supersymmetric fields.

\section{Emergence of dynamical laws}
\label{sec_why_the}

\Sct{sec_first_observation} proved that if the quantum superposition principle held absolutely then at least some experimentally measurable constant numerical parameters of the physical Hamiltonian would instead vary randomly in time and space. We are ready to address how, despite the arguments of \sct{sec_first_observation}, physical evolution can proceed by unchanged dynamical laws. 
The current section will demonstrate that the generically emergent locally supersymmetric fields evolve by fixed laws.\footnote{
   Spontaneous breaking of local symmetry is, of course, possible. Yet, despite the name, it does not destroy the local symmetry of the dynamics or of the wave function. In particular, spontaneous breaking of local symmetry does not change the number of independent degrees of freedom or the field constraints.
}   
The laws differ among various locally supersymmetric systems but do not change in any of them.  

\subsection{Preservation of inherent local symmetries}
\label{subsec_locality}

Let~$\psi(f,\nf)$ be an emergent wave function of bosonic and fermionic fields $\nbrk{F(\xv)=(f(\xv),\chi(\xv))}$. As in \sct{sec_fermions}, its bit arguments  $\nbrk{\nf=(\nf_1,\nf_2,\dots)}$ are the occupation numbers\rf{nf_def} of modes of the fermionic fields~$\chi(\xv)$. 

We call a transformation $\nbrk{F(\xv)\to F'(\xv)}$ of the fields at a fixed current time an \emph{inherent symmetry} if it corresponds to a symmetry of the current-time \emph{wave function}. A symmetric wave function of bosonic fields returns the same value for symmetry-equivalent field configurations. Hence, the wave function depends only on the classes of symmetry-equivalent field configurations. 
An example is gauge symmetry, \sct{subsec_abelian} or \apx~\ref{apx_gauge}. The spatial diffeomorphism subgroup of diffeomorphism symmetry is another inherent symmetry, Ref.\ct{DeWitt} and \sct{subsec_emergent_gravity}. 
More generally, by \sct{subsec_wf_constant}, any restriction of local symmetry transformations to the fields at the current time that forms a group of their transformations is an inherent symmetry. 

Let an emergent wave function~$\psi(f,\nf)$ of a field system have inherent local symmetry.
This wave function is fundamentally represented by objectively existing, satisfying the condition $\nbrk{\langle i|i\rangle \gg \delta\chi^2_{\rm min}}$ [cf.\rf{norm_unit}], Everett's branches of presentations of the smoothed generic distribution~$\dist(Q)$, eq.\rf{psi_general}. We now show that the arguments of \sct{sec_first_observation}, to prove that the strict superposition principle results in noticeable variation of physical laws, are inapplicable to the emergent system with inherent local symmetry. (This does not contradict the conclusion of \sct{sec_first_observation} because the superposition principle holds for the emergent systems only up to certain limits.) 

Consider an infinitesimal unitary transformation $\nbrk{\psi(f,\nf)\to\psi'(f,\nf)}$ that destroys the inherent symmetry of~$\psi(f,\nf)$. The resulting wave function~$\psi'(f,\nf)$ needs a material representation~$\psi'(q)$ with substantially increased dimensionality. A new dimension should be added to the system's configuration space for every gauge and constrained mode of the destroyed inherent symmetry. 

These can be continuous bosonic or bit fermionic dimensions. In either case, the configuration space for the dynamical degrees of freedom of~$\psi'(f,\nf)$ has many more dimensions. In it, the same underlying finite discrete distribution (e.g., same number of dots in \fig{fig_A}) cannot represent the transformed smooth wave function with the original accuracy. This degradation of the accuracy of representing the wave function would push the majority of its Everett branches below their objective existence threshold\rf{norm_unit}. Consequently this would destroy most or all of the material presentations of the current physical state (see \sct{sec_probabilities} for specifics).

Thus evolution transformation of elementary physical fields that breaks their inherent local symmetry is admissible for abstract continuous wave functions but is impossible for most of its material implementations that emerge from a finite discrete structure.

\subsection{Mechanism of Hamiltonian fixing}
\label{subsec_fixing}

By the previous subsection, evolution of the generically emergent quantum fields preserves their gauge and spatial diffeomorphism symmetries on the current-time hypersurface. This however does not confine their dynamics to definite physical laws. For example, the local evolution generator $\cH_\lapse$~may equal a sum of gauge and 3-diffeomorphism covariant operators that are multiplied by time-dependent coefficients (couplings).  Adiabatic change of the couplings would not disrupt the physical objects or excite field modes on length scales smaller than the non-adiabaticity scale. This scale, in turn, can be set to any macroscopic size. Hence the system's internal observers should then generically experience physical evolution where couplings change randomly over time.

Local supersymmetry, on the contrary, fixes the evolution of an emergent system. For a proof, remember that the wave function of a locally supersymmetric system is invariant under a local supersymmetry transformation:
\be
\delta\psi = -i\lf(\!\int d^3x~\xi^\alpha \cH_\alpha\rt) \psi = 0\,
\lb{delta_psi_susy}
\ee
by\rf{secondary_constraint_A}. A commutator~$[\delta_1,\delta_2]$ of two supersymmetry transformations with infinitesimal fermionic transformation parameters~$\xi_1$ and~$\xi_2$ is an even  (bosonic) element of the graded supersymmetry algebra. Hence it is a linear combination of the bosonic generators~$\cH_A$ of the symmetry group:
\be
[\delta_1,\delta_2] = -i\int d^3x~\veps^A(\xi_1,\xi_2)\,\cH_A\,.
\lb{delta_com_lin_comb}
\ee 
By the supersymmetry algebra\rf{SUSY_com_standard}, this linear combination for typical~$\xi_1$ and~$\xi_2$ includes the generators~$\cH_\lapse(\xv)$ of local temporal translations, entering\rf{delta_com_lin_comb} with nonzero coefficients.  By\rf{delta_psi_susy}, the entire operator\rf{delta_com_lin_comb} annihilates the wave function: $\nbrk{[\delta_1,\delta_2]\psi = 0}$.
All the bosonic~$\nbrk{\cH_A\neq\cH_\lapse}$, generating either spatial diffeomorphism or gauge transformations, on the right-hand side of\rf{delta_com_lin_comb} also annihilate the wave function. Consequently, an emergent locally supersymmetric wave function satisfies the Hamiltonian constraint
\be
\cH_\lapse\psi = 0\,
\lb{Hamiltonian_constr}
\ee
automatically.

Thus for locally supersymmetric fields the Hamiltonian constraint is a consequence of the inherent symmetries of the wave function, i.e., is its inherent property. For intuitive understanding, recall that, as seen from the supersymmetry anticommutator\rf{susy_anticom_global}, a supersymmetry transformation is a ``half-step" of a displacement in spacetime\ct{Smolin_book_Sec_6}. The constancy of the wave function at the half-step\rf{delta_psi_susy} ensures its constancy at the full step of the temporal displacement.

Let us determine how a current state of a locally supersymmetric system encodes its future evolution. Let us modify the supersymmetry generators~$\cH_\alpha$ infinitesimally. By the reasons described in the previous subsection, we should avoid introducing new independent degrees of freedom. Let us therefore simultaneously modify the bosonic symmetry generators~$\cH_A$, including~$\cH_\lapse$, so that the modified generators continue to form a complete basis of a closed algebra with unchanged dimensionality. 

Let~$\psi'$ be the wave function that is annihilated by the modified generators and that coincides with the original wave function~$\psi$ in the top component~$Y$ of\rf{psi_n_L}. By\rf{psi_n_L_constr}, $\psi'$~should differ from~$\psi$ in the bottom component~$Y_0$. 
The bottom component~$Y_0$ of the emergent wave function\rf{psi_n_L} has no independent material representation. It is instead a transformation\rf{psi_n_L_constr} of the top entry~$Y$. Therefore, although for a given~$Y$ in\rf{psi_n_L} the bottom component~$Y_0$ is rigidly related to the supersymmetry generators and the respective Hamiltonian\rf{H_from_HA}, $Y_0$~does not exist on its own to specify particular dynamics. Rather, it is a reflection of the dynamics.
 
Nonetheless, physical objects and other physical constituents of a physical state do exist materially, even if only as features of a presentation of the generic basic structure. They are inseparable from their specific wave function\rf{psi_n_L}, with its top and bottom components combined. Hence the physical objects themselves ingrain the information about the unique dynamics of their locally supersymmetric wave function.

For deeper understanding of how they do, transform the emergent wave function\rf{psi_n_L} with a Hamiltonian~$H'$ that is \emph{not} a linear combination\rf{H_from_HA} of the original generators~$\cH_A$. If the alternative Hamiltonian~$H'$ does not describe a locally supersymmetric system then transition to evolution by~$H'$ would require discontinuous and substantial increase of the number of the independent physical degrees of freedom. \Sct{subsec_locality} showed that it would destroy most implementations of the typical emergent physical states with the original symmetry.

The alternative Hamiltonian~$H'$ could instead be composed of local generators~$\{\cH'_A\}$ that form a basis of an algebra whose dimensionality equals the dimensionality of the original symmetry algebra. An emergent wave function~$\psi'$ constructed with these~$\{\cH'_A\}$ represents a system that evolves by different (primed) physical laws.  The memory of an intelligent observer in the emergent world with the wave function~$\psi'$, however, reflects past events in the system whose dynamics has been generated by~$\nbrk{H'\neq H}$. The other characteristics of its any subsystem also causally continue past states of the system with the inherent primed physical laws. The observers in the primed system therefore cannot remember or otherwise experience consequences of the events in the original system, and vice versa. Nothing would change in either of the two systems if the other one were somehow erased from physical existence. The inhabitants and environment in either of the systems thus evolve by the fixed dynamics of their own system. They are unaffected by whatever happens in the other system.  

Let us highlight qualitative difference between the described coexistence of independent systems with different yet specific dynamical laws from the situation of \sct{sec_first_observation}, where the dynamics could vary randomly during evolution. For the fixed-dynamics emergent systems considered here, a quantum state evolves by its specific inherent physical laws that are unchanged throughout its evolution. In particular, its internal observer sees that the current dynamical laws are the same as the laws that described the past events as retained in the observer's memory and other historical records. In contrast, for the randomly varying dynamics found in \sct{sec_first_observation} a~given quantum state would spawn many different paths of evolution. A future state along the typical path would reflect consequences, including observers' recollection, of the dynamics having been different in the past.

\section{Planck-scale physics}
\label{sec_Planck_scale}

A major effort of high-energy theoretical physics of the last half-century was attempts to unify the Standard Model of elementary particles and the Einstein gravity by a theory whose dynamics is regular in the ultraviolet limit. We will however show that the simplest realistic emergent quantum fields evolve unambiguously (for certain initial conditions) despite ultraviolet \emph{incompleteness} of their dynamics. At the Planck scale the emergent fields are not superseded by more fundamental dynamical degrees of freedom: strings, loops, etc. All the physical degrees of freedom of the emergent systems are sub-Planckian. Moreover, as described below, singularity of dynamics for all \emph{but very special initial conditions} lets us resolve several outstanding problems of particle physics and cosmology.

To work with a specific theory, we assume that no new gravitational degrees of freedom, besides those of the metric tensor, become dynamically relevant at high but still sub-Planckian energy. 
We suppose that general relativity with the Hilbert gravitational action\rf{action_Hilbert}, complemented by its gravitino counterpart, applies almost until the Planck scale. We will see in part~\ref{subsec_Planck_scale} that this assumption is reasonable and sufficiently generic.

\subsection{Are trans-Planckian modes fine-tuned?}
\label{subsec_inflating_worlds}
 
Abundant empirical evidence indicates that our universe underwent cosmological inflation. The paradigm of inflation agrees with\ct{Guth_inflation} the observed nearly homogeneous, spatially flat, low-entropy cosmological initial conditions. The simplest---single-field, slow-roll---models of inflation were also impressively successful in \emph{predicting} a variety of cosmological properties of the universe that since have been confirmed observationally. This includes ``adiabaticity" of the primordial inhomogeneities, their approximate Gaussianity, and the small and almost scale-independent tilt of the primordial power spectral index~$\nbrk{n_s-1}$\ct{Boomerang01,WMAP_12Final,Planck_15_Constraints_on_inflation}.

Yet various authors\ct{Penrose_Difficulties_88,Hollands_Wald_02,Steinhardt_infl_Sc_Am} have stressed that theoretical justification for inflation, despite its undeniable empirical success, is by no means as solid as originally believed. For example, any light field increases energy density~$\varepsilon$ of radiation in the early universe by
\be
\Delta\varepsilon_{\rm rad} = \int\!\dbar^3k\,k\,n(k)\sim \bar n\, m_P^4 \,,
\lb{inflation_rad_pressure}
\ee
where~$n(k)$ are the occupation numbers of the field's sub-Planckian modes. In the last part of\rf{inflation_rad_pressure}, $\bar n$~is the energy-weighted average occupation number of the modes, and $m_P$~is the Planck mass\rf{m_P_def}. Due to the radiation pressure $\nbrk{p_{\rm rad}=\varepsilon_{\rm rad}/3}$, the necessary for inflation condition that the total pressure satisfies $\nbrk{p\leqslant -\varepsilon/3}$ (e.g.,\ct{LiddleLyth_book}) holds only if
\be
\bar n\ll 1\,.
\lb{inflation_necessary_condition}
\ee
Thus inflation is sustainable only if the high-frequency modes, evolving on subhorizon scales adiabatically, emerge from the Planck scale essentially in the ground state. 

Why should the universe satisfy the highly special low-entropy initial conditions\rf{inflation_necessary_condition}\ct{Penrose_Difficulties_88,Hollands_Wald_02}?  Someone might respond that inflation resolves the historical problem of the seemingly fine-tuned cosmological initial conditions by starting spontaneously in an acceptably smooth region when the universe had about the Planckian energy density. However, this implicitly assumes that such a region in the earliest universe is more generic than other, extremely \emph{inhomogeneous} initial conditions that lead to anthropically suitable worlds. Various researches have highlighted\ct{Penrose_Difficulties_88,Hollands_Wald_02,Steinhardt_infl_Sc_Am} that for unitary quantum evolution under nonsingular dynamics the opposite should be expected.

Indeed, consider a smooth spatial region that is suitable for the start of inflation. In this region all the trans-Planckian modes that will evolve into the vastly---perhaps infinitely\ct{Linde_eternal}---expanded universe should be prepacked to exquisitely the right state, whose total energy and entropy do not exceed the Planck mass~$\nbrk{m_P\sim 10^{-5}}$\,g.
Under unitary quantum evolution, finding such a special state of the universe at the beginning of inflation remains as unlikely as finding its observed---low-entropy, flat, and smooth---state today. Many more configurations, highly inhomogeneous and initially extending over a large spatial domain, should be compatible with natural development of intelligent life without inflation. Any single realization of the inhomogeneous but anthropically suitable initial conditions might seem fine-tuned. Yet altogether they result in apparently more generic\ct{Penrose_Difficulties_88,Hollands_Wald_02,Steinhardt_infl_Sc_Am} anthropically suitable universes than those created by inflation.\footnote{
   These arguments also lead to the ``Boltzmann brain'' problem:
   seemingly, an overwhelmingly higher probability 
   of observing a universe whose state is random beyond the regions currently 
   occupied by localized intelligent subsystems, like us. 
   The described reasoning should therefore be incomplete or inapplicable 
   to our world. Nevertheless, the question of what resolves 
   the Boltzmann brain problem 
   and makes the inflationary initial conditions 
   preferable remains. The generically 
   emergent quantum worlds offer a resolution, presented in \sct{sec_init_conds}.
} 
From this perspective, inflation appears highly unnatural. 

And yet, unlike the other fundamental physics theories of the recent decades, the inflationary paradigm has confronted the observed reality with spectacular success. Notably, numerous independent observations of fluctuations in the cosmic microwave background and distribution of cosmic matter confirmed such inflationary expectations as equal primordial number density perturbations in all the cosmological species (``adiabaticity" of initial conditions), their approximate Gaussianity, approximate scale-invariance of the primordial power spectrum, and its small, almost scale-independent ``tilt''. It is worth mentioning that these properties prior to their observational discovery were disfavored by then the leading competitive models for seeding the cosmological structure, e.g., phase transitions or cosmic strings.

\begin{figure}[t]
\centering
\includegraphics[width=0.3\textwidth]{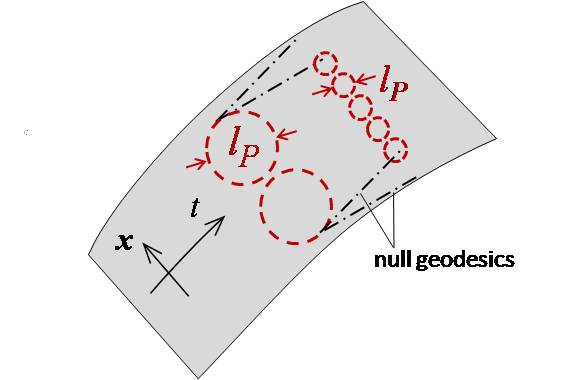}
\caption{Planck length~$\ell_P$ (set to~1 in our units) in spacetime that undergoes cosmological expansion. The figure shows that in any coordinate frame where over a superhorizon spatial region ~$\nbrk{}g_{00}<0$ (i.e., $dx^0$~is timelike) cosmological expansion necessarily creates new sub-Planckian modes.}
\label{fig_l_P}
\end{figure}

Another puzzle of inflation is why it does not ``run out'' of the degrees of freedom for the universe that has expanded tremendously and is predicted to inflate eternally\ct{Linde_eternal}. The change in the number of physical degrees of freedom in a fixed coordinate volume depends on the applied spacetime coordinates.
However, in any \emph{superhorizon} region\footnote{
  A spatial region is ``superhorizon'' when its characteristic extent satisfies $\nbrk{L\gg \Hub^{-1}}$, 
  where $\Hub$~is the cosmological Hubble expansion rate.   
  Likewise, a region is ``subhorizon''  when $\nbrk{L\ll \Hub^{-1}}$.
} 
of an expanding universe this number changes approximately equally any coordinate frame where the time coordinate~$x^0$ is timelike\ct{my_PRD}. This should be evident from \fig{fig_l_P}. Thus in any coordinates that cover superhorizon distances and provide the normal causal relation between coordinate past and future, inflation necessarily creates new sub-Planckian degrees of freedom.

These puzzles disappear in the presented physical picture. The next subsection~\ref{subsec_Planck_scale} shows that cosmological expansion of the emergent fields creates new dynamical modes around the Planck scale. The created modes initially have the ground state, regardless of how complicated the fields' ground-state wave function is. Then further \scts{sec_probabilities} and~\ref{sec_init_conds} show that physical worlds can be composed of only the emergent states with sufficiently long regular past evolution. The emergent systems with inflationary past satisfy this criterion, whereas typical random anthropically suitable configurations, e.g., Boltzmann brain worlds, do not.

\subsection{Emergent physics near the Planck scale}
\label{subsec_Planck_scale}

Consider emergent wave functions of fields with specific dynamical laws. The laws can be fixed, e.g., by local supersymmetry (\sct{sec_why_the}). Let the fields' Hamiltonian permit phenomenologically acceptable cosmological inflation. Many emergent systems with suitable (even if fine-tuned) Hamiltonians exist. 
In some of such emergent wave functions the field modes with large wavelengths evolve quasiclassically. They create a classical background~$\bar f\cl(x)$ of an inflating or post-inflationary world (\sct{subsec_emergent_classical}). 	Short-scale modes of the quantum fields are described by a reduced wave function~$\pert{\psi}[\pert{f}(\xv),t]$ (\sct{subsec_tilde_Hamiltonian}). 

We explore coupled dynamics of $\bar f\cl$ and~$\pert{\psi}$ in a region where the curvature invariants---$R$, $R_{\mu\nu\lambda\chi}R^{\mu\nu\lambda\chi}$, etc.---are sub-Planckian, i.e., their absolute values are well below $m_P^2$, $m_P^4$, etc.  For the discussed general relativistic gravitational action, this implies sub-Planckian background energy density:~$\nbrk{\bar\varepsilon\ll m_P^4}$ in the \emph{local rest frame}, defined as the local Minkowski frame where the overall momentum in the studied local volume is zero.

Whenever gravitational mass density\ct{Tolman_30}
\be
\quad \mu\equiv \veps + \sum_i p_i 
~~~(\mbox{where}~\veps\equiv-T^0_0,~p_i\equiv T^i_i )\quad
\lb{mu_def}
\ee 
integrated over proper volume of a region exceeds a critical mass for this region, in general relativity the matter in the region collapses gravitationally. Under the classical Einstein equations, the collapse leads to singularity if during the subsequent evolution  $\nbrk{\mu\geq0}$ at every point of the region\ct{Penrose_65,Hawking_Penrose_singularities}. 

Consider emergent fields whose evolution at high density up to the Planck scale is not qualitatively different from the respective evolution in classical general relativity. Specifically,  there appears no obstacle for identifying quantum field systems where for the modes of wavelengths $\nbrk{\lambda\sim \ell_P\equiv m_P^{-1}}$:
\ben
\item[a.]
The \emph{ground} state has $\nbrk{|\mu|\ll m_P^4}$, and
\item[b.]
The \emph{excited} states of these modes have $\nbrk{\mu\sim m_P^4}$ and develop a dynamical singularity over about the Planck time. 
\een

By the earlier sections, the amplitudes of the dynamical modes of the emergent fields are arguments~$q^m$ of smooth presentations~$\psi(q)$ of the generic basic distribution. Consider field systems that satisfy the above conditions a.\ and~b. For them, regular regions of a physical world can be represented only by~$\psi(q)$ that place the modes with $\nbrk{\lambda\sim \ell_P}$ to the ground state of the fields' Hamiltonian. The prospective emergent wave functions that are composed of the other presentations~$\psi(q)$ cannot evolve regularly for longer than about the Planck time. Thereafter their inherent evolution terminates in singularity. Those presentations therefore are not wave functions of physical systems.

\begin{figure}[t]
\centering
\includegraphics[width=0.45\textwidth]{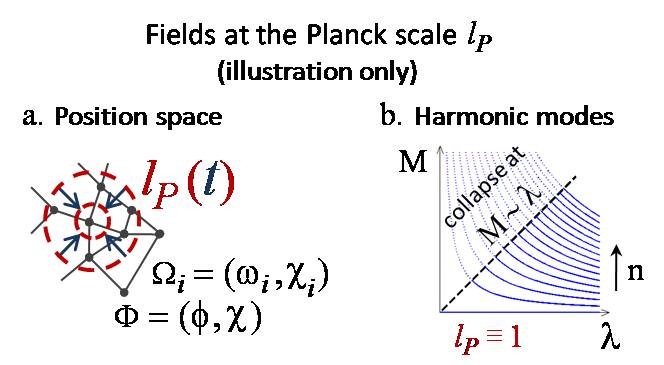}
\caption{(a).~The coordinate value of the Planck length~$\ell_P$ (of the constant proper value\rf{m_P_def}) contracts during inflation in any spacetime coordinates  where over superhorizon scales the time coordinate is timelike ($\nbrk{}g_{00}<0$), \fig{fig_l_P}. 
(b).~An excited state of fields with a finite characteristic spatial scale~$\lambda$ collapses to a black hole when its occupation number~$n$ becomes sufficiently high, making it sufficiently massive ($\nbrk{M\sim\lambda}$, in units with $\nbrk{\ell_P\equiv1}$). Then only the ground state may be stable for the fields' modes with~$\lambda\sim \ell_P$. These modes in a regular physical region are thus restricted to the ground state. As $\ell_P$~during inflation becomes much smaller than~$\lambda$, as depicted in~(a), the modes can start to evolve non-trivially.}
\label{fig_l_P_grid}
\end{figure}

\begin{figure*}[t]
\centering
\includegraphics[width=0.7\textwidth]{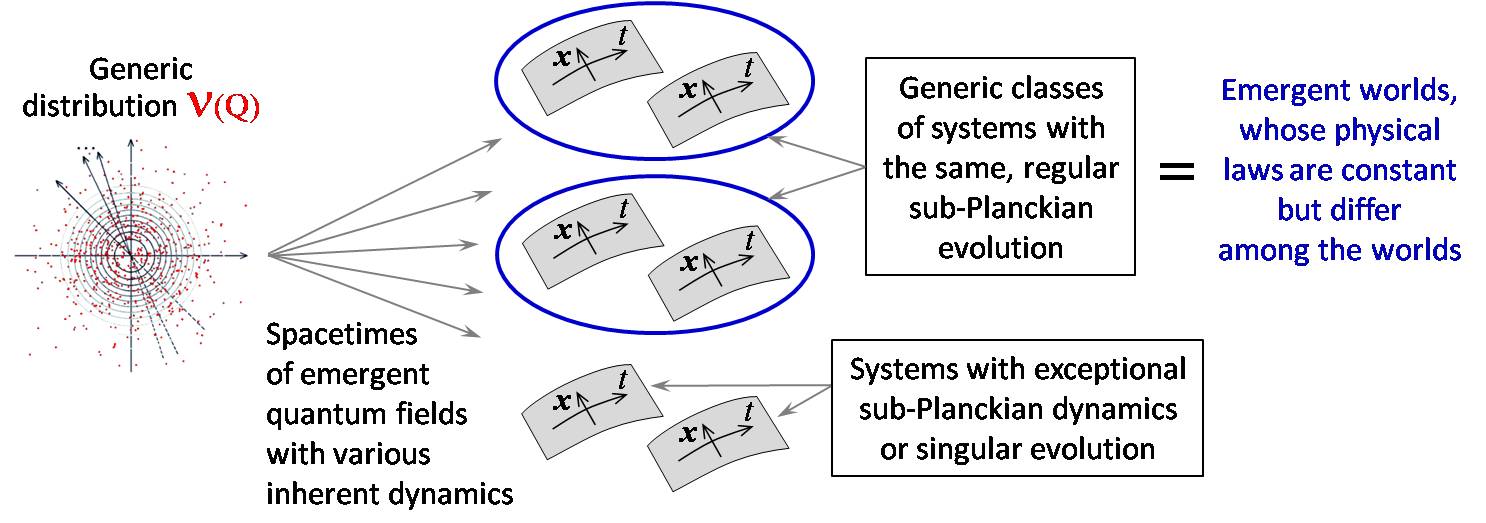}          
\caption{An emergent physical world is a \emph{generic class} of emergent quantum states. It encompasses emergent states with physically indistinguishable evolution of sub-Planckian modes. In addition, it is a concentration point in the set of all the emergent systems.}
\label{fig_classes}
\end{figure*}

During cosmological inflation the coordinate value of the Planck length~$\ell_P$ in comoving coordinates, extending regularly to superhorizon scales, contracts. The coordinate wavelength~$\lambda$ of the modes is time-independent.  The wavelength of any mode starts with $\nbrk{\lambda\sim \ell_P}$ and redshifts to $\nbrk{\lambda\gg \ell_P}$, \fig{fig_l_P_grid}.a. 

Consider inflation whose energy scale has fallen well below the Planck energy.\footnote{
  This characterized inflation when the observationally accessible scales were 
  exiting the Hubble horizon, as follows from the observational upper 
  constraints on the primordial tensor fluctuations\ct{WMAP_12Final,Planck_15_Constraints_on_inflation,Planck_15_cosmo_params}. 
} 
Then the field modes evolve from $\nbrk{\lambda\sim \ell_P}$ to larger subhorizon scales,~$\nbrk{\ell_P\lesssim \lambda \ll \Hub^{-1}}$, nearly adiabatically. As described above, a physical wave function places the modes with $\nbrk{\lambda\sim \ell_P}$ to the ground state. During their subsequent adiabatic evolution on subhorizon scales, the modes remain in the ground state, until excited by reheating or another non-adiabatic process. The inflation condition\rf{inflation_necessary_condition} thus applies automatically.
 
The dimensionality~$N$ of the fundamental distribution~$\dist(Q)$, where $\nbrk{Q=(Q^1,\dots,Q^N)}$, and correspondingly the number of the independent arguments~$q^n$ of its smooth presentations~$\psi(q)$ may be finite. 
Then during eternal inflation the amplitudes of new short-scale modes that appear from the Planck scale will eventually be represented by~$q^n$ that already participate in representing other modes on larger scales, including superhorizon ones. Inflation can therefore continue perpetually even for a finite number of the basic entities~$E_a$ (\sct{subsec_smooth_pres}) and their independent properties~$Q^n_a$. Then the basic-level information \emph{recycles} from superhorizon to the newly emerging microscopic scales.
The ``recycling'' pertains to the basic information in the discrete distribution~$\dist(Q)$ but not to the quantum information\ct{Everett} in the physical wave function. The initial smooth wave function of the new sub-Planckian modes equals the ground-state wave function of the Hamiltonian of the emergent system. It carries no information about the physical objects on larger scales.
 
In the presence of interaction, even only gravitational one, the ground-state wave function of an inflating field is not Gaussian\ct{Maldacena_non_Gaussian}. The deviation from the Gaussian wave function~$\psi\gaus(q)$ of\rf{psi_0} can be substantial when interaction is significant. This applies even to the standard gravity on scales close to Planckian. Yet the initial wave function of the modes can be any of the equivalent representations of~$\psi\gaus$, e.g., $\nbrk{\psi_0=\hat U_0\,\psi\gaus}$ where~$\hat U_0$ is a non-degenerate operator. To produce a physically viable emergent system, the operator~$\hat U_0$ should transform the Gaussian function~$\psi\gaus$ into its equivalent presentation
\be
\psi_0(q)=\int dq~ U_0(q,q')\,\psi\gaus(q')
\lb{vac_prime}
\ee 
that sets the Planck-scale modes to the ground state of the system's Hamiltonian.
To maintain the canonical form of the Hermitian product\rf{Hermitian_product_canonical}, choose an operator~$\hat U_0$ that yields the desired~$\psi_0$ and is unitary. For example, 
\be
U_0(q,q') = \sum_{n=0}^\infty \psi_n(q)\,\psi_n'^*(q')
\lb{U_0_explicit}
\ee
where $\{\psi_n\}_{n=1}^\infty$ and $\{\psi_n'\}_{n=1}^\infty$ are any functions that together with respectively $\psi_0$ and $\nbrk{\psi_0'=\psi\gaus}$ provide two alternate complete sets of basis functions orthonormal in the canonical Hermitian product\rf{Hermitian_product_canonical}.

Instead of the presentation\rf{vac_prime} of the generic distribution, we could pick a special, non-typical subset of the basic fundamental entities~$\{E_a\}$ for which the smoothed distribution of their properties would directly equal the desired~$\psi_0$. There is, however, no justification to limit the physical world to such exclusive subsets of the basic entities. In many more other sets of generic entities the same physical world emerges as the representation\rf{vac_prime}.

The requirement of definite dynamical laws, fulfilled for the emergent locally supersymmetric fields, permits a variety of these laws. The systems may differ, e.g., by superpotentials, field multiplets, and other local symmetries. Regardless of how tightly experiments and observations will constrain the low-energy effective field theory, they will generally admit many choices of Planck-scale dynamics. We argued at the beginning of this subsection and will present additional arguments at the end of the next subsection that the degrees of freedom for scales comparable to the Planck length should be in the ground state. This fixes the dependence of the wave function on these degrees of freedom for a given Hamiltonian. We can then describe evolution on larger spatial scales by an effective Hamiltonian for only sub-Planckian modes of the physical fields.   At sub-Planckian energies the same effective action describes many systems whose full actions differ around the Planck energy by non-renormalizable terms for matter and by higher mass-dimension terms for metric dynamics. 

With a detailed description to follow in \sct{sec_probabilities}, we expect that the low-energy physical world is represented \emph{simultaneously} by all the observationally indistinguishable emergent systems. This justifies considering only renormalizable or, for the metric dynamics, the closest to renormalizable terms in the action up to the Planck scale. The respective actions specify \emph{concentration points in the set of all the emergent systems}. The concentration points of emergent field systems that evolve by specific physical laws are various emergent physical worlds, \fig{fig_classes}. The state and dynamics of \emph{trans-Planckian modes} (of wavelength $\nbrk{\lambda\lesssim \ell_P}$) in these emergent worlds are undefined. 

\subsection{Macroscopic gravitational collapse}

According to general relativity, some regions of the universe should gravitationally collapse to a singularity of classical dynamics. The collapse is unstoppable\ct{Penrose_65,Hawking_Penrose_singularities} under conditions that are commonly expected in the central parts of massive stars, galactic cores, and other cosmic overdense regions. Physical laws, to qualify for such, should unambiguously specify if not necessarily evolution of the complete interior of these regions then, at least, interaction of their boundaries with the remaining universe throughout its observed history.

Companion paper\ct{my_bh} presents detailed study of gravitational collapse and of the Hawking evaporation\ct{Hawking_radiation} of the resulting black hole for the generically emergent quantum fields. In short, evolution of the emergent quantum fields with a general relativistic gravitational action is unambiguous outside the black hole's event horizon. The physical---sub-Planckian---field modes in appropriate coordinates evolve smoothly outside the horizon through the complete evaporation of the black hole. The matter that has fallen across the horizon and most of the physical information about it forever disappear from the outside physical world\ct{my_bh}. Inside the collapsing regions evolution of the emergent fields continues unambiguously until some of the local curvature invariants approach their Planck values. Then in these regions, surrounded by the event horizon, the emergent fields and  composed of them physical objects cease to exist\ct{my_bh}.

Although the information in the collapsed matter does not return in full with the Hawking radiation, the total energy and momentum remain constant. Despite the concerns raised in\ct{Banks_Susskind_Peskin_84_EM_viol} (counterexamples to which have been long known\ct{Unruh_Wald_95,Liu_pure_into_mixed_93,Unruh_12,Nikolic_nonUn_vs_EM_15}), the violation of unitarity in black hole evaporation is compatible with energy-momentum conservation. The covariant conservation of energy and momentum is a consequence of the diffeomorphism symmetry\ct{my_bh}. The latter is an inherent property of the described emergent fields with fixed dynamics (\sct{subsec_fixing}). The reader is referred to\ct{my_bh} for details.

In the presented view, singularity of quantum field dynamics near the Planck scale is not a nuisance to avoid with a deeper dynamical theory. Rather, it is an essential property of the physical world. It provides specific initial conditions for physical evolution. 
If the dynamics were regular for all initial conditions then, as discussed in \sct{subsec_inflating_worlds}, the states with inflationary past would be a minor subset of the field states that contain intelligent life arbitrarily similar to us. 
\Sct{sec_init_conds} will show that only the emergent states with \emph{regular past} evolution can be habitable worlds. Then, as \sct{sec_init_conds} demonstrates, the inflationary initial conditions become preferable to arbitrary initial inhomogeneities that accidentally produce an intelligent observer.

\section{Probability and the Born rule}
\label{sec_probabilities}

\subsection{Local mixed states}
\label{subsec_mixed_states}

The generically emergent quantum-field worlds that resemble or, possibly, compose our observed universe are mixed quantum states. Indeed, the degrees of freedom beyond the event horizons of black holes and accelerated cosmological expansion should be integrated out. This yields the physically relevant density matrix of the causally accessible universe. 

\Sct{subsec_norm} demonstrated that an decoherent branch~$\psi_i$ of an emergent wave function~$\psi$ is an objectively existing entity only when the squared norm $\nbrk{\langle\psi_i|\psi_i\rangle}$ of the branch substantially exceeds a positive threshold~$\delta\chi^2_{\rm min}$. Yet when~$\psi_i$ represents the state of the universe then its norm depends strongly on how we extend the current-time spatial hypersurface over superhorizon scales. Many proposed ``reasonable'' choices\ct{Lindes_Mezhlumian_93,Vilenkin_95,Vilenkin_98,Guth_07,DeSimone_Guth_Salem_Vilenkin_08,Linde_Noorbala_10} differ drastically from each other. 

The ambiguity of defining equal--time hypersurfaces on superhorizon scales disappears once the degrees of freedom beyond the local Hubble volume are integrated out. For the resulting mixed state of the matter and spacetime metric in  the local Hubble volume, the various physically relevant choices of equal-time hypersurfaces in the weakly perturbed Friedmann-Lemaitre-Robertson--Walker spacetime are qualitatively similar.

Akin to the decoherent branches~$\psi_i$ of an emergent wave function~$\psi$, an decoherent branch~$\rho_i$ of an emergent density matrix~$\rho$ exists objectively only when $\trace\rho_i$ exceeds the threshold~$\delta\chi^2_{\rm min}$ from\rf{norm_unit}. To see this, consider an emergent wave function~$\psi(q_{\rm loc},q_{\rm ext})$. Integrate out~$q_{\rm ext}$ to obtain the emergent density matrix
\be
\rho(q_{\rm loc},q_{\rm loc}')= \int d q_{\rm ext}\,\psi(q_{\rm loc},q_{\rm ext})\,\psi^*(q'_{\rm loc},q_{\rm ext})\,.\quad 
\lb{rho_def}
\ee
Let~$\nbrk{\rho_i(q_{\rm loc},q_{\rm loc}')}$ be one of its decoherent branches. 
Consider a basis formed by einselected  ``pointer'' states\ct{Zurek_03}. In it, $\rho$~is block-diagonal while $\rho_i$~matches to diagonal block and has zero elements outside it\ct{Zurek_03}. Let in the present context $\delta\Psi$~in\rf{delta_chi2} correspond [via\rf{Psi_i_def}] to the part~$\psi_i$ of the emergent wave function that composes via\rf{rho_def} the mixed state~$\rho_i$. This part consists of the $i$'s components of the global wave function~$\psi$ in the specified basis. The corresponding right-hand side of\rf{delta_chi2}  equals~$\trace\rho_i$.  
Thus, by\rf{delta_chi2}, the branch~$\rho_i(q_{\rm loc},q_{\rm loc}')$ stands at a statistically significant level above the intrinsic uncertainty in the emergent~$\rho(q_{\rm loc},q_{\rm loc}')$  if and only if
\be
\trace\rho_i=\delta\chi^2(\delta\Psi)\gg \delta\chi^2_{\rm min}\,.
\lb{w_min_def} 
\ee

Evolution of the global wave function~$\psi(q_{\rm loc},q_{\rm ext})$ outside the causally accessible spatial region intermixes the variables  $\nbrk{q_{\rm ext}=(q_{\rm ext}^1,q_{\rm ext}^2,\dots)}$ without affecting the local density matrix~$\nbrk{\rho(q_{\rm loc},q_{\rm loc}')}$ of\rf{rho_def}. Hence the local density matrix~$\rho$ cannot split into decoherent branches through evolution of the degrees of freedom~$q_{\rm ext}$. Since branching of the global~$\psi$ due to evolution of the fields beyond the cosmological and black hole event horizons does not split~$\rho$, it cannot reduce the squared norm of an emergent local mixed state below the threshold~$\delta\chi^2_{\rm min}$ of its objective existence.

\subsection{Physical world as an ensemble of branches}
\label{subsec_Born_rule}

Consider an emergent physical universe similar to and possibly including our one. It contains many intercoupled ``environmental'' degrees of freedom. Let its internal observer perform a multiple-outcome quantum experiment, e.g., the Stern-Gerlach, double-slit, or Schrodinger cat experiment. The experimental object, apparatus, experimentalist, local environment, and other objects and intelligent observers that are causally accessible to the experimentalist are described by a density matrix\rf{rho_def}. It arises by integrating out the inaccessible degrees of freedom~$q_{\rm ext}$, e.g., those beyond the cosmological and black hole event horizons.
After the experiment, the density matrix splits into a sum of rapidly decohering terms:
\be
\rho= \sum_i\rho_{i}\,.
\lb{split_rho}
\ee
This equation displays no cross terms $\int d q_{\rm ext}\,\psi_i^{}\,\psi^*_j$ with $\nbrk{i\neq j}$ in\rf{rho_def} because they vanish upon decoherence of the macroscopically distinct outcomes $i$~and~$j$\ct{Zurek_03}. 

Let
\be
\W= \{\rho_r(q_{\rm loc},q_{\rm loc}')\}\,
\lb{W_def}
\ee
be the set of the macroscopically similar Everett's branches of the emergent density matrices for the discussed causally accessible region of the universe. Typically, most of the other decoherent branches of multi-branch emergent quantum systems are physical configurations dissimilar to the considered one.
To the author's knowledge, the first description of a physical world as an equilibrium ensemble of local mixed states that are decoherent branches of a global state (of an eternally inflating universe) appeared in Ref.\ct{Starob_branches_86}. An important distinction from\ct{Starob_branches_86} is that the ensemble~$\W$ of\rf{W_def} contains branches of \emph{many} global emergent systems. As seen below, globally these systems can substantially differ from each other.
 
We adopt the following terminology: 
\begin{itemize}
\item
 ``\emph{Emergent quantum system}'' is the entire global emergent wave function. 
\item
 ``\emph{Emergent quantum state}'' is a decoherent branch~$\rho_r$ of the density matrix for the mixed state of the spatial region that is causally accessible to a given internal observer.  
\item
  ``\emph{Emergent physical world}'' is an ensemble~$\W$ of macroscopically similar emergent quantum states~$\rho_r$. These states~$\rho_r$ will  be called ``\emph{realizations}'' of the world~$\W$.
\end{itemize}

The realizations (density matrices)~$\rho_r$ from an ensemble~$\W$ represent physically similar local worlds. Hence $\rho_r$~have similar functional form. However, they generally differ substantially in their squared norm (to be called \emph{weight})
\be
w_r \equiv  \int d q_{\rm loc}\ \rho_r(q_{\rm loc},q_{\rm loc})\equiv \trace\rho_r\,.
\lb{w_r_def}
\ee

After the experiment with several decoherent outcomes~$i$, the density matrix~$\rho_r$ of every realization of a world~$\W$ splits similarly to\rf{split_rho} into the sum of the density matrices for the branches~$i$:
\be
\rho_r= \sum_i\rho_{ir}\,.
\lb{split_rho_r}
\ee
For the outcomes' weights $\nbrk{w_{ir}=\trace \rho_{ir}}$,
eq.\rf{split_rho_r} yields
\be
w_r= \sum_i w_{ir}\,.
\lb{sum_w_ir}
\ee

Let the design of the experiment set unambiguously at its beginning the relative wave function~$\psi_{\rm obj}(q_{\rm obj}^{})$ of the probed object.  The density matrices of the realizations of the respective world have the factorizable form
\be
\rho_r^{}(q_{\rm loc},q_{\rm loc}') = \psi_{\rm obj}^{}(q_{\rm obj}^{})\,\psi_{\rm obj^{}}^*(q'_{\rm obj})\,\rho_r' (q_{\rm rest}^{},q'_{\rm rest}) \,.
\nn
\ee
Here, $\nbrk{q_{\rm loc}=(q_{\rm obj},q_{\rm rest})}$ and~$\psi_{\rm obj}$ is normalized by $\nbrk{\int dq_{\rm obj}\,|\psi_{\rm obj}(q_{\rm obj})|^2=1}$.  Under the specified conditions, the initial wave function of the probed object~$\psi_{\rm obj}$ for these realizations~$\rho_r$ is the same. A ``clean'' experiment, whose results are determined only by its design and the initial state of the experimental object but are unaffected by the environment, then yields the same ratios
\be
\alpha_i = w_{ir}/w_r
\lb{alpha_i_def}
\ee
in every realization~$\rho_r$. By\rf{sum_w_ir},
\be
\sum_i \alpha_i = 1\,.
\ee

Describe the ensemble\rf{W_def} for an emergent physical world~$\W$ by \emph{cumulative distribution~$F(w)$ of weights}\rf{w_r_def} of its members, the macroscopically similar realizations~$\rho_r$:
\be
F(w)\equiv \lf(\ba{c}\mbox{number of realizations}~\rho_r\\
	\mbox{with}~w_r\equiv \trace\rho_r \geqslant w\ea
	\rt).
\lb{Fw_def}
\ee
Let~$F(w)$ be the cumulative distribution for the realizations of the world at the beginning of the considered experiment. Let~$F_i(w)$ be the similarly defined cumulative distribution for the realizations with an outcome~$i$ after the branches~$\rho_{ir}$ for the various outcomes of the experiment decohere.  From\rf{alpha_i_def}, a realization of the outcome~$i$ with weight~$w$ is created by a realization of the initial configuration with weight~$w/\alpha_i$. Hence $F_i(w)$ and the initial distribution~$F(w)$ are related as
\be
F_i(w)= F\lf({w}/{\alpha_i}\rt).
\lb{Fwi_prime}
\ee

As evident from the definition of cumulative distribution\rf{Fw_def},  the \emph{total number} of the objectively existing realizations with an outcome~$i$ is
\be
N_i=F_i(w_{\rm min})\,,
\lb{N_i}
\ee 
where
\be
w_{\rm min}\equiv  \delta\chi^2_{\rm min}
\ee
is the weight below which a realization stops being an objectively existing entity, eq.\rf{w_min_def}.
The frequentist probability of the outcome~$i$ thus equals
\be
\prob_i = \frac{N_i}{\sum_i N_i}= \frac{F_i(w_{\rm min})}{\sum_i F_i(w_{\rm min})} \,.
\lb{probability}
\ee
In particular, if an outcome~$i$ is so unlikely that the respective~$N_i$ of\rf{N_i} diminishes below unity then there are no physical outcomes of that type. This distinguishes the \emph{objective physical probability}~$\prob_i$  from alternative formal assignments of probability, even if the latter are ``rational" in the sense of Refs.\ct{Deutsch_probability_decisions99,Barnum_etal_99,Saunders_02,Wallace_02}.

Return to the experiment that splits an initial state into decoherent branches~$i$ of relative weights $\alpha_i$, eq.\rf{alpha_i_def}. The probability\rf{probability} of its outcome~$i$ by\rf{Fwi_prime} becomes
\be
\prob_i = \frac{F(w_{\rm min}/\alpha_i)}{\sum_i F(w_{\rm min}/\alpha_i)}\,.
\lb{prob_i_gen}
\ee
A power-law cumulative distribution
\be
F(w)= \frac{A}{w^p}
\ee
[with $\nbrk{p\geqslant0}$ so that $\nbrk{dF/dw\leqslant0}$ by $F(w)$~definition\rf{Fw_def}] yields
\be
\prob_i = \frac{\alpha_i^p}{\sum_i \alpha_i^p} \,.
\lb{prob_power_law}
\ee
Given $\alpha_i$ definition\rf{alpha_i_def},
the Born rule requires $\nbrk{\prob_i = \alpha_i}$. Thus\rf{prob_power_law} recovers the Born rule only when~$\nbrk{p=1}$. Eq.\rf{prob_i_gen} shows similarly that the Born rule fails for any cumulative distribution~$F(w)$ that is not a power law.

\subsection{Deriving the Born rule for typical branches}
\label{subsec_Born_rule1}

We now prove that the Born rule arises under the following generic conditions, which produce the required power-law form of~$F(w)$ with $\nbrk{p=1}$. Let an ensemble~$\W(t)$ of emergent mixed states~$\{\rho_r\}$, each being a decoherent branch of an emergent density matrix, represent a local universe that qualitatively resembles our universe at a cosmological time~$t$. This time may, for example, be counted from the end of inflation or be defined by the temperature of the cosmic microwave background. Unlike the previous subsection, allow the states~$\{\rho_r\}$ in the ensemble~$\W(t)$ differ macroscopically from each other. It is sufficient that they share certain qualitative features and have the specified value of the time~$t$.

\begin{figure}[t]
\centering
\includegraphics[width=0.25\textwidth]{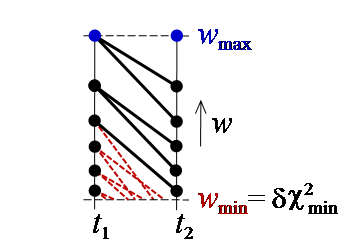}
\caption{The weights (squared norms)~$\nbrk{w_r\equiv\trace\rho_r}$ of the realizations of an evolving emergent world. Its realizations are the decoherent branches of the emergent density matrices that represent the local universe. They form an ensemble~$\nbrk{\W(t)\equiv\{\rho_r(t)\}}$. The picture describes two nearby moments~$t_1$ and~$t_2$ of equilibrium evolution of the ensemble~$\W(t)$.}
\label{fig_F_evol}
\end{figure}

During evolution of the states~$\rho_r$ in their internal cosmological time~$t$, their Everett's splitting can increase their total number by creating new decoherent branches. The splitting can also decrease their number by some of the descendent branches thinning below the threshold of their objective existence\rf{norm_unit}. \Fig{fig_F_evol} illustrates these competing processes by respectively solid black and dashed red lines.

The weight of any followed realization~$\rho_r$ of a physical world~$\W(t)$ monotonically decreases in~$t$. On the other hand, $\W(t)$~at any~$t$  contains realizations with the largest weight,~$w_{\rm max}$, possible~for the generic distribution~$\dist(Q)$. For example, they include pure states whose entire emergent wave function is a single Everett's branch. They are produced as described by  eq.\rf{vac_prime}. 
These realizations typically do not have an extended periods of regular past evolution. Yet for the cosmological time~$t$ they exist; therefore, they also contribute to~$\W(t)$.
\Fig{fig_F_evol} depicts these realizations as the uppermost blue dots.

In many respects an evolving emergent physical world~$\W(t)$ is analogous to a thermodynamic system. Both are ensembles of microscopically different realizations with common values of relatively few macroscopic parameters. A realization~$\rho_r$ of an emergent world is analogous to a microscopically detailed realization of a thermodynamic system. The macroscopic characteristics of the emergent world, including its cosmological time~$t$, are analogous to the macroscopic thermodynamic parameters: temperature, density, etc. A distribution of the microstates of a thermodynamic system can be compared to the \emph{weight distribution density}~$f(w)$ of the realizations of the emergent world. This distribution density is related to the cumulative distribution~$F(w)$ of\rf{Fw_def} as
\be
f(w)\equiv \frac{dN}{dw}=-\frac{dF}{dw}\,.
\lb{fw_def}
\ee

The historically successful Copernican principle suggests that the observed universe is one of the emergent worlds~$\{\W(t)\}$ that are typical among the anthropically suitable worlds. The typical worlds may still vary substantially in their macroscopic parameters. Next we obtain the Born rule of quantum mechanics for the typical worlds where the majority of the realizations have extended regular past. 

Let $t$ and~$t'$ be the values of the local cosmological time at respectively the beginning and the end of the considered experiment with several alternate outcomes. Let the interval $\nbrk{[t,t']}$ be minute relative to the temporal scales of qualitative global changes in the causally accessible universe. This is true for typical natural microscopic processes, e.g., scattering or decays of elementary particles. This is applicable to a laboratory experiment, testing the Born rule, \emph{only if} the observed experimental outcome does not place the experimentalist in a non-generic Everett's branch. When it does, the observer's world may change drastically during the experiment. We will explore that possibility in the next subsection.

Consider an emergent world that is a \emph{typical} ensemble of regularly evolving emergent states.
The mentioned thermodynamic analogy suggests that its weight distribution~$F(w)$ has settled to an equilibrium form. Then in the specified temporal interval $\nbrk{[t,t']}$ the competing processes of creation and destruction of the world's realizations, visualized in \fig{fig_F_evol}, balance each other. Then the resulting equilibrium weight distribution at~$t'$, after the experiment, equals to the distribution at~$t$, before it.

Although the experiment's outcomes~$i$ differ for the experimentalist, let the respective realizations of the local universe be qualitatively similar. Let them contain at many places similar experimentalists who find any of the same plausible outcomes~$\{i\}$. This allows us to place the branches with all the outcomes to the same ensemble~$\W(t)$ for a single typical world.

Under the described conditions, the equilibrium distribution~$F(w)$ should not change by the Everett splitting caused by the experiment. The weight distribution for the ensemble of the typical realizations at the time~$t'$ equals\footnote{
  Possible untypical branches~$j$ with negligible 
  weight,~$\nbrk{\sum_{j}\alpha_{j}\ll 1}$, 
  contribute negligibly to the sum~$\nbrk{\sum_i F_i(w)}$ 
  over all the created branches. Indeed,
  by\rf{Fwi_prime}, $\nbrk{F_j(w)= F\lf({w}/{\alpha_j}\rt)}$,
  becoming negligible for $\nbrk{\alpha_j\to0}$. This lets us
  replace the sum in\rf{Fw_equilibrium} over the typical 
  branches by the sum over all the branches.
} 
$\nbrk{\sum_i F_i(w)}$. Thus the requirement that the distribution remains unchanged reads:
\be
F(w)= \sum_i F_i(w)\,.
\lb{Fw_equilibrium}
\ee
Then by\rf{Fwi_prime} 
\be
F(w)= \sum_i F\lf({w}/{\alpha_i}\rt)\,.
\lb{Fw_equilibrium1}
\ee
For an experiment with two alternate outcomes, of relative weights $\nbrk{\alpha_1\equiv\alpha}$ and $\nbrk{\alpha_2=1-\alpha}$, condition\rf{Fw_equilibrium} becomes
\be
F(w)= F\lf(\fr{w}{\alpha}\rt) + F\lf(\fr{w}{1-\alpha}\rt)\,.
\lb{F_condition}
\ee

This equation for the equilibrium distribution~$F(w)$ applies to many quantum processes with various values of~$\alpha$. It therefore should hold identically in~$\alpha$. This lets us differentiate its both sides over~$\alpha$ to obtain
\be
\fr1{\alpha^2}\,f\lf(\fr{w}{\alpha}\rt) = \fr1{(1-\alpha)^2}\,f\lf(\fr{w}{(1-\alpha)^2}\rt),
\lb{f_condition}
\ee
where~$f(w)$ is the weight distribution density\rf{fw_def}.
 The identity\rf{f_condition} gives $\nbrk{f=A/w^2}$, where $\nbrk{A =\const}$. Integrating this result and fixing the integration constant by\rf{F_condition}, we determine the equilibrium cumulative distribution:
\be
F(w)= \frac{A}{w}\,.
\lb{F_dist_p1}
\ee

We noted after eq.\rf{prob_power_law} that the distribution\rf{F_dist_p1} and only this distribution yields the Born rule. We have thus derived the Born rule from the first principles for a typical ensemble of emergent quantum states most of which have extended regular past. On the other hand, eq.\rf{prob_power_law} shows that the Born rule fails for a non-typical, out-of-equilibrium emergent world whose weight distribution differs from\rf{F_dist_p1}.  The inapplicability of the Born rule in the general situation does not contradict Gleason's theorem\ct{Gleason_theorem} for non-negative measures on a Hilbert space by the reasons explained in footnote~\ref{ftn_Gleason} on p.~\pageref{ftn_Gleason}.

\subsection{Important real-world consequences}
\label{subsec_implications}

A mechanism for objective collapse of a wave function still has not been found in nature. For the described generically emergent quantum-field systems it does not exist. Without such a mechanism and barring superdeterminism (see footnote~\ref{ftn_superdet} on p.~\pageref{ftn_superdet}), the quantum state of our universe incessantly splits under numerous natural and artificial processes. The natural processes include radioactive decays, stellar nuclear fusion, high-energy collisions of particles in space, etc. A universe where inflation is ending undergoes an immense number of splits during particle creation in reheating. The state of a universe that inflates also splits continuously into branches with different classical values of the field modes that become superhorizon\ct{Starob_branches_86}.

The demonstrated connection between the norm of an emergent quantum state and the probability of its observation relied on the existence of the positive least value for the norms of the objectively existing Everett branches of the emergent systems (\sct{subsec_norm}).  Then any followed physical branch of our world, continuously splitting into progressively ``thinner" branches, is bound to terminate upon its norm diminishing to\rf{norm_unit}.

Our universe has nonetheless survived a tremendous amount of splits due to natural processes, including those listed above. We thus face two crucial questions:
\ben
\item
Why has our universe so far sustained natural splitting of its state?
\item
Can the splits due to recent or future artificial processes be more dangerous than the natural ones?
\een

The previous subsection~\ref{subsec_Born_rule} partially answered the first question. Namely, an emergent world that is \emph{typical} within not too restrictive anthropic constraints is an evolving equilibrium ensemble of emergent states. In it, termination of some of the existing branches is balanced by creation of new branches. 

However, the condition $\nbrk{\langle i|i\rangle\gg\delta\chi^2_{\rm min}}$, while necessary, is \emph{not sufficient} for an Everett's branch~$i$ that is formally predicted by the Schrodinger equation to be physical.
Consider the following example. 

An arbitrary wave function~$\psi(q)$ can be formally constructed by transforming the smoothed generic distribution~$\Psi(Q)$ with\rfs{vac_prime}{U_0_explicit}. The resulting realization of $\psi(q)$ has the largest possible norm, whose square thus well exceeds~$\delta\chi^2_{\rm min}$.  Nevertheless, this~$\psi(q)$ cannot materially represent the physical objects in a region~$X$ of the configuration space~$\nbrk{\{q\}}$ where
\be
\Delta_X \Psi(Q)\equiv \int_X\! dq~\psi(q)\,B_q(Q)
\ee
is not objectively represented by the underlying discrete~$\dist(Q)$ in\rf{psi_general}. Indeed, a fixed finite number of elements of the basic discrete distribution~$\dist(Q)$ cannot objectively represent an arbitrarily complicated physical configuration. 

This may but does not have to ever interfere with the future of intelligent life in a generically emergent world. For example, the causally accessible volume of our future physical universe is already limited by the event horizon of its accelerating cosmological expansion. The respective upper bound on the complexity of the physical structures in our causal future may be restrictive enough for their objective material presentation to be available forever.

The quantum splits from recent, in cosmological timeframe, human activity should be a minute fraction of all the splits. 
Yet certain artificial splits, e.g., those caused by some high-energy experiments or quantum computing, possibly have never happened before. It is thus crucial if our branch remains sufficiently typical for the emergent world to continue its existence over additional satisfactory duration. 

One can easily devise practically realizable selection of the branches that throws the observer to untypical, non-generic branches. For example, the Schrodinger cat can be arranged to survive only under an extremely unlikely quantum fluctuation that is yet allowed by the Schrodinger equation. If we accept the Everett view of quantum evolution literally and believe that every outcome, with an arbitrarily small norm of its branch, exists then from the cat's perspective the animal would be safe. It is clearly not the case in the fundamental picture of physical evolution that is described in this paper. By it, when the norm becomes too tiny, no physical presentation of a live cat remains.

\section{Preference for inflation; no~Boltzmann brain problem}
\label{sec_init_conds}

 \Sct{sec_structure} introduced~$\Psi(Q)$ as a smooth presentation of the coarse-grained generic discrete distribution. Transformation\rf{vac_prime} converts~$\Psi(Q)$ to an arbitrary smooth wave function~$\psi(q)$ of the dynamical modes of emergent fields. By the earlier sections, the fields' gauge modes are transformations of this~$\psi(q)$. Their generators are related to the generators of the symmetry transformations of the resulting emergent field system. Superficially, we could thus identify emergent locally supersymmetric wave functions that reproduce any conceivable configuration of dynamical degrees of freedom of any locally supersymmetric fields. 

Consider various emergent wave functions constructed as described and representing a world with internal intelligent life. This construction does not require the life to develop naturally via biological evolution. Nor does it need the wave functions to evolve regularly in the past. The overwhelming majority of these configurations have never experienced cosmological inflation. The typical configurations with internal life would be suitable for~life only within a region just big enough to encompass a single self-aware being. Their environment elsewhere would have enormous entropy density. I.e., these configurations would be ``Boltzmann brain'' worlds\ct{Dyson_Kleban_Sussk_02,BB_DeSimone_etal_08,BB_Carroll_17}.
This section describes a natural mechanism that lets the generically emergent quantum fields avoid the Boltzmann brain problem.

\subsection{A physical world should be predictable}
\label{subsec_physical_def}

What, if anything, distinguishes a random continuous path in various arrangements of some elements from
a worldline of a physical system?  During observed evolution of our quantum-classical world  the same causes inevitably lead to effects in a relatively narrow range (which, of course, depends on the causes). It is easy to select continuously changing configurations, for example, cartoon frames or other subsets of a larger set, so that they follow some artificial cause-effect structure up to a given moment. Yet if nothing compels us to retain the same cause-effect structure past that moment then the future of the arranged path remains unpredictable.  

A distinctive property of a potentially inhabitable physical world is that its internal observer can predict many future events. Here, ``predicting'' refers not to calculating probabilities of future events but rather to the ability to anticipate the likely future evolution itself for at least macroscopic objects. Without it, macroscopic evolution would appear random to an observer and biological evolution would be impossible (\sct{subsubsec_basic}).

\subsection{Emergence of classicality requires extended regular past}
\lb{subsec_need_reg_past}

Let us describe how the considered emergent worlds avoid the Boltzmann brain problem. First, we argue that an emergent physical world needs an extended period of regular past evolution. 

By \sct{subsec_Born_rule1}, the Born rule is absent unless during a sufficiently long period of regular evolution the competing processes of creation and destruction of the Everett branches have established detailed equilibrium. A wave function of a locally supersymmetric system far from the equilibrium still transforms by specific laws. However, its branches with relatively low norm then remain physical. Hence then the respective paths of macroscopic evolution are not much less probable than the paths represented by branches with a large norm. Typical non-equilibrium collections of emergent states thus are not worldlines of almost deterministic evolution of macroscopic quasiclassical objects.  Their specific quantum dynamics fails to translate into roughly predictable macroscopic evolution. Without the Born rule, effectively eliminating ``exotic'' branches with very low norm, the evolving quantum states are not evolving physical worlds suitable for life (\sct{subsubsec_basic}). Rather, they constitute what has been vividly called ``multimess''\ct{multimess_Sc_Am}.  

\subsection{Singular past of random states vs inflation}

Second, let us  show that evolution paths that contain a period of cosmological inflation can be the most typical worldlines of emergent systems with the required long period of regular evolution.  

The dynamics of the studied emergent fields satisfies the discussed in \sct{subsec_Planck_scale} requirement of singular gravitational collapse of sufficiently dense regions. The known quantum field theories with locally Lorentz-invariant and unitary dynamics are $CPT$-invariant. For them, the reverse evolution of a typical emergent system backward in time qualitatively resembles the normal, future-directed evolution of a typical system with the same action. Either the future-directed or past-directed evolution of the typical emergent state quickly turns singular due to the gravitational instability. The state's evolution does not continue beyond the singularity (\sct{subsec_Planck_scale}). 

The typical time of regular evolution of a random emergent wave function with fixed dynamics may or may~not be sufficient for establishing the detailed equilibrium\rf{Fw_equilibrium}. This depends on the range~$\nbrk{[w_{\rm min},w_{\rm max}]}$ of the norms\rf{w_r_def} for the objectively existing decoherent branches of the emergent wave functions (\sct{sec_probabilities}). The range~$\nbrk{[w_{\rm min},w_{\rm max}]}$ in turn depends on the fundamental structure that gives rise to the discrete basic distribution~$\dist(Q)$.
This range should be huge for~$\dist(Q)$ that can represent the known physical universe. Then the typical time of regular evolution of a random emergent system with self-aware life may be \emph{insufficient} for achieving the detailed equilibrium\rf{Fw_equilibrium}. Assume this to be the case.

When the distribution\rf{Fw_def} of the norms of phenomenologically similar emergent states has not converged to its equilibrium form, the Born rule is absent (\sct{sec_probabilities}). The previous part~\ref{subsec_need_reg_past} showed that then the collection of the emergent states is multimess, not a potentially inhabitable physical world. 

The systems whose Lagrangian permits inflation are exceptions. Some of their emergent states begin to inflate. Eternal inflation\ct{Linde_eternal} then continues regularly indefinitely. This guarantees that eventually the detailed equilibrium\rf{Fw_equilibrium} will be established. Then the numerous Everett branches of the inflating\ct{Starob_branches_86} and post-inflationary universes will obey the Born rule (\sct{subsec_Born_rule1}) and will thus compose a physical world with roughly predictable macroscopic evolution, necessary for life (\sct{subsubsec_basic}). 

The described fundamental picture manifestly avoids the Boltzmann brain problem. Its typical worlds with roughly predictable macroscopic evolution are composed of the decoherent branches of the emergent density matrices for a universe that has underwent inflation, creating the standard nearly flat and homogeneous cosmological initial conditions.

\section{Conclusion}
\label{sec_conclusion}

The sole material structure behind the described emergent quantum-field worlds is the generic discrete distribution~$\dist(Q)$ of many real properties $\nbrk{Q\equiv (Q^1,Q^2,\dots)}$ over a large number of fixed material entities. For example, let numerous independent properties~$\lf\{Q^p\rt\}$ characterize each of many various entities~$\lf\{a\rt\}$ by real values~$Q^p_a$.  Then let $\dist(Q)$~be the distribution over the entities~$\lf\{a\rt\}$ of their generic properties $\nbrk{Q^n_a = \sum_p c^n_p\, Q^p_a}$, where $c^n_p$ are the generic real coefficients.

Some descriptions of the smoothed distribution~$\dist(Q)$, appearing continuous at imperfect resolution, compose wave functions of evolving bosonic and fermionic fields with local symmetries, including gauge and diffeomorphism symmetry. In compliance with PBR~theorem\ct{PBR_theorem}, the wave function for the \emph{dynamical modes} of the fields exists materially. Their wave function is identified with the coefficients~$\psi(q)$ in the superposition\rf{psi_general} of linearly independent smooth basis functions~$B_q(Q)$ that approximates the discrete~$\dist(Q)$. More descriptively, $\psi(q)$~are thus the amplitudes of modes that compose the smoothed~$\dist(Q)$, \fig{fig_outline}.   

The gauge and constrained degrees of freedom of the field system do not span independent dimensions of its configuration space. Correspondingly, the fields' gauge and constrained modes do not map to the arguments~$q$ of the presentations~$\psi(q)$ of the smoothed~$\dist(Q)$. The amplitudes of the \emph{gauge modes} are instead the variation parameters of the $\psi(q)$'s~transformations (reflecting other choices of the basis functions~$\{B_q(Q)\}$) from the group of the local symmetries of the emergent fields. 

The distribution~$\dist(Q)$ is static. Various temporal moments of evolution of the emergent wave functions correspond to different choices of the smooth basis functions~$B_q(Q)$, superposed with the respectively basis-dependent coefficients~$\psi(q)$ to fit~$\dist(Q)$, eq.\rf{psi_general}. The physical spacetime for the emergent fields arises by constructive interference of their quasiclassical large-scale modes, \sct{sec_physical_world} and \figs{fig_summary} or~\ref{fig_interference}. 

Discreteness of~$\dist(Q)$ is essential for the existence of specific physical laws of the fields' evolution (\sct{sec_why_the}, reviewed below). 
Discreteness of~$\dist(Q)$ also leads to unambiguous values of probability for alternate outcomes of quantum processes in the emergent worlds. Finally, it provides a mechanism for resolving two long-standing questions about cosmological initial conditions: the extremely low entropy of the initial patch that evolves via inflation into the contemporary universe and the Boltzmann brain problem.

\Sct{sec_first_observation} considered a hypothetical quantum system of any fundamental origin for which the quantum superposition principle is an absolute law. Then \sct{sec_first_observation} showed that any current state of the system should experience physically real alternate branches of its future evolution by various Hamiltonians whose decoherence-stable ``pointer'' states\ct{Zurek_81,Zurek_03} continue to evolve smoothly. An example is a smooth change of the cosmological constant or of some of the Standard Model couplings.  Besides contradicting numerous observations and experiments, this prevents evolutionary development in the system of \emph{any} internal life (\sct{subsubsec_basic}). 

For a resolution, we recalled that the wave function~$\psi[f(\xv),t]$ of a field system at a fixed time~$t$ is unchanged under any transformation of its arguments~$f(\xv)$ that is a local symmetry of the fields' action, i.e., of their \emph{dynamics}. In particular, the wave function is constant on the orbits of gauge and spatial diffeomorphism local symmetries of the action\ct{DeWitt}.
Dynamics that violates wave-function constancy on these orbits cannot itself be symmetric. It then requires a large number of new dynamical degrees of freedom. Yet for systems that emerge from a finite structure~$\dist(Q)$, the number of independent dynamical variables is capped. Then the numerous additional degrees of freedom are unavailable.

Gauge and spatial diffeomorphism symmetries do not fix physical laws yet. Alternate paths of symmetry-preserving evolution could be generated by symmetric Hamiltonians whose couplings arbitrarily change in time. 

In contrast, evolution of a given state of fields with local supersymmetry is unique. Its Hamiltonian is encoded in the inherent symmetries of the current-time wave function, \sct{subsec_fixing}.  Complementary locally supersymmetric systems emerge. Their dynamical laws differ among the systems but cannot change in any of them during its evolution. 

\emph{Quantum entanglement} over arbitrarily large distances in the emergent systems is a trivial phenomenon. Entangled characteristics of two physical objects are fundamentally different coordinates, e.g., $q^1$ and~$q^2$, in a decoherent term~$\nbrk{\psi_i(q)\equiv\psi_i(q^1,q^2,\dots)}$ of a presentation of the generic basic distribution. The variables $q^1$ and~$q^2$ are entangled when they are not statistically independent with respect to the probability distribution $\nbrk{dP=|\psi_i(q)|^2dq}$. Dynamical variables for regions that are separated by enormous, even cosmological distances are thus various characteristics of the \emph{same feature} in a presentation~$\psi(q)$ of the underlying basic distribution~$\dist(Q)$. Distance in the emergent physical spacetime is meaningful only with respect to the internal local dynamics of the emergent fields. 

\emph{Locality of dynamics} of the emergent fields originates as follows. Their dynamics is a manifestation of local supersymmetry transformations, inherent to their current-time wave function (\sct{sec_why_the}). Basis generators for the graded Lie group of the symmetry, i.e., basis vectors of its algebra, can be labeled by continuous spatial coordinates~$\xv$. A step of evolution transformation for the fields equals a sequence of inherent symmetry transformations of the wave function. They are all local. The dynamics is thus necessarily local. 

Alternative formulations of quantum mechanics---for example, nine are described in\ct{nine_formulations_of_QM}---are indistinguishable for most of their testable predictions. Yet many of the formulations are {\em not mutually equivalent\/}. They suggest different objective, observer-independent organization of nature. They differ in their predictions for certain experiments that may be contemplated in principle.  

Due to discreteness of the fundamental distribution~$\dist(Q)$, the number of the objectively existing decoherent branches of an emergent wave function is finite, although possibly huge. Hence the probabilities for the macroscopic outcomes of a quantum process are well-defined. Importantly, the branches whose norm diminishes below a fixed positive threshold cease to exist. Therefore, some outcomes that would be permitted by the standard axiomatic quantum mechanics as an unlikely but allowed quantum fluctuation \emph{do not materialize} as physical reality.

The \emph{Born rule} of quantum mechanics arises dynamically for \emph{typical} emergent states with \emph{extended regular past}. Among such states, the probability for an observer to follow a given branch is proportional to the square of the branch's norm  (\scts{sec_structure} and~\ref{sec_probabilities}).  The relevant norm and related Hermitian product are identified in \scts{sec_structure}--\ref{subsec_Hermitian_product}. 

The described fundamental picture of quantum evolution is thus a middle ground between the strict Everett's view, by which essentially all the futures compatible with the conservation laws are realized, and the quantum-mechanics interpretations that suggest a single branch of macroscopic evolution. In the presented picture, only a relatively narrow subset of the dynamically possible configurations and future outcomes is realized physically. On the other hand, branching into objectively existing alternate futures does occur (within limits). In particular, branching is essential for establishing the dynamical equilibrium that gives rise to the Born rule (\sct{sec_probabilities}).

This study thus provides concrete answers to the questions about origin, interpretation, and self-consistency for a complete quantum view of a physical universe with the known particles and forces, including gravity. For example, it explains what produces unchanging fundamental dynamical laws (\sct{sec_first_observation}) and the Born rule. Besides, this picture---not merely a mathematical construction but a description of structures generic in nature---offers resolution to several conundrums of high-energy physics and cosmology:
\begin{itemize}
\itm
Ultraviolet divergences in perturbatively formulated quantum field theories;
\itm
Apparent preference\ct{Penrose_Difficulties_88,Hollands_Wald_02,Carroll_Tam_10,Steinhardt_infl_Sc_Am,Carroll_Fine-Tuning_14} of Boltzmann-brain states with accidental appearance of intelligent observers over a universe that underwent inflation;
\itm
The central singularity and the end of evaporation of a black hole; its information paradox\ct{Hawking_Breakdown_of_Predictability_76}.
\end{itemize}
Let us review the solutions.

A generic inhabitable physical world is comprised of decoherent branches of emergent wave functions with:
\ben
\itm[(a)]
inherent specific dynamics\\ (fixed, e.g., by local supersymmetry);
\itm[(b)]
the dynamics becoming singular and local evolution ending upon gravitational collapse of sufficiently dense regions;
\itm[(c)]
extended regular past evolution\\ (e.g., during an initial stage of inflation).
\een
Property~(c)~is required for establishing the detailed balance that produces the Born rule. It creates distinct quasiclassical branches of macroscopic evolution (\sct{subsec_need_reg_past}). Given~(c), property~(b) leads to specific initial conditions. They and specific dynamics---property~(a)---are necessary for predictable evolution in a world with internal life (\sct{sec_why_the}). 

Physical laws in the studied \emph{emergent} worlds break down locally during gravitational collapse of sufficiently dense regions\ct{my_bh}. Event horizons then isolate the centers of the formed black holes from the outside space, where the evolution continues regularly. Ref.\ct{my_bh} analyzes in detail full quantum evolution of gravitational black holes in the generically emergent quantum worlds. It explains physics at the central singularity, describes the final stage of the Hawking evaporation, resolves the black hole information paradox, and is free from the firewall paradox.  

To reconcile termination of physical evolution in isolated regions with our common sense, note the following. 
First, the underlying fundamental structure~$\dist(Q)$ is static. It therefore remains non-singular. Temporal states of evolving physical systems are alternate presentations of the smoothed~$\dist(Q)$. Physical evolution is the continuous transformation that relates them. Nothing demands this transformation to be regularly extendable indefinitely. Second, the singularities are covered by event horizons. The emergent dynamics remains well-defined outside them\ct{my_bh}. The emergent spacetime also extends regularly under the horizons until local evolution becomes undefined near the black hole's center\ct{my_bh}. Third, evolution is unitary in the regular regions of the emergent spacetime. Globally for systems with black holes it is non-unitary\ct{my_bh}. Fourth, covariant energy-momentum conservation remains a strict law of the emergent evolution. It follows from the inherent symmetries of the emergent systems\ct{my_bh}.

Cosmological inflation proceeds only when the degrees of freedom for distance scales of the order of the Planck length occupy the ground state (\sct{subsec_inflating_worlds}). That would be highly unlikely for the general unitarily evolving system even among its anthropically suitable states\ct{Penrose_Difficulties_88,Hollands_Wald_02,Carroll_Tam_10,Steinhardt_infl_Sc_Am,Carroll_Fine-Tuning_14}. In contrast, for the described emergent systems it is automatic.  By property~(b) above, only the emergent wave functions that place these degrees of freedom to the ground state can represent a regular physical region. 

For these systems, 3+1 dimensional quantum field theory is \emph{not} superseded around the Planck energy by deeper regular dynamics. An effective field theory with symmetry-preserving counterterms governs the dynamics of the sub-Planckian modes of the emergent fields in the regular regions.  To fix the physical laws to a specific form, i.e., to avoid the paradox highlighted in \sct{sec_first_observation}, the Standard Model should still be extended, e.g., by supergravity, above the currently probed energy.

Our results demonstrate that, contrary to suggestions\ct{Deutsch_probability_decisions99,Barnum_etal_99,Saunders_02,Wallace_02,Zurek_03,Tegmark_MWinContext_08}, the Born rule is not an automatic consequence of unitary evolution of a physical wave function. A material wave function can evolve unitarily and yet represent quantum evolution where the Born rule fails; see the end of \sct{subsec_Born_rule} for various examples.  The probabilities of macroscopic outcomes are determined by the law of evolution of the wave function and an additional law---the Born rule (or another principle leading to it, e.g.,\ct{Carroll_Sebens_14}). This requires an underlying material structure (e.g., the studied generic discrete distribution) that produces the Born rule.\footnote{\label{ftn_superdet}
  An alternative---superdeterministic evolution with appropriate initial conditions---is considered in\ct{tHooft_superdet_chall_17}. It yet requires highly special dynamics and special initial conditions\ct{tHooft_Born_orig_11}. The laws or even possibility of superdeterministic dynamics that may produce physical evolution similar to the one that we observe remain unknown\ct{tHooft_superdet_chall_17}.
}

Unlike the previous theoretical approaches, whether leading to Everett's many-world picture or to a single branch of macroscopic evolution, the presented study does not postulate fundamental physical principles, dynamics, or initial conditions. They all are what they are for the emergent quantum field systems that objectively exist in nature. The dynamics and initial conditions for a generic class of the emergent systems are indistinguishable from those observed in our world. Concluding paragraphs of\ct{my_bh} provide other indications that our universe is perhaps one of the emergent systems. 

As noted above, the described fundamental picture turns out to be a middle ground between the Everett's many-world view and approaches that suggest a unique macroscopic world. Likewise, it appears to lead to a variety of non-communicating universes with different core physical laws, perhaps justifying anthropic explanation\ct{Weinberg_cc_87} for the tiny but non-zero cosmological constant. On the other hand, it is very distant from the proposition to regard any anthropically viable Lagrangian as describing some physical world\ct{Tegmark_mathematical_universe,Linde_foundational_conference}. The possible physical laws are, on the contrary, tightly restricted by the requirement of not changing unpredictably on every dynamical scale. Of the present physical theories, only those with local supersymmetry, e.g., supergravity, fulfill this requirement. This restriction and the arguments of \sct{sec_first_observation} stand regardless of the fundamental origin of quantum principles.

There are innumerable examples of material or mathematical structures that can provide a suitable basic distribution~$\dist(Q)$. If our physical world originates as described, it should be represented simultaneously by all the materially existing information carriers, not limited to the objects of our or other emergent physical worlds. It is therefore difficult to speculate on the origin of the fundamental elements that provide the generic basic distribution~$\dist(Q)$. On the other hand, we may already possess all the tools for theoretical study of the physical systems that emerge generically. Together with data that is accessible experimentally, this may turn out to be sufficient for predicting the probabilities of the outcomes in every feasible physical process.

\begin{acknowledgments}

I thank Pavel Grigoriev, Piotr Puzynia, and Kostas Savvidis for conversations important for developing the presented ideas. I~also thank Dmitry Novikov and Valery Rubakov for comments on the first preprint.  This work was funded partly by private contribution from Denis Danilov.  I~acknowledge essential contribution, including clarifying discussions and partial financial support, by Rabiyat Bashinskaya.

\end{acknowledgments}

\appendix

\section{Invariance of wave function for non-abelian gauge symmetry in curved spacetime}
\label{apx_gauge}

This \apx shows that the wave functions for abelian or non-abelian gauge field theories in curved spacetime are unchanged under gauge transformation of the fields. For this purpose, it is sufficient to treat the spacetime metric~$g_{\mu\nu}(x)$, unaffected by gauge transformation, as an external field. This \apx and the next~one address bosonic fields. Then the arguments of the wave function are the configurations~$f(\xv)$ of the theory fields on a 3-dimensional spacelike slice of spacetime. Appendix~\ref{apx_fermions} will analyze and prove similar statements for the general systems of bosonic and fermionic fields with the general local symmetry, including local supersymmetry. 

Gauge-invariant Lagrangian density\footnote{
	Here, displaying dependence on~$D_\mu\phi$ and~$\phi$ also implies possible dependence on~$(D_\mu\phi)^\dagger$ and~$\phi^\dagger$.
	}
\be
\cL= \cL(D_\mu\phi,F_{\mu\nu}^r,\phi,g_{\mu\nu})
\lb{Lagrangian_density_gauge_generic}
\ee
contains spacetime derivatives of covariant matter fields\footnote{
   In \apxs~\ref{apx_gauge} and~\ref{apx_Hamiltonian} 
   for clarity and 
   relevance to phenomenology we set bosonic matter fields to Lorenz 
   scalars. In the general case, studied in 
   \apx~\ref{apx_fermions}, we regard any covariant 
   matter fields, bosonic or fermionic, of any spin as world 
   scalars~$\Phi^\iota$, possibly carrying spinor or tensor indices of   
   local Lorentz transformations in the tangent space.
}~$\nbrk{\phi\equiv\{\phi^\iota\}}$ only as a part of the covariant derivative 
\be
D_\mu\phi = (\pd_\mu - A_\mu^r t_r)\phi\,.
\lb{D_nonab_def}
\ee
Here~$t_r$ are square matrices, anti-Hermitian for a unitary representation of a compact Lie group. They multiply the column~$\phi$ of the matter fields~$\phi^\iota$ and generate their gauge transformation. For infinitesimal transformation parameters~$\delta\varphi^r(x)$, 
\be
\delta\phi^\iota = \delta\varphi^r t^\iota_{r\,\kappa}\phi^\kappa\,.
\lb{delta_phi_non_abelian}
\ee
The generators~$t_r$ compose a basis of a representation of the gauge Lie algebra:
\be
[t_r,t_s]= f_{rs}{}^t t_t\,,
\lb{def_structure_const}
\ee
where $f_{st}{}^r$ are the algebra's structure constants.

Likewise, the gauge-invariant Lagrangian density~$\cL$ involves spacetime derivatives of the gauge fields~$A_\mu^r$ only through the covariant field-strength tensor~$F_{\mu\nu}^r$, defined as
\be
[D_\mu,D_\nu]= -F_{\mu\nu}^r t_r\,,
\ee
or equivalently,
\be
F_{\mu\nu}^r= \pd_\mu^{} A_\nu^r - \pd_\nu^{} A_\mu^r - A_\mu^s A_\nu^t f_{st}{}^r\,.
\lb{field_strength_tensor}
\ee
For a formal proof of these intuitively natural assertions about the general form of Lagrangian density\rf{Lagrangian_density_gauge_generic} that is invariant under the general local symmetry see ``Theorem on covariant derivatives'' in\ct{Proeyen_deriv_thm_83,Bergshoeff_deriv_thm_86,Proeyen_SUSY_approach}.

Manifestly, $F_{\mu\nu}^r$ and, hence, the Lagrangian do not contain~$\pd^{}_0 A_0^r$. The equation of motion for~$A_0^r$ is then replaced by the \emph{primary constraint}
\be
\pi^0_r\,\psi = \frac{\pd \cL}{\pd (\pd^{}_0 A_0^r)}\,\psi = 0\,.
\lb{primary_constraints_gauge}
\ee
By it,
\be
\frac{\delta}{\delta A_0^r}\,\psi = i\pi^0_r\,\psi = 0\,.
\lb{psi_A0_independence}
\ee
Thus the wave function does not depend on~$A_0^r$, in agreement with the construction of \sct{sec_gauge_fields}.

The constancy of the wave function~$\psi(\phi,A_i^r)$ on the gauge orbits follows from the secondary constraint, to be now derived. 
The Lagrangian density~$\cL$ corresponds to the Hamiltonian density
\be
\cH= \dot\phi\cdot\pi + \dot A_i^r\pi^i_r - \cL\,,
\lb{Hamiltonian_from_Lagrangian_gauge}
\ee
where $\nbrk{\pi\equiv\{\pi_\iota\}}$ are the momenta fields canonically conjugate to $\nbrk{\phi=\{\phi^\iota\}}$. The dot product in\rf{Hamiltonian_from_Lagrangian_gauge} and below denotes contraction over~$\iota$.
By\rf{D_nonab_def}, 
\be
\dot\phi\cdot\pi= D_0\phi\cdot\pi + A_0^r{}j^0_r\,,
\lb{dot_phi_pi}
\ee
where
\be
j^0_r\equiv \frac{\delta \phi}{\delta \varphi^r}\cdot\frac{\pd \cL}{\pd (\pd_0 \phi)}= (t_r\phi)\cdot\pi\,.
\ee
By\rf{field_strength_tensor},
\be
\dot A_i^r\pi^i_r= F_{0i}^r\pi^i_r+(\pd_iA_0^r+A_0^s A_i^t f_{st}{}^r)\pi^i_r\,.
\lb{dot_A_pi}
\ee
For the Lagrangian density\rf{Lagrangian_density_gauge_generic}, the canonical momenta fields are:
\be
\pi_\iota \eqa  \frac{\pd \cL}{\pd \dot\phi^\iota}=\frac{\pd \cL(D_\mu\phi,F_{\mu\nu}^r,\phi,g_{\mu\nu})}{\pd (D_0\phi)^\iota}\,,
\lb{pi_gauge_generic}\\
\pi^i_r\eqa  \frac{\pd \cL}{\pd \dot A_i^r}=2\,\frac{\pd \cL(D_\mu\phi,F_{\mu\nu}^r,\phi,g_{\mu\nu})}{\pd F_{0i}^r}\,.
\lb{pii_gauge_generic}
\ee
Eqs.\rfs{pi_gauge_generic}{pii_gauge_generic} determine~$D_0\phi$ and~$F_{0i}$ as functions of $(\pi,\pi^i_r,D_i\phi,F_{ij}^r,\phi,g_{\mu\nu})$. Then substitution of\rfd{dot_phi_pi}{dot_A_pi} to\rf{Hamiltonian_from_Lagrangian_gauge} yields:
\be
\cH\eqa  \cH_N(\pi,\pi^i_r,D_i\phi,F_{ij}^r,\phi,g_{\mu\nu})+A_0^r\cH_r\,
\lb{Hamiltonian_density_gauge}
\ee
with
\be
\cH_r \eqa j^0_r -\pd_i\pi^i_r+f_{rs}{}^t A_i^s\pi^i_t= j^0_r -(D_i\pi^i)_r\,.\quad
\lb{H_r}
\ee
In\rf{Hamiltonian_density_gauge} we dropped the total derivative~$\pd_i(A_0^r\pi^i_r)$ since it does not contribute to the Hamiltonian~$\nbrk{H=\int d^3x\,\cH}$. For the second equality in\rf{H_r} we remembered that $\pi^i_r$~transform by the adjoint representation, for which the generators~$t_r$ in its covariant derivative\rf{D_nonab_def} are the structure constants~$f_{rs}{}^t$. 

Evolution $\nbrk{\pd_t\psi= -i H\psi}$ with the Hamiltonian density\rf{Hamiltonian_density_gauge} should not introduce the dependence of~$\psi$ on~$A_0^r$, forbidden by the primary constraint\rf{psi_A0_independence}. Hence the operator that multiplies~$A_0^r$ in\rf{Hamiltonian_density_gauge} should annihilate~$\psi$:
\be
\cH_r\psi = 0\,.
\lb{secondary_constraint_gauge}
\ee
It is the \emph{secondary constraint}.

The wave function changes under a gauge transformation of the dynamical fields~$\nbrk{f=(\phi,A_i)}$ as
\be
\delta\psi \eqa \psi(f-\delta f)-\psi(f)\,= \nn\\
\eqa-\int\! d^3x\,\lf(\delta\phi\cdot \frac{\delta\psi}{\delta \phi(\xv)} + \delta A_i^r \frac{\delta\psi}{\delta A_i^r(\xv)}\rt)=\nn\\
\eqa  -\,i\!\int d^3x\,\lf(\delta\phi\cdot \pi + \delta A_i^r \pi^i_r\rt)\psi\,.
\ee
Substitution of the gauge variations~$\delta\phi$ from\rf{delta_phi_non_abelian} and $\delta A_i^r$ from 
\be
\delta A_i^r = \pd_i\delta\varphi^r + A_i^s\delta\varphi^t f_{st}{}^r=(D_i \delta\varphi)^r\,
\lb{delta_gauge_A_nonab}
\ee
yields
\be
\delta\psi = -i\!\int d^3x\,\delta\varphi^r \cH_r\psi\,.
\ee
By the secondary constraint\rf{secondary_constraint_gauge},
\be
\delta\psi = 0\,.
\lb{delta_gauge_psi_is_0}
\ee
Thus the wave function is constant along the orbits of the gauge group.

\section{Generally covariant Hamiltonian of bosonic matter}
\label{apx_Hamiltonian}  

This \apx derives the Hamiltonian density for the general gauge- and diffeomorphism-invariant bosonic action\rf{action_full} with renormalizable non-gravitational part. The \apx also demonstrates that the respective operator $\int d^3x\,N^i \cH_i$ generates Lie translation of the fields in space. 

The Lagrangian density of gauge fields is
\be
\cL^A = \int d^4x \frac{\sqrt{-g}}{4e^2}\,F_{\mu\nu}^r F^{r\,\mu\nu}\,.
\ee
It assumes the standard normalization $\nbrk{\trace(t_r t_s)=\frac12\delta_{rs}}$ for generators of the fundamental representation of the gauge Lie group. 

Direct calculation determines the corresponding Hamiltonian density as:
\be        
\cH^A & = & \dot A^r_i \pi^i_r - \cL^A\, = \nn\\
 & = & N^a\cH^A_\alpha + (\pd_i A^r_0- A^s_iA^t_0 f_{st}{}^r)\pi^i_r \vphantom{\fr{I}{I}}
\lb{HA}
\ee
where $\nbrk{N^a\equiv(N,N^i)}$ are the lapse and shift functions\rf{gADM}, and
\be
\cH^A_N& = &  \fr{e^2}{2\sqrt{\gamma}}\,\pi^i_r\gamma_{ij}\pi^j_r+ 
	\fr{\sqrt{\gamma}}{4e^2}\,\gamma^{ik}\gamma^{jl}F_{ij}^r F_{kl}^r\,,
	\lb{HA_N}\\
\cH^A_i & = & F_{ij}^r\pi^j_r \, . \lb{HAi}
\ee

For scalar fields, the Lagrangian density of the renormalizable gauge- and diffeomorphism-symmetric action is
\be
\cL^\phi = -\sqrt{-g}\lf[\fr{g^{\mu\nu}}2\,(D_\mu\phi) \pdot D_\nu\phi + V(\phi)\rt].
\ee
Here, $\phi$ is a column of real scalar fields~$\phi^\iota$ (expressing complex fields as pairs of real ones), $D_\mu$~is the gauge-covariant derivative\rf{D_nonab_def}, and $V(\phi)$ is a potential that is invariant under gauge transformation of~$\phi$. The corresponding Hamiltonian density equals
\be
\cH^\phi = \dot\phi \cdot \pi - \cL^\phi = 
 N^a \cH^\phi_a + A^r_0{} j^0_r   \quad
\lb{Hphi}
\ee
with
\be
\cH^\phi_N& = &   \fr{\pi\cdot\pi}{2\sqrt{\gamma}}  + \sqrt{\gamma}\lf[\fr12\gamma^{ij} (D_i\phi)\pdot D_j\phi + V(\phi)\rt],\quad\\
{}\cH^\phi_i & = & (D_i\phi) \cdot \pi \, \lb{Hphii} 
\ee
and, by eqs.\rfd{j_phi}{delta_phi_non_abelian},
\be
j^0_r=(t_r\phi)\cdot\pi\,.
\ee
Due to the secondary gauge constraint\rf{secondary_constraint_gauge} and antisymmetry of~$f_{st}{}^r$, the sum of the terms from the right-hand sides of\rf{HA} and\rf{Hphi} that contain~$A_0$ annihilates~$\psi$. The remaining terms and the gravitational part of the Hamiltonian from\rfs{H_g}{Hgi} give for the total Hamiltonian of the action\rf{action_full}:
\be
H\psi = \int d^3x \, N^a \cH_a\psi
\lb{H_apx_B}
\ee
where
\be
\cH_a = \cH_a^g  + \cH_a^A + \cH_a^\phi \, ,  
\ee
with $\cH_a^g\equiv(\cH^{gN},\gamma_{ij}\cH^{gj})$.

Eq.\rf{eta_variation} suggests that $\int d^3x\,N^i \cH_i$, where $i$~runs over $\nbrk{\{1,2,3\}}$, generates Lie translation of the physical fields in space.
Let us verify this for the Hamiltonian density  above. Indeed, by\rf{Hgi} and\rfd{diffeomorphism_transformation_3metric}{pi_ij_def},
\be
\int d^3x \, N^i \cH_i^g = \int d^3x \, N_{(i|j)}\,\pi^{ij} \, = \nn\\*
=\,-i\int d^3x \, L_\Nv\gamma_{ij}\, \fr{\delta}{\delta\gamma_{ij}}\,,
\lb{shift_gamma}
\ee
where~$L_\Nv$ is the Lie derivative along the spatial vector field~$N^i(\xv)$, eq.\rf{diffeomorphism_transformation_def}. 
By\rf{HAi} and\rf{pi_i_def},
\be
\int\! d^3x \, N^i \cH_i^A = \int d^3x \, N^i F_{ij}^r\,\pi^j_r  \, = \qquad\qquad \nn\\*
		= \int d^3x \,[-i L_\Nv A^r_i\, \fr{\delta}{\delta A^r_i} + N^i A^r_i j^0_r +C]\,,~
\ee
where $C\psi = 0$. For the last equality, we integrated the term $\nbrk{\int d^3x \, N^i (-\pd_j A^r_i )\pi^{aj}}$ by parts and applied the secondary gauge constraint\rfs {H_r}{secondary_constraint_gauge}. Finally, from\rf{Hphii} and\rf{pi_def},
\be
\int d^3x \, N^i \cH_i^\phi = \int d^3x \, N^i (\phi_{,i}\cdot\pi - A^r_i j^0_r) \, = \nn\\*
		=\int d^3x \,[-i (L_\Nv\phi)\cdot\fr{\delta}{\delta \phi} - N^i A^r_i j^0_r]\,.
\lb{shift_phi}
\ee
Adding up eqs.\rfs{shift_gamma}{shift_phi}, we obtain
\be
\int d^3x \, N^i \cH_i\psi = -i \int d^3x \!\!\!\!\sum_{f=\gamma_{ij},A^r_i,\phi}\!\! L_\Nv f\, \fr{\delta}{\delta f(\xv)}\,\psi \,.  ~\qquad
\lb{spatial_Lie_generator}
\ee
This confirms that the operator $\nbrk{\int d^3x \, N^i \cH_i}$ generates Lie translations of the fields in space:
\be
e^{-i \int d^3x \, N^i \cH_i}\psi(f)=\psi(f-L_\Nv f) \,.
\ee

\section{Structure of local (super)\,symmetry}
\label{apx_fermions}

\subsection{Covariant fields and connections}

Given a quantum field theory, let $F^\iota$~denote collectively its bosonic and fermionic field operators, $f^\iota(x)$~and~$\chi^\iota(x)$ respectively. Let the theory action be invariant and, hence, its field equations be covariant under a group of local symmetry transformations
\be
F^\iota(\xv) \to \tilde F{}^\iota(\xv) = U^{-1}F^\iota(\xv)\, U\,.
\lb{transf_general_local}
\ee
The unitary operator~$U$ for local transformations\rf{transf_general_local} with infinitesimal variation parameters~$\var^A(\xv)$ should take the form
\be
U = \exp\lf[-i\int d^3x~\var^A(\xv)\,\cH_A(\xv)\rt].
\lb{U_general_local}
\ee
Here $\cH_A(\xv)$ are functions of the field operators $\nbrk{F^\iota(\xv)= (f^\iota(\xv),\chi^\iota(\xv))}$ and of their canonically conjugate momenta $\nbrk{\Pi_\iota(\xv)= (\pi_\iota(\xv),i\chi_\iota^\dagger(\xv))}$, obeying
\be
[F^\iota(\xv),\Pi_\kappa(\yv)\}= i\delta^\iota_\kappa\,\delta^{(3)}(\xv-\yv)\,.
\lb{canonical_brackets}
\ee 
As usual, $[A,B\}$ stands for the anticommutator when both $A$ and $B$ are fermionic operators and for the commutator when at least one of  $A$ or $B$ is bosonic.
The functions $\cH_A(F(\xv),\Pi(\xv))$ in\rf{U_general_local} may contain spatial but not temporal derivatives of their argument fields~$(F,\Pi)$. 

Let~$\Phi$ denote the covariant matter fields, defined in \sct{subsec_supersym}, eq.\rf{covar_fields}. Their change under an infinitesimal symmetry transformation\rfs{transf_general_local}{U_general_local} can be parameterized as
\be
\delta\Phi^\iota(x)\equiv \tilde\Phi^\iota(x)-\Phi^\iota(x) = \var^A(x)\,(\tau_A\Phi)^\iota\,.
\lb{delta_F}
\ee
For covariant matter fields, by their definition, the right-hand side of\rf{delta_F} contains
no spacetime derivatives of the variation parameters~$\var^A(x)$\ct{Proeyen_SUSY_approach,Freedman_Proeyen_book}.

In the \emph{Hamiltonian formulation}, the operators~$(\tau_A\Phi)^\iota$ in\rf{delta_F} are linear combinations of $\{\Phi^\kappa\}$, $\{\Pi_\kappa\}$, and their spatial but not temporal derivatives, all evaluated at the considered~$x$. Their relation to the more conventional Lorentz-covariant form of local symmetry transformation\rf{delta_F} of the Lagrangian formulation is discussed below [after eq.\rf{dotPhi2Pi}].

Let diffeomorphism be symmetry of the theory. Then it is a subgroup of the overall local symmetry group\rfs{transf_general_local}{U_general_local}. Let $\nbrk{\var^A_{\rm diff}(\veps)}$ be the variation parameters of the transformation\rfs{transf_general_local}{U_general_local} that Lie-transports the fields by an infinitesimal displacement vector~$\veps^\mu(x)$, eq.\rf{diffeomorphism_transformation_def}. 

Following\ct{Proeyen_SUSY_approach,Freedman_Proeyen_book}, we regard any covariant matter field~$\Phi$ of any spin as a world scalar, possibly carrying spinor or tensor indices ($\alpha$ or~$a$) for their local Lorentz transformation in the tangent space.
Then an infinitesimal diffeomorphism transformation changes the field by
\be
\delta\Phi=\veps^\mu \partial_\mu\Phi\,. 
\ee
Since it is a special case of the arbitrary local symmetry transformation, this change has the form\rf{delta_F}, with $\nbrk{\var^A=\var^A_{\rm diff}(\veps)}$. Thus
\be
\veps^\mu \partial_\mu\Phi = \var^A_{\rm diff} \tau_A\Phi\,.
\lb{delta_dif_decomposition}
\ee 
For the infinitesimal transformations, the functions $\nbrk{\var^A_{\rm diff}(\veps)}$ are linear:
\be
\var^A_{\rm diff}(x)= \Omega^A_\mu(x)\,\veps^\mu(x)\,.
\lb{OmegaA_mu_def}
\ee
The above relation defines the \emph{connections}~$\Omega^A_\mu$.

Eqs.\rfs{delta_dif_decomposition}{OmegaA_mu_def} should be valid for any infinitesimal~$\veps^\mu(x)$. They hence require
\be
\partial_\mu\Phi = \Omega^A_\mu \tau_A \Phi\,.
\lb{Omega_def}
\ee
This formula relates variation of matter fields in spacetime to their equivalent variation in the tangent and internal spaces.

Denote
\be
\var(x)\eqva \var^A(x)\, \tau_A
\lb{var_def}
\\
\Omega_\mu(x)\eqva \Omega^A_\mu(x)\, \tau_A\,.
\lb{omega_mu_def}
\ee
By\rf{Omega_def}, diffeomorphism displacement over a vector field~$\veps^\mu(x)$ transforms covariant matter fields as
\be
\Phi \diff  \Phi' = e^{\veps^\mu \partial_\mu}\Phi = e^{\veps^\mu \Omega_\mu}\Phi\,.
\lb{Phi_prime}
\ee
The general local symmetry transformation of matter fields\rf{delta_F} in notation\rf{var_def} reads
\be
\Phi(x)\to \tilde\Phi(x) = e^{\var(x)}\Phi(x)\,.
\lb{gen_loc_tr}
\ee

Transform $\Phi$ and $\Phi'$ from\rf{Phi_prime} by~$\var$ and~$\var'$ respectively:
\be
\Phi \to  \tilde\Phi  = e^{\var}\Phi\,, \quad
\Phi' \to \tilde\Phi' = e^{\var'}\Phi\,. 
\lb{Phi_tilde}
\ee
The diffeomorphism transformation\rf{Phi_prime} for the gauge-transformed $\tilde\Phi$ and~$\tilde\Phi'$ of\rf{Phi_tilde} becomes
\be
\tilde\Phi \diff \tilde\Phi'= e^{\var'}e^{\veps^\mu \Omega_\mu}e^{-\var}\,\tilde\Phi =
                e^{\veps^\mu \tilde\Omega_\mu}\tilde\Phi\,.
\lb{tildePhi_diff}
\ee
This equation, with $\nbrk{\var'=\var+\veps^\mu \partial_\mu\var}$, determines the connection in the new gauge:
\be
\tilde\Omega_\mu = \Omega_\mu + \partial_\mu\var + [\var,\Omega_\mu]\,.
\lb{omega_mu_transform}
\ee
Thus the components~$\Omega^A_\mu$ of the connection field\rf{omega_mu_def} change under the general local symmetry transformation as
\be
\delta\Omega^A_\mu= \partial_\mu\var^A + \Omega^C_\mu \var^B f_{BC}{}^A =
(\D_\mu \var)^A.
\lb{delta_Omega_A}
\ee
Here, $f_{BC}{}^A$ are the structure constants\rf{structure_const_def}, and $\D_\mu$~is the covariant derivative of the vector~$\var^A$, belonging to the adjoint representation of the symmetry algebra.

Let us show that $\nbrk{\partial_\mu-\Omega_\mu}$~is a covariant operator.
It changes covariantly under transformation\rf{gen_loc_tr},
\be
\partial_\mu-\Omega_\mu \to \, e^{\var} (\partial_\mu-\Omega_\mu) e^{-\var},
\ee 
whenever
\be
\Omega_\mu \to \, e^{\var}\Omega_\mu e^{-\var}- e^{\var}\partial_\mu e^{-\var}.
\lb{Omega_mu_prime}
\ee
For an infinitesimal~$\var$ the last formula reduces to\rf{omega_mu_transform}, which was already proven.

\subsection{Curvatures}

Let 
\be
\tau_a \equiv D_a \equiv iP_a\,
\ee 
be the generators of translation in the tangent Minkowski space at a spacetime point~$x$. Identify the tangent space with an infinitesimal region of spacetime viewed in a locally inertial frame and gauge with locally vanishing connections. In the general frame and gauge, by\rf{Omega_def} and\rf{Omega_general},
\be
\partial_\mu\Phi = \Omega^A_\mu \tau_A\Phi = e^a_\mu D_a\Phi + A^I_\mu \tau_I\Phi\,.
\lb{Omega_A_expand}
\ee 
Here, $e^a_\mu(x)$~are the vierbein fields and
$A^I_\mu(x)$~are the remaining connection fields. Their index~$I$ ranges over the non-translation directions of the symmetry algebra. For the connections\rf{Omega_general}, $\nbrk{A^I_\mu = (\omega^{ab}_\mu,\chi^\alpha_\mu,A^r_\mu)}$. 

Part~\ref{apx_vierbein} of this \apx extends the ADM framework\rf{ADM} to the vierbein~$e^a_\mu$. This helps identify its gauge and dynamical degrees of freedom. The field~$\Phi$
on the left-hand side of\rf{Omega_A_expand} is a function of the spacetime world coordinates~$x$. Regard~$\Phi$ on the right-hand side of\rf{Omega_A_expand} as a function of the local Minkowski coordinates
\be
\xi^a= e^a_\mu dx^\mu
\lb{Minkowski_coo}
\ee
of the tangent space, in which
\be
D_a = \partial/\partial\xi^a\,.
\lb{D_a_as_dxi}
\ee

The covariant operator
\be
[\partial_\mu-\Omega_\mu,\partial_\nu-\Omega_\nu]= -\,\partial_{[\mu} \Omega_{\nu]} + [\Omega_\mu,\Omega_\nu]\,
\lb{Omegas_commutator}
\ee
expands with\rf{omega_mu_def} and the commutators\rf{structure_const_def} into a linear combination of the generators~$\tau_A$.
By\rf{Omega_def}, the operator\rf{Omegas_commutator} returns zero when acting on any covariant field. Hence all the coefficients of its expansion over~$\tau_A$ vanish. This operator therefore equals zero.
Expanding connections~$\Omega_\mu$ on the right-hand side of\rf{Omegas_commutator} as $\nbrk{e^a_\mu D_a + A^I_\mu \tau_I}$ and equating the result to zero, we calculate \emph{curvature}:
\be                  
\R_{\mu\nu}^A \tau_A \equiv e^a_\mu e^b_\nu [D_a,D_b] &=&
\lb{R_A_def}\\
=\, \partial_{[\mu}^{} e^a_{\nu]}D_a + \partial_{[\mu}^{} A^I_{\nu]}\tau_I \!\! &-&\!
 e^a_{[\mu}A^I_{\nu]}[D_a,\tau_I]- [A^I_\mu \tau_I, A^J_\nu \tau_J].  \nn
\ee
The only nonzero~$[D_a,\tau_I]$ in the semifinal term with the generators\rf{A_list} are 
\be
[iP_a,iJ_{bc}]= \eta_{a[b}iP_{c]}\,.
\ee
The components of the curvature\rf{R_A_def} thus equal:
\be
\R_{\mu\nu}^a \eqa \partial_{[\mu}^{} e^a_{\nu]} - 
 \omega^{ab}_{[\mu} e_{\nu]b}^{} - \bar\chi_{[\mu}\gamma^a \chi_{\nu]}
\lb{R_a}\\
\R_{\mu\nu}^I \eqa \partial_{[\mu}^{} A^I_{\nu]} - A^K_\nu A^J_\mu f_{JK}{}^I.
\lb{R_I}
\ee
Eq.\rf{R_a} used that for the supersymmetry connections and generators
\be
A^\alpha_\mu \tau_\alpha = \bar\chi_\mu Q =
\bar Q\,\chi_\mu\,
\ee
and
\be
\{Q, \bar Q\}=\gamma^a iP_a\,.
\lb{susy_anticom_global}
\ee
[Footnote~\ref{ftnote_susy_norm} on p.~\pageref{ftnote_susy_norm} explains the normalization of $Q$ and~$\bar Q$ for\rf{susy_anticom_global}.]

The spin connections~$\omega^{ab}_\mu$ in both general relativity and supergravity are not independent degrees of freedom. They are functions of the vierbein~$e^a_\mu$ and (in supergravity) of the gravitino field~$\chi^\alpha_\mu$ that fulfill the zero-torsion constraint
\be
\R_{\mu\nu}^a= 0\,,
\lb{zero_torsion_constraint}
\ee
e.g., Refs.\ct{Proeyen_SUSY_approach,Freedman_Proeyen_book}.

\subsection{Constructing a symmetric Hamiltonian}
\label{apx_supersymmetric_evolution}

Consider a locally supersymmetric action\footnote{
  To present the action of supergravity\ct{sugra_orig_76,Freedman_Proeyen_book} in the form\rf{S_supersymmetric_apx}, write its part for the bosonic fields as
\be
S_g =  \int d^4x\lf[\fr{(\det e_\mu^a)}{2}\,R + c^{\mu\nu}_a\R_{\mu\nu}^a \rt].
\nn
\ee
Here $R$~is the Riemann curvature that is composed of~$\R_{\mu\nu}^{ab}$ with the spin connections~$\omega^{ab}_\mu$ regarded as independent fields. In the last term, $c^{\mu\nu}_a$~are non-dynamical auxiliary fields. Then $\nbrk{\delta S/\delta c^{\mu\nu}_a=0}$ yields the zero-torsion constraint\rf{zero_torsion_constraint}. The remaining part of the supergravity action\ct{sugra_orig_76,Freedman_Proeyen_book}, describing gravitino dynamics,
\be
S_{\pert{g}} =  -\int d^4x\,\frac{(\det e_\mu^a)}{2}\,(\bar\chi_\mu \gamma^{\mu\nu\rho})_\alpha\R_{\nu\rho}^\alpha\,,
\nn
\ee
likewise conforms to the form\rf{S_supersymmetric_apx}.
}
\be
S = \int d^4x~\cL(\R_{ab}^A,(D_a\Phi)^\iota,F^\iota),
\lb{S_supersymmetric_apx}
\ee
where $\nbrk{F=(\Phi^\iota,\Omega^A_\mu)}$ and
\be
\R_{ab}^A \tau_A \equiv [D_a,D_b] = e_a^\mu e_b^\nu \R_{\mu\nu}^A \tau_A\,. 
\ee
View its Lagrangian density~$\cL$ in a gauge where at a studied point the connections~$A^I_\mu$ vanish. 
Then $D_a$ and the spacetime derivatives in other generators~$\tau_I$ (e.g., in $Q_\alpha$ for supersymmetry) reduce to\rf{D_a_as_dxi}. I.e., they are derivatives with respect to the local Minkowski coordinates~$\xi^a$\rf{Minkowski_coo}. 

Any of the generators~$\tau_A$ for the symmetries\rf{A_list} contains no more than one derivative:
\be
(\tau_A\Phi)^\iota= t^\iota_{A\,\kappa}\Phi^\kappa + s^{\iota\,a}_{A\,\kappa} \fr{\partial\Phi^\kappa}{\partial\xi^a} \,.
\ee
In the general coordinates and arbitrary gauge $\partial\Phi/\partial\xi^a$ of the last term equals~$D_a\Phi$:
\be
\fr{\partial\Phi^\kappa}{\partial\xi^a} = (D_a\Phi)^\kappa\equiv D^\kappa_{a\,\lambda}\Phi^\lambda\,. 
\ee
Then\rf{Omega_A_expand}~yields
\be
\partial_\mu\Phi= (e^a_\mu+ A^I_\mu s^a_I)D_a\Phi + A^I_\mu t_I\Phi\,.
\ee
This equation lets us express~$D_a\Phi$ in the action\rf{S_supersymmetric_apx} via~$\partial_\mu\Phi$ and the gauge fields~$(e^a_\mu,A^I_\mu)$. 

Solve the equations\footnote{
  This subsection, constructing Hamiltonians with desired symmetries, 
  treats fermionic fields as Grassmann variables.
}

\be
\Pi_\iota = \fr {\partial\cL}{\partial(\partial_0\Phi^\iota)}
\lb{dotPhi2Pi}
\ee
to express~$\partial_0\Phi$ as functions of the canonical momenta~$\Pi$, fields~$F$, and spatial derivatives of $F$ and~$\Pi$ at the current time. This lets us convert a covariant expression for symmetry transformations~$\delta\Phi^\iota$ into\rf{delta_F} where $\tau_A\Phi$ contained no temporal derivatives but included momenta~$\Pi$, along with~$\Phi$ and spatial derivatives.

Let the structure constants~$f_{AB}{}^C$ of the symmetry Lie group contain no spacetime derivatives. Then the Lagrangian density in\rf{S_supersymmetric_apx} with curvatures\rfs{R_a}{R_I} does not contain~$\partial_{0}^{} \Omega^A_{0}$. Hence all~$\Omega^A_0$ are non-dynamical fields. Correspondingly, they are not arguments of the system's wave function. 
The wave function thus satisfies the \emph{primary constraint}
\be
\frac{\pd \cL}{\pd (\pd_0^{} \Omega^A_0)}\,\psi = \Pi^0_A\,\psi = 0\,.
\lb{primary_constraint_A}
\ee

In particular, the wave function does not depend on the temporal components~$\omega^A_0$ of bosonic connections ($\nbrk{\Omega^A_\mu=\omega^A_\mu}$):
\be
\frac{\delta}{\delta \omega^A_0}\,\psi = i\pi^0_A\,\psi = 0\,.
\ee
Likewise, the constraint\rf{primary_constraint_A} states that the temporal components of the fermionic connections ($\nbrk{\Omega^\alpha_\mu=\chi^\alpha_\mu}$) are not physical degrees of freedom. Indeed, since $\nbrk{(\chi^\alpha_0)^2=0}$, in some basis for a given~$\alpha$ and a given mode or spatial point
\be
\chi^\alpha_0 = \lf(\ba{cc}0 & 0 \\c & 0\ea\rt),\quad
\pi_\alpha^0 = \lf(\ba{cc}0 & d \\0 & 0\ea\rt)
\ee
with nonzero $c$ and~$d$. In this basis, the wave function---constrained as $\nbrk{\pi_\alpha^0\psi=0}$ by\rf{primary_constraint_A}---takes the form:
\be
\psi = \lf(\ba{c}\mbox{function of dynamical variables} \\ 0\ea\rt).
\ee

The secondary constraint follows similarly to the case of gauge symmetry [eq.\rf{secondary_constraint_gauge_abelian} or\rf{secondary_constraint_gauge}]. Namely, an infinitesimal step of evolution by the Hamiltonian\rf{H_from_HA} changes the Schrodinger wave function by
\be
\delta\psi = -i\int d^3x\,dt\,\Omega^A_0 \cH_A\, \psi \,.
\lb{delta_psi_evol_A}
\ee
The operators~$\cH_A$ where defined by\rf{U_general_local}; they are independent of the transformation parameters~$\var^A$. Hence $\cH_A$~do not depend on~$\Omega^A_0$ of\rf{Omega0}. We can set them to be also independent of~$\Pi^0_A$ because, due to the primary constraint\rf{primary_constraint_A}, terms with~$\Pi^0_A$ do not contribute to~$\cH_A \psi$. With~$\cH_A$ being independent of both $\Omega^A_0$~and~$\Pi^0_A$, the wave function $\nbrk{\psi+\delta\psi}$ that has evolved by\rf{delta_psi_evol_A} obeys the primary constraint\rf{primary_constraint_A} only under the \emph{secondary constraint}
\be
\cH_A \psi = 0\,.
\lb{secondary_constraint_A}
\ee

Finally, let us determine~$\cH_A$ that generate a given symmetry transformation\rfs{transf_general_local}{U_general_local}.  To preserve the action, a local symmetry transformation\rfs{transf_general_local}{U_general_local} with parameters~$\var^A(x)$ can change the Lagrangian density only by a total divergence:
\be
\delta\cL= \partial_\mu (\var^A K_A^\mu)\,.
\lb{K_mu_def}
\ee
Then the Lagrangian changes by a total time derivative
\be
\delta L= \partial_0(\int d^3x~\var^A K_A^0)\,.
\ee
Let~$Q$ be an operator that generates the considered symmetry transformation as
\be
\delta F^\iota =e^{iQ}F^\iota e^{-iQ}-F^\iota = i[Q,F^\iota]~,\quad
\delta\dot F^\iota = i[Q,\dot F^\iota]  \nn
\ee
for every dynamical field~$\nbrk{F^\iota=(\Phi^\iota,\Omega^A_i)}$. This generator~$Q$ can be shown\ct{Weinberg_bookIII,Gracia_Pons_92} to equal
\be
Q = \int d^3x\,\lf(\delta F^\iota\Pi_\iota-\var^A K_A^0\rt).
\lb{Q_intro}
\ee
With the variations $\nbrk{\delta F^\iota=(\delta\Phi^\iota,\delta\Omega^A_i)}$ from\rf{delta_F} and\rf{delta_Omega_A}, we then obtain:
\be
Q= \int d^3x\,\var^A \cH_A
\ee
where
\be
\cH_A = (\tau_A\Phi)^\iota\Pi_\iota - \D_i\Pi^i_A - K_A^0\,.
\lb{H_A_explicit}
\ee
Here
\be
\D_i\Pi^i_A= \partial_i\Pi^i_A - (-1)^{|AB|}\Omega^B_i f_{AB}{}^C \Pi^i_C\,,
\lb{D_i_Pi}
\ee
where $|AB|$ by definition equals~1 when both $A$~and $B$ are odd (fermionic) directions of the supersymmetry algebra and equals 0~otherwise.

\subsection{Vierbein in the ADM decomposition}
\label{apx_vierbein}

Vierbein vectors~$e^\mu_a(x)$ by definition satisfy
\be
e^\mu_a g_{\mu\nu} e^\nu_b = \eta_{ab}\,,
\ee
where~$\eta_{ab}$ is the canonical Minkowski metric tensor.  Considering the fields on an arbitrary current-time spatial hypersurface, it is convenient to choose one of the vierbein vectors,~$e^\mu_N(\xv)$, as the unit vector~$n^\mu(\xv)$ that is normal to this hypersurface. Then, by the explicit expression\rf{n_mu} for~$n^\mu$ and orthogonality of~$e^\mu_a$ with different~$a$'s,
\be
e^\mu_N \eqva n^\mu = \lf(\fr1\lapse,-\frac{~\shift^i}{\lapse}\rt)
\lb{emu_N}\\
e^\mu_I \eqa  \lf(0,\, e^i_I \rt), \qquad I\in \{1,2,3\}.
\lb{emu_I}
\ee
Since by\rf{emu_I} $\nbrk{e^0_I=0}$, the spatial vectors $\nbrk{{\bm e}_I\equiv\{e^i_I\}}$ are orthonormal:
\be
{\bm e}_I\cdot {\bm e}_J \equiv e^i_I \gamma_{ij} e^j_J= 
e^\mu_I g_{\mu\nu} e^\nu_J = \delta_{IJ}\,.
\ee

With the ADM-parameterized metric tensor~$g_{\mu\nu}$\rf{gADM}, the vierbein reciprocal $\nbrk{e_\mu^a=g_{\mu\nu}\eta^{ab}e^\nu_b}$ equals:
\be
e_\mu^N \eqa -n_\mu = ( \lapse,\, {\bm 0} )
\lb{eN_mu}\\
e_\mu^I \eqa  ( \shift_i e^i_I,\, \gamma_{ij} e^j_I )\,.
\lb{eI_mu}
\ee
The ADM shift vector~$N^i$ matches to the vector
\be
N^I= e_i^I N^i\,
\ee
in the tangent Minkowski space\rf{Minkowski_coo}.
Also remembering that the index~$I$ is raised with $\nbrk{\delta^{IJ}}$,  eqs.\rfs{eN_mu}{eI_mu} for the vierbein reciprocal then become
\be
e_\mu^a = \lf(\ba{cc}N & {\bm 0}\\N^I & e_i^I\ea\rt).
\ee

The spatial metric~$\gamma_{ij}$ and its inverse~$\gamma^{ij}$ are simply expressed via $e_i^I$ and~$e^i_I$ respectively:
\be
\gamma_{ij}\eqa g_{ij} = e_i^a e_j^b \eta_{ab} = \sum_{I=1,2,3}e_i^I e_j^I\,,
\lb{gamma_from_e}\\
\gamma^{ij}\eqa g^{ij}+\frac{~\shift^i\shift^j}{\lapse^2}=\sum_{I=1,2,3}e^i_I e^j_I\,.
\ee
Thus the spatial vectors~$e^i_I$ form a triad of orthonormal basis vectors in space. Their reciprocal $\nbrk{e_i^I=\gamma_{ij} e^j_I}$ form the dual orthonormal basis 1-forms $\nbrk{{\bm de}^I=e_i^Idx^i}$. Of the 9 components of the triad~$e^i_I$, 3~components parameterize its arbitrary orientation in space and 6~are the components of~$\gamma_{ij}$.

\clearpage


\bibliographystyle{apsrev4-1}       
\bibliography{biblio}  

\end{document}